\documentclass[11pt]{article}
\usepackage{graphicx}
\usepackage{amssymb}
\usepackage{amscd}
\usepackage{mathrsfs}
\usepackage{longtable,lscape}
\usepackage{amsthm}
\usepackage{amsfonts}
\usepackage{amsmath}
\usepackage{bbm}
\usepackage{float}
\usepackage{url}
\usepackage{hyperref}
\usepackage[margin=0.5cm,font=footnotesize]{caption}
\usepackage{lineno}
\usepackage[round]{natbib}
\usepackage{appendix}
\usepackage{subfig}

\begin{document}
\title{Quantum Structure in Cognition: Human Language \\ as a Boson Gas of Entangled Words}
\author{Diederik Aerts and Lester Beltran  \vspace{0.5 cm} \\ 
        \normalsize\itshape
        Center Leo Apostel for Interdisciplinary Studies
        \\ 
        \normalsize\itshape
         Brussels Free University, Krijgskundestraat 33 \\ 
        \normalsize\itshape
         1160 Brussels, Belgium \\
        \normalsize
        E-Mails: \url{diraerts@vub.ac.be, diraerts@gmail.com},
           \\ \url{lbeltran@vub.ac.be, lestercc21@yahoo.com}
              	\\
              }
\date{}
\maketitle
\begin{abstract}
\noindent
We model a piece of text of human language telling a story by means of the quantum structure describing a Bose gas in a state close to a Bose-Einstein condensate near absolute zero temperature. For this we introduce energy levels for the words (concepts) used in the story and we also introduce the new notion of `cogniton' as the quantum of human thought. Words (concepts) are then cognitons in different energy states as it is the case for photons in different energy states, or states of different radiative frequency, when the considered boson gas is that of the quanta of the electromagnetic field. We show that Bose-Einstein statistics delivers a very good model for these pieces of texts telling stories, both for short stories and for long stories of the size of novels. We analyze an unexpected connection with Zipf's law in human language, the Zipf ranking relating to the energy levels of the words, and the Bose-Einstein graph coinciding with the Zipf graph. We investigate the issue of `identity and indistinguishability' from this new perspective and conjecture that the way one can easily understand how two of `the same concepts' are `absolutely identical and indistinguishable' in human language is also the way in which quantum particles are absolutely identical and indistinguishable in physical reality, providing in this way new evidence for our conceptuality interpretation of quantum theory. 
\end{abstract}
\medskip
{\bf Keywords}: human language, Bose-Einstein statistics, Zipf's law, identity, indistinguishability, Bose gas

\section{Introduction \label{introduction}}
Human language is a substance consisting of combinations of concepts giving rise to meaning. We will show that a good model for this substance is the one of a gas of entangled bosonic quantum particles such as they appear in physics in the situation close to a Bose-Einstein condensate. In this respect we also introduce the new notion of `cogniton' as the entity playing the same role within human language of the `bosonic quantum particle' for the `quantum gas'. There is a gas of bosonic quantum particles that we all know very well, and that is the electromagnetic field, which we will also briefly call `light', which is a substance of photons. Often we will use `light' as an example and inspiration of how we will talk and reason about human language where `concepts' (words), as `states of the cogniton', are then like `photons of different energies (frequencies, wave lengths)'. With the new findings we present here, we also make an essential and new step forward in the elaboration of our `conceptuality interpretation of quantum theory', where quantum particles are the concepts of a proto-language, in a similar way that human concepts (words), are the quantum particles (cognitons) of human language \citep{aerts2009a,aerts2010a,aerts2010b,aerts2013,aerts2014,aertsetal2018d,aertsetal2019c}.

There are several new results and insights that we will put forward in the coming sections. We summarize them here, referring also to earlier work on which they are built, guaranteeing however that the article is self-contained, so that it is not necessary to have studied these earlier works for understanding its content. The reason we can present here a self-contained theory of human language is because most of our earlier results take a simple and transparent form in the model of a boson gas that we elaborate here for human language. Since we also introduce the basics of the physics of a boson gas, our presentation will remain self-contained also from a physics' perspective. In the article, we will use the terms `words' and `concepts' interchangeably because their difference does not play a role in the aspects of language we study.

We will see that the state of the gas of bosonic quantum particles which we identify explicitly to also be the state of a piece of text such as that of a story is one of very low temperature, i.e. a temperature in the neighborhood of where also the fifth state of matter appears, namely the Bose-Einstein condensate. This means that the interactions between `words', which are the boson particles of language in our description, is mainly one of `quantum superposition' and `quantum entanglement', or more precisely one of `overlapping de Broglie wave functions'. This corresponds well with some of our earlier findings, when studying the combinations of concepts in human language, namely that superposition and entanglement are abundant, and the type of entanglement is deep, namely it also violates additionally to Bell's inequality the marginal laws \citep{aerts2009b,aertsbroekaertgabora2011,aertssozzo2011,aertssozzo2014,aertssozzoveloz2015a,aertssozzoveloz2016,aertsetal2012,aertsetal2018a,aertsetal2018b,aertsetal2018c,aertsetal2019a,aertsetal2019b,aertsarguelles2018,beltrangeriente2019}. 

When we present our model in the next sections, we will see that it contains several new explanations of aspects of human language which we brought up in earlier work. For example, we elaborated an axiomatic quantum model for human concepts, which we called SCoP (state context property system), and in which different exemplars of a specific concept are considered as different states of this concept \citep{gaboraaerts2002,aertsgabora2005a,aertsgabora2005b,aerts2009b,aertsetal2013,aertsgaborasozzo2013}. In the theory of the boson gas for human language that we develop here, we will not only introduce these states explicitly, but also introduce them as eigenstates for specific values of the energy and a detailed energy scale for all the words appearing in a considered piece of text will be introduced. If we compare this with the quantum description of light, it means that the cognitons of our piece of text of human language will radiate their meaning with different frequencies to the human mind, engaging in the meaning of this piece of text.

Let us consider an example of a text, namely the Winnie the Pooh story entitled `In Which Piglet Meets a Haffalump' \citep{milne1926}, to make this introduction of `energy' in our theory of language more concrete. We define the `energy level' of a word (concept, cogniton) in the story by looking at the number of times this word appears in that story. The most often appearing word, namely 133 times, is the concept {\it And} (we will denote concepts or words when they are looked upon as states of a cogniton in italics and with a capital letter, like in our earlier works we have denoted concepts) and we attribute to it (for reasons that will become clearer later) the lowest energy level $E_0$. The second most often appearing word, 111 times, is the concept {\it He}, and we attribute to it the second lowest energy level $E_1$, and so on, till we reach words such as {\it Able}, which only appears once. In other words, if we think of a story as a `gas of bosonic particles' in `thermal equilibrium with its environment', these `number of times of appearance in the story' indicate different energy levels of the particles of the gas, following the `energy distribution law governing the gas', and this is our inspiration for the introduction of `energy' in human language. Remember indeed that each of these words (concepts) is a `state of the cogniton', exactly like different energy levels of photons (different wave lengths of light) are each `states of the photon'. Proceeding in this way we arrive at 452 energy levels for the story `In Which Piglet Meets a Haffalump', the values of which are taken to be
\begin{eqnarray} \label{Ei}
 \{E_i = i \  |\ i \in [0, 1, \ldots, 451, 452]\}
 \end{eqnarray}
We denote $N(E_i)$ the `number of appearances' of the word (concept, cogniton) with energy level $E_i$, and if we denote $n$ the total number of energy levels, we have that
\begin{eqnarray}
N = \sum_{i=0}^n N(E_i)
\end{eqnarray}
is the total number of words (concepts, cognitons) of the considered piece of text, which is 2655 for the story `In Which Piglet Meets a Haffalump'.

For each of the energy levels $E_i$, $N(E_i)E_i$ is the amount of energy `radiated' by the story `In Which Piglet Meets a Haffalump' with the `frequency or wave length' connected to this energy level. For example, the energy level $E_{54} = 54$ is populated by the concept {\it Thought} and the word {\it Thought} appears $N(E_{54})=10$ times in the story `In Which Piglet Meets a Haffalump'. Each of the 10 appearances of {\it Thought} radiates with energy value 54, which means that the total radiation with the wave length connected to {\it Thought} of the story `In Which Piglet Meets a Haffalump' equals $N_{54}E_{54} = 10 \cdot 54 = 540$.

The total energy $E$ radiated by the considered piece of text is therefore 
\begin{eqnarray}
E = \sum_{i=0}^n N(E_i) E_i
\end{eqnarray}
For the story `In Which Piglet Meets a Haffalump' we have $E = 242891$. Let us represent now some of the other findings that we will describe more in detail in the following sections.

When we applied the Bose-Einstein distribution
\begin{eqnarray}
N(E_i) = {1 \over {Ae^{{E_i \over B}}-1}}
\end{eqnarray}
to model the data we collected on the story `In Which Piglet Meets a Haffalump', determining the parameters $A$ and $B$ by the two requirements
\begin{eqnarray}
\sum_{i=0}^n N(E_i) =  2655 \quad \sum_{i=0}^n N(E_i) E_i = 242891
\end{eqnarray}
we found an almost complete fit with the data (see Section \ref{languagebosegas}, Table \ref{piglethaffalunmp}, Figure \ref{piglethaffalunmpgraphpiglethaffalunmploggraph} (a), Figure \ref{piglethaffalunmpgraphpiglethaffalunmploggraph} (b) and Figure \ref{piglethaffalunmpenergygraph}. We tested numerous other texts, short stories (see Section \ref{storiesboseeinsteincondensates}, Table \ref{magicshop}, Figure \ref{magicshopgraphmagicshoploggraph}, Figure \ref{magicshopenergygraph}) and long stories of the size of novels (see Section \ref{boseeinsteinzipf}, Figure \ref{gulliverstravelsloggraphgulliverstravelspowerloggraph} (b)), and each time it showed that a modeling by means of a Bose-Einstein statistical energy distribution, like explained above, gives rise to an almost complete fit with the data.

We started this investigation with the idea that `concepts within human language behave like bosonic entities', an idea we expressed earlier as one of the basic pieces of evidence for the `conceptuality interpretation' \citep{aerts2009a}. The origin of the idea is the simple direct understanding that if one considers, for example, the concept combination {\it Eleven Animals}, then, on the level of the `conceptual realm' each one of the eleven animals is completely `identical with' and `indistinguishable from' each other of the eleven animals. It is also a simple direct understanding that in the case of `eleven physical animals', there will always be differences between each one of the eleven animals, because as `objects' present in the physical world, they have an individuality, and as individuals, with spatially localized physical bodies, none of them will be really identical with the other ones, which means that each one of them will also always be able to be distinguished from the others. Even if all the animals are horses, simply because they are `objects' and not `concepts', they will not be completely identical and hence they will be distinguishable. The idea is that it is `this not being completely identical and hence being distinguishable' which makes the Maxwell-Boltzmann statistics being applicable to them. However, when we consider `eleven animals' as concepts, such that their ontological nature is conceptual, they are all `completely identical and hence intrinsically indistinguishable'. Within the conceptuality interpretation of quantum theory, where we put forward the hypothesis that quantum entities are `conceptual' and hence are not `objects', their `being completely identical and hence intrinsically indistinguishable', would also be due to their being conceptual instead of objectual entities.

In earlier work we already investigated this idea by looking at simple combinations of concepts with numerals, such as indeed {\it Eleven Animals} and then considering two states of {\it Animal}, namely {\it Cat} and {\it Dog}. We then checked whether the twelve different exemplars of them that form in these two states, namely {\it Eleven Dogs}, {\it One Cat And Ten Dogs}, {\it Two Cats And Nine Dogs}, \ldots, {\it Ten Cats And One Dog}, {\it Eleven Cats}, in their appearance in texts follow a Maxwell-Boltzmann or rather a Bose-Einstein statistical pattern. In a less convincing way because of a collection of limited data \citep{aerts2009a,aertssozzoveloz2015b}, but with an abundance of data and very convincingly \citet{beltran2019}, it was shown that indeed the Bose-Einstein statistics delivers a better model for the data as compared to the Maxwell-Boltzmann statistics.

The result that we put forward in the present article, namely that the Bose-Einstein statistics as explained above models entire texts of any size, is a much stronger one, although it expresses the same idea. Consider any text, and then consider two instances of the word {\it Cat} appearing in the text, if then one of the concepts {\it Cat} is exchanged with the other concept {\it Cat}, absolutely nothing changes in the text. Hence, a text contains a perfect symmetry for the exchange of cognitons (concepts, words) in the same state. This is not true for physical reality and its physical objects. Suppose one considers a physical landscape where two cats are within the landscape, exchanging the two cats will always change the landscape, because the cats are not identical and are distinguishable as physical objects. If we introduce a quantum description of the text, the wave function must be invariant for the exchange of the two cats, which would again be not the case if the wave function would describe the physical landscape containing  two cats as objects. This is the result we will present in Section \ref{languagebosegas}.

Section \ref{storiesboseeinsteincondensates} is devoted to a self-contained presentation of the phenomenon of Bose-Einstein condensation in physics. We illustrate the different aspects of the Bose-Einstein condensation valuable for our discussion, by means of two examples of Bose gases, the rubidium 87 atom gas and the sodium atom gas, that also originally where the first ones to be used to realize a Bose-Einstein condensate \citep{andersonetal1995,davisetal1995}. We compare the Bose-Einstein condensates of the gases and how their energy level distribution is modeled by the Bose-Einstein distribution function with our Bose-Einstein modeling of pieces of texts of stories and point out the points of correspondence.

Another finding that we will put forward, in Section \ref{boseeinsteinzipf}, was completely unexpected. The method of attributing an energy level to a word depending on the number of appearances of the word in a text, introduces the typical ranking considered in the well-known Zipf's law analysis of this text \citep{zipf1935,zipf1949}. When we look at the $\log/\log$ graph of ranking in function of the number of appearances, we indeed see the linear function, or a slight deviation of it, which represents the most common version of Zipf's law. Zipf's law is an experimental law, which has not yet been given any theoretical foundation, hence perhaps our finding, of its unexpected connection with Bose-Einstein statistics, might provide such a foundation. We also show, in Section \ref{boseeinsteinzipf}, how the connection with Zipf's law allows us to develop more in depth the Bose-Einstein model of texts of different sizes, short stories and long stories of the size of novels.

In Section \ref{identifyindistinguishability}, we reflect about the issue of `identity and indistinguishability' 
from the perspective we developed in the foregoing sections, taking into account the conundrum this issue actually still is in quantum theory with respect to quantum particles \citep{diekslubberdink2019}. Confronting the theoretical view where bosons and fermions are considered to be identical and indistinguishable even if they are in different states, we note that experimentalists take another stance in this respect considering, for example, photons of different frequencies as distinguishable. A recent experiment shows that if this experimentally accepted possibility to distinguish them is erased by means of a quantum eraser, these different frequency photons behave as indistinguishable \citep{zhaoetal2014}. This makes us put forward the proposal that `the way in which we clearly see and understand the identity and indistinguishability of concepts (words, cognitons) in human language' is also `the way in which identity and indistinguishability for quantum particles can be understood'. More specifically, it shows that `identity and indistinguishability' are contextual notions for a quantum particle, depending on the way a measuring apparatus or a heat bath interacts with the quantum particle, similarly  to how `identity and indistinguishability' are contextual notions for a human concept, depending on how a mind interacts with the concept. We elaborate with examples this new way of interpreting `identity and indistinguishability' and show how it is a strong confirmation of our conceptuality interpretation of quantum theory.

\section{Human language as a Bose gas \label{languagebosegas}}
Let us consider again the Winnie the Pooh story `In Which Piglet Meets a Haffalump' as published in \citet{milne1926}. In Table \ref{piglethaffalunmp}, we have presented the list of all words that appear in the story (in the column `Words concepts cognitons'), with their `number of appearances' (in the column `Appearance numbers $N(E_i)$'), ordered from lowest energy level to highest energy level (in the column `Energy levels $E_i$'), where the energy levels are attributed according to these numbers of appearances,  lower energy levels to higher number of appearances, and their values are given as proposed in (\ref{Ei}).

The word {\it And} is the most often appearing word, namely 133 times, hence the cognitons in this state populate the ground state energy level $E_0$, which as per (\ref{Ei}) we put equal to zero. The word {\it He} is the second most often appearing word, namely 111 times, hence the cognitons in this state populate the first energy level $E_1$, which following (\ref{Ei}) we put equal to 1. Hence, the `words', their `energy levels' and their `numbers of appearances' are in the first three columns of Table \ref{piglethaffalunmp}.

The question can be asked `what is the unity of energy in this model that we put forward?', is the number `1' that we choose for energy level $E_1$ a quantity expressed in joules, or in electronvolts, or still in another unity? This question gives us the opportunity to reveal already one of the very new aspects of our approach. Energy will not be expressed in `${\rm kg m}^2/{\rm s}^2$' like it is the case in physics. Why not? Well, a human language is not situated somewhere in space, like we believe it to be the case with a physical boson gas of atoms, or a photon gas of light. Hence, `energy' is here in our approach a basic quantity, and if we manage to introduce -- this is one of our aims in further work -- what the `human language equivalent' of `physical space' is, then it will be oppositely, namely this `equivalent of space' will be expressed in unities where `energy appears as a fundamental unit'. Hence, the `1' indicating that `{\it He} radiates with energy 1', or `the cogniton in state {\it He} carries energy 1', stands with a basic measure of energy, just like `distance (length)' is a basic measure in `the physics of space and objects inside space', not to be expressed as a combination of other physical quantities. We used the expressions `{\it He} radiates with energy 1', and `the cogniton in state {\it He} carries energy 1', and we will use this way of speaking about `human language within the view of a boson gas of entangled cognitons that we develop here', in similarity with how we speak in physics about light and photons. 

The words {\it The}, {\it It}, {\it A} and {\it To}, are the four next most often appearing words of the Winnie the Pooh story, and hence the energy levels $E_2$, $E_3$, $E_4$ and $E_5$ are populated by cognitons respectively in the states {\it The}, {\it It}, {\it A} and {\it To} carrying respectively 2, 3, 4 and 5 basic energy units.
Hence, the first three columns in Table \ref{piglethaffalunmp} describe the experimental data that we extracted from the Winnie the Pooh story `In Which Piglet Meets a Haffalump'. As we said, the story contains in total 2655 words, which give rise to 542 energy levels, where energy levels are connected with words, hence different words radiate with different energies, and the size of the energies are determined by `the number of appearances of the words in the story', the most often appearing words being states of lowest energy of the cogniton and the least often appearing words being states of highest energy of the cogniton. In Table \ref{piglethaffalunmp}, we have not presented all 542 energy levels, because that would lead to a too long table, but we have presented the most important part of the energy spectrum, with respect to the further aspects we will point out.

More concretely, we have represented the range from energy level $E_0$, the ground state of the cogniton, which is the cogniton in state {\it And}, to energy level $E_{78}$, which is the cogniton in state {\it Put}. Then we have represented the energy level from $E_{538}$, which is the cogniton in state {\it Whishing}, to the highest energy level $E_{542}$ of the Winnie the Pooh story, which is the cogniton in state {\it You've}.

These last five highest energy levels, from $E_{538}$ to $E_{542}$, corresponding respectively to the cogniton in states {\it Whishing}, {\it Word}, {\it Worse}, {\it Year} and {\it You've}, all have a number of appearance of `one time' in the story. They do however radiate with different energies, but the story is not giving us enough information to determine whether {\it Whishing} is radiating with lower energy as compared to {\it Year} or vice versa. Since this does not play a role in our actual analysis, we have ordered them alphabetically. So, different words which radiate with different energies that appear an equal number of times in this specific Winnie the Pooh story will be classified from lower to higher energy level alphabetically.

In the column `Energies from data $E(E_i)$', we represent $E(E_i)$, the `amount of energy radiated by the Winnie the Pooh story by the cognitons of a specific word, hence of a specific energy level $E_i$'. As we mentioned already in the previous section, the formula for this amount is given by
\begin{eqnarray}
E(E_i) = N(E_i) E_i
\end{eqnarray}
the product of the number $N(E_i)$ of cognitons in the state of the word with energy level $E_i$ multiplied by the amount of energy $E_i$ radiated by such a cogniton in that state. In the last row of Table \ref{piglethaffalunmp}, we give the {\it Totalities}, namely in the column `Appearance numbers $N(E_i)$' of this last row the total number of words 
\begin{eqnarray}
\sum_{i=0}^n N(E_i) = N = 2655
\end{eqnarray}
and in the column `Energies from data $E(E_i)$' of the last row we give the total amount of energy 
\begin{eqnarray}
\sum_{i=0}^n E(E_i) = \sum_{i=0}^n N(E_i) E_i = E = 242891
\end{eqnarray}
radiated by the Winnie the Pooh story `In Which Piglet Meets a Haffalump'. Hence, columns `Words concepts cognitons', `Energy levels $E_i$', Appearance numbers $N(E_i)$ and `Energies from data $E(E_i)$' contain all the experimental data of the Winnie the Pooh story `In Which Piglet Meets a Haffalump'.

\small
\begin{longtable}{p{1.5cm}p{1.5cm}p{1.5cm}p{1.5cm}p{1.5cm}p{1.5cm}p{1.5cm}p{1.5cm}}
\label{piglethaffalunmp}
  Words concepts cognitons & Energy levels $E_i$ & Appearance numbers $N(E_i)$ & Bose-Einstein modeling &  Maxwell-Boltzmann modeling & Energies from data $E(E_i)$ & Energies Bose-Einstein & Energies Maxwell-Boltzmann  \\
  \hline
      {\it And}  & 0 & 133 & 129.05 & 28.29 & 0    & 0     & 0  \\
         {\it He} & 1 & 111 & 105.84 & 28.00 & 111 & 105.84 & 28.00  \\
        {\it The} & 2 & 91  & 89.68   & 27.69 & 182 & 179.36 & 55.38  \\
            {\it It} & 3 & 85  & 77.79   & 27.40 & 255 & 233.36 & 82.19  \\
            {\it A} & 4 & 70  & 68.66   & 27.11 & 280 & 274.65 & 108.43  \\
           {\it To} & 5 & 69  & 61.45   & 26.82 & 345 & 307.23 & 234.09  \\
        {\it Said} & 6 & 61  & 55.59   & 26.53 & 366 & 333.55 & 159.20  \\
        {\it Was} & 7 & 59  & 50.75   & 26.25 & 413 & 355.24 & 183.76  \\
      {\it Piglet} & 8 & 47  & 46.68   & 25.97 & 376 & 373.40 & 207.78  \\
              {\it I} & 9 & 46  & 43.20   & 25.70 & 414 & 388.82 & 231.27  \\
        {\it That} & 10 & 41 & 40.21  & 25.42 & 410 & 402.05 & 254.24  \\
       {\it Pooh} & 11 & 40 & 37.59  & 25.15 & 440 & 413.52 & 276.69  \\
         {\it Of}   & 12 & 39 & 35.30  & 24.89 & 468 & 423.55 & 298.64  \\
         {\it Had} & 13 & 28 & 33.26  & 24.62 & 364 & 432.38 & 320.09  \\
      {\it Would} & 14 & 26 & 31.44  & 24.36 & 364 & 440.21 & 341.05  \\
         {\it As}    & 15 & 25 & 29.81  & 24.10 & 375 & 447.19 & 361.53  \\
          {\it In}    & 16 & 25 & 28.34  & 23.86 & 400 & 453.44 & 381.53  \\
        {\it But}    & 17 & 23 & 27.00  & 23.59 & 391 & 459.07 & 401.07  \\
 {\it Haffalump}& 18 & 23 & 25.79  & 23.34 & 414 & 464.15 & 420.15  \\
        {\it His}    & 19 & 23 & 24.67  & 23.09 & 437 & 468.77 & 438.78  \\
      {\it Very}    & 20 & 23 & 23.65  & 22.85 & 460 & 472.96 & 456.97  \\
       {\it You}     & 21 & 23 & 22.70  & 22.61 & 483 & 476.79 & 474.72  \\
      {\it Then}    & 22 & 21 & 21.83  & 22.37 & 462 & 480.30 & 492.05  \\
    {\it Honey}    & 23 & 20 & 21.02  & 22.13 & 460 & 483.51 & 508.95  \\
          {\it So}    & 24 & 20 & 20.27  & 21.89 & 480 & 486.47 & 525.43  \\
          {\it Up}    & 25 & 20 & 19.57  & 21.66 & 500 & 489.19 & 541.51  \\
       {\it They}    & 26 & 19 & 18.91  & 21.43 & 494 & 491.71 & 557.19  \\
             {\it If}    & 27 & 18 & 18.30  & 21.20 & 486 & 494.03 & 572.47  \\
          {\it Jar}    & 28 & 18 & 17.72  & 20.98 & 504 & 496.18 & 587.37  \\
      {\it There}    & 29 & 18 & 17.18  & 20.75 & 522 & 498.18 & 601.89  \\
            {\it At}    & 30 & 17 & 16.67  & 20.53 & 510 & 500.03 & 616.03  \\
           {\it Be}    & 31 & 15 & 16.19  & 20.32 & 465 & 501.75 & 629.80  \\
          {\it Got}    & 32 & 15 & 15.73  & 20.10 & 480 & 503.34 & 643.21  \\
         {\it Just}    & 33 & 15 & 15.30  & 19.89 & 495 & 504.83 & 656.26  \\
        {\it What}    & 34 & 15 & 14.89  & 19.68 & 510 & 506.22 & 668.97  \\
 {\it Christopher} & 35 & 14 & 14.50  & 19.47 & 490 & 507.51 & 681.33  \\
             {\it This} & 36 & 14 & 14.13  & 19.26 & 504 & 508.71 & 693.35  \\
             {\it Trap} & 37 & 14 & 13.78  & 19.06 & 518 & 509.83 & 705.03  \\
           {\it About} & 38 & 13 & 13.44  & 18.85 & 494 & 510.88 & 716.40  \\
                {\it All} & 39 & 13 & 13.12  & 18.65 & 507 & 511.86 & 727.44  \\
         {\it Should} & 40 & 13 & 12.82  & 18.45 & 520 & 512.77 & 738.17  \\
               {\it For} & 41 & 12 & 12.53  & 18.26 & 492 & 513.62 & 748.59  \\
              {\it Like} & 42 & 12 & 12.25  & 18.06 & 504 & 514.41 & 758.70  \\
            {\it Robin} & 43 & 12 & 11.98  & 17.87 & 516 & 515.15 & 768.51  \\
               {\it See} & 44 & 12 & 11.72  & 17.68 & 528 & 515.84 & 778.03  \\
            {\it When} & 45 & 12 & 11.48  & 17.49 & 540 & 516.48 & 778.26  \\
            {\it Down} & 46 & 11 & 11.24  & 17.31 & 506 & 517.08 & 796.20  \\
   {\it Heffalumps} & 47 & 11 & 11.01  & 17.12 & 517 & 517.64 & 804.87  \\
              {\it With} & 48 & 11 & 10.79  & 16.94 & 528 & 518.15 & 813.26  \\
                 {\it Do} & 49 & 10 & 10.58  & 16.76 & 490 & 518.63 & 821.39  \\
                 {\it Go} & 50 & 10 & 10.38  & 16.58 & 500 & 519.08 & 829.25  \\
                 {\it Off} & 51 & 10 & 10.19  & 16.41 & 510 & 519.49 & 836.85  \\
                 {\it On} & 52 & 10 & 10.00  & 16.23 & 520 & 519.87 & 844.19  \\
             {\it Think} & 53 & 10 & 9.82  & 16.06 & 530 & 520.22 & 851.29  \\
         {\it Thought} & 54 & 10 & 9.64  & 15.89 & 540 & 520.54 & 858.13  \\
              {\it More} & 55 & 9   & 9.47    & 15.72 & 495 & 520.83 & 864.74  \\
                  {\it No} & 56 & 9   & 9.31    & 15.56 & 504 & 521.10 & 871.11  \\
                 {\it Out} & 57 & 9   & 9.15    & 15.39 & 513 & 521.35 & 877.25  \\
                   {\it Pit} & 58 & 9   & 8.99    & 15.23 & 522 & 521.57 & 883.15  \\
               {\it Went} & 59 & 9   & 8.84    & 15.07 & 531 & 521.77 & 888.84  \\
               {\it Don't} & 60 & 8   & 8.70    & 14.91 & 480 & 521.95 & 894.30  \\
               {\it Good} & 61 & 8   & 8.56    & 14.75 & 488 & 522.11 & 899.55  \\
               {\it Head} & 62 & 8   & 8.43    & 14.59 & 496 & 522.25 & 904.58  \\
               {\it Know} & 63 & 8   & 8.29    & 14.44 & 504 & 522.37 & 909.41  \\
                   {\it Oh} & 64 & 8   & 8.16    & 14.28 & 512 & 522.48 & 914.03  \\
                {\it Right} & 65 & 8   & 8.04    & 14.13 & 520 & 522.57 & 918.45  \\
                  {\it Well} & 66 & 8   & 7.92    & 13.98 & 528 & 522.64 & 922.67  \\
                  {\it Bed} & 67 & 7   & 7.80    & 13.83 & 469 & 522.70 & 926.70  \\
                {\it Could} & 68 & 7   & 7.69    & 13.68 & 476 & 522.74 & 930.54  \\
                 {\it Deep} & 69 & 7   & 7.58    & 13.54 & 483 & 522.77 & 934.20  \\
                    {\it Did} & 70 & 7   & 7.47    & 13.40 & 490 & 522.78 & 937.67  \\
 {\it \underline {First}} & {\bf \underline {71}} & 7   & 7.36    & 13.25 & 497 & {\bf \underline{522.79}} & 940.96  \\ 
                  {\it Have} & 72 & 7   & 7.26    & 13.11 & 504 & 522.78 & 944.08  \\ 
                   {\it Help} & 73 & 7   & 7.16    & 12.97 & 511 & 522.76 & 947.02  \\ 
               {\it Himself} & 74 & 7   & 7.06    & 12.84 & 518 & 522.72 & 949.79  \\ 
                    {\it How} & 75 & 7   & 6.97    & 12.70 & 525 & 522.68 & 952.40  \\
                {\it Looked} & 76 & 7   & 6.88    & 12.56 & 532 & 522.63 & 954.85  \\
                     {\it Now} & 77 & 7   & 6.79    & 12.43 & 539 & 522.56 & 957.13  \\
                       {\it Put} & 78 & 7   & 6.70    & 12.30 & 546 & 522.49 & 959.27  \\
        \ldots &  \ldots &  \ldots   &  \ldots    & \ldots &  \ldots &  \ldots &  \ldots  \\
                                   &     &      &      &     &      &      &       \\
                                   &     &      &      &     &      &      &       \\
       \ldots &  \ldots &  \ldots   &  \ldots    & \ldots &  \ldots &  \ldots &  \ldots  \\
                 {\it Wishing} & 538 & 1   & 0.67    & 0.09 & 538 & 359.92 & 48.65  \\
                     {\it Word} & 539 & 1   & 0.67    & 0.09 & 539 & 359.58 & 48.22  \\
                   {\it Worse} & 540 & 1   & 0.67    & 0.09 & 540 & 359.24 & 47.80  \\
                      {\it Year} & 541 & 1   & 0.66    & 0.09 & 541 & 358.90 & 47.38  \\
                   {\it You've} & 542 & 1   & 0.66    & 0.09 & 542 & 358.55 & 46.96  \\
                   \hline
                                     &      &2655   &2655.00    &2654.96 & 242891 & 242891.01 & 242889.76  \\
\caption{{\footnotesize An energy scale representation of the words of the Winnie the Pooh story `In Which Piglet Meets a Haffalump' by A. A. Milne as published in \citet{milne1926}. The words are in the column `Words concepts cognitons' and the energy levels are in the column `Energy levels $E_i$', and are attributed according to the `numbers of appearances' in the column `Appearance numbers $N(E_i)$', such that  lower energy levels correspond to higher order of appearances, and the value of the energy levels is determined according to (\ref{Ei}). The `amounts of energies radiated by the words of energy level $E_i$' are in the column `Energies from data $E(E_i)$'. In the columns `Bose-Einstein modeling', `Maxwell-Boltzmann modeling', `Energies Bose-Einstein' and `Energies Maxwell-Boltzmann' are respectively the predicted values of the Bose-Einstein and the Maxwell-Boltzmann model of the `numbers of appearances', and of the `radiated energies'.}}
\end{longtable}
\normalsize
\noindent
In columns `Bose-Einstein modeling' and `Maxwell-Boltzmann modeling' of Table \ref{piglethaffalunmp}, we give the values of the populations of the different energy states for,
respectively, a Bose-Einstein and a Maxwell-Boltzmann model of the data of the considered story. Let us explain what these two models are. 
As we recalled in the introduction, the Bose-Einstein distribution function is given by
\begin{eqnarray} \label{boseeinsteindistribution}
N(E_i) = {1 \over {Ae^{{E_i \over B}}-1}}
\end{eqnarray}
where $N(E_i)$ is the number of bosons obeying the Bose-Einstein statistics in energy level $E_i$ and $A$ and $B$ are two constants  that are determined by expressing that the total number of bosons equals the total number of words, and that the total energy radiated equals the total energy of the Winnie the Pooh story `In Which Piglet Meets a Haffalump', hence by the two conditions
\begin{eqnarray} \label{BEconstraint01}
&&\sum_{i=0}^n {1 \over {Ae^{{E_i \over B}}-1}} = N = 2655 \\ \label{BEconstraint02}
&&\sum_{i=0}^n {E_i \over {Ae^{{E_i \over B}}-1}} = E = 242891
\end{eqnarray}
We remark that the Bose-Einstein distribution function is derived in quantum statistical mechanics for a gas of bosonic quantum particles where the notions of `identity and indistinguishability' play the specific role they are attributed in quantum theory \citep{huang1987}. We will come back to this in Section \ref{identifyindistinguishability}, when we will analyze what our findings and our aim are, given our conceptuality interpretation of quantum theory, to understand better how `identity and indistinguishability' can be explained for a physical Bose gas using our understanding of it in human language.

\begin{figure}%
    \centering
    \subfloat[Numbers of appearances distribution graphs]{{\includegraphics[width=8cm]{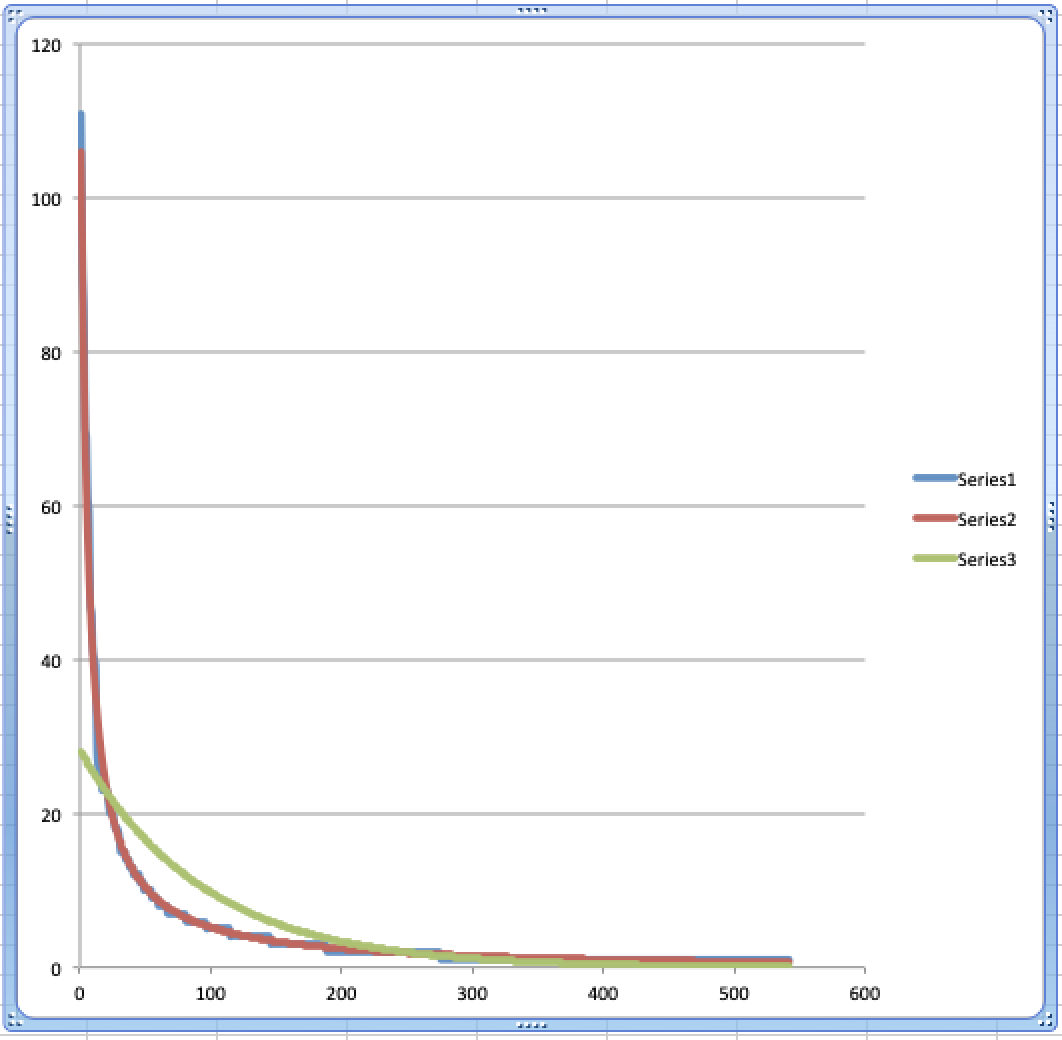} }}%
    \qquad
    \subfloat[$\log/\log$ graphs of numbers of appearances distributions]{{\includegraphics[width=8cm]{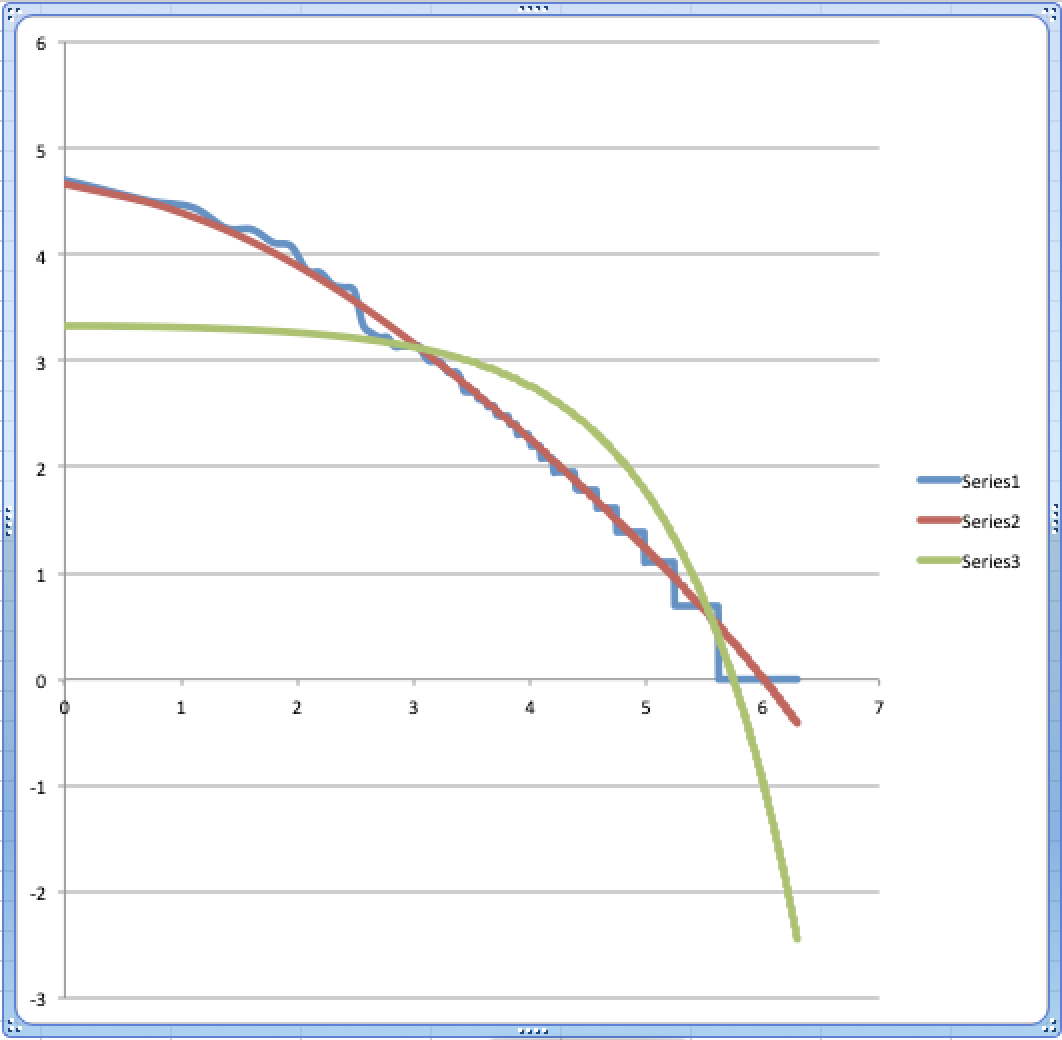} }}%
    \caption{In (a) we represent the `number of appearances' of words in the Winnie the Pooh story `In Which Piglet Meets a Haffalump' \citep{milne1926}, ranked from lowest energy level, corresponding to the most often appearing word, to highest energy level, corresponding to the least often appearing word as listed in Table \ref{piglethaffalunmp}. The blue graph (Series 1) represents the data, i.e. the collected numbers of appearances from the story (column `Appearance numbers $N(E_i)$' of Table \ref{piglethaffalunmp}), the red graph (Series 2) is a Bose-Einstein distribution model for these numbers of appearances (column `Bose-Einstein modeling' of Table \ref{piglethaffalunmp}), and the green graph (Series 3)  is a Maxwell-Boltzmann distribution model (column `Maxwell-Boltzmann modeling' of Table \ref{piglethaffalunmp}). In (b) we represent the $\log / \log$ graphs of the `numbers of appearances' and their Bose-Einstein and Maxwell-Boltzmann models. The red and blue graphs coincide almost completely in both (a) and (b) while the green graph does not coincide at all with the blue graph of the data. This shows that the Bose-Einstein distribution is a good model for the numbers of appearances, while the Maxwell-Boltzmann distribution is not.}%
    \label{piglethaffalunmpgraphpiglethaffalunmploggraph}%
\end{figure}

Since we want to show the validity of the Bose-Einstein statistics for concepts in human language, we compared our Bose-Einstein distribution model with a Maxwell-Boltzmann distribution model, hence we introduce also the Maxwell-Boltzmann distribution explicitly. It is the distribution described by the following function 
\begin{eqnarray} \label{maxwellboltzmanndistribution}
N(E_i) = {1 \over Ce^{{E_i \over D}}}
\end{eqnarray} 
where $N(E_i)$ is the number of classical identical particles obeying the Maxwell-Boltzmann statistics in energy level $E_i$ and $C$ and $D$ are two constants that will be determined, like in the case of the Bose-Einstein statistics, by the two conditions
\begin{eqnarray} \label{MBconstraint01}
&&\sum_{i=0}^n {1 \over Ce^{{E_i \over D}}} = N = 2655 \\ \label{MBconstraint02}
&&\sum_{i=0}^n {E_i \over Ce^{{E_i \over D}}} = E = 242891
\end{eqnarray}
The Maxwell-Boltzmann distribution function is derived for `classical identical and distinguishable' particles, and can also be shown in quantum statistical mechanics to be a good approximation if the quantum particles are such that their `the Broglie waves' do not overlap  \citep{huang1987}. In the last two columns `Energies Bose-Einstein' and `Energies Maxwell-Boltzmann' of Table \ref{piglethaffalunmp}, we show the `energies' related to the Bose-Einstein modeling and to the Maxwell-Boltzmann modeling, respectively.

We have now introduced all what is necessary to announce the principle result of our investigation. 

\begin{figure}
\centering
\includegraphics[width=8cm]{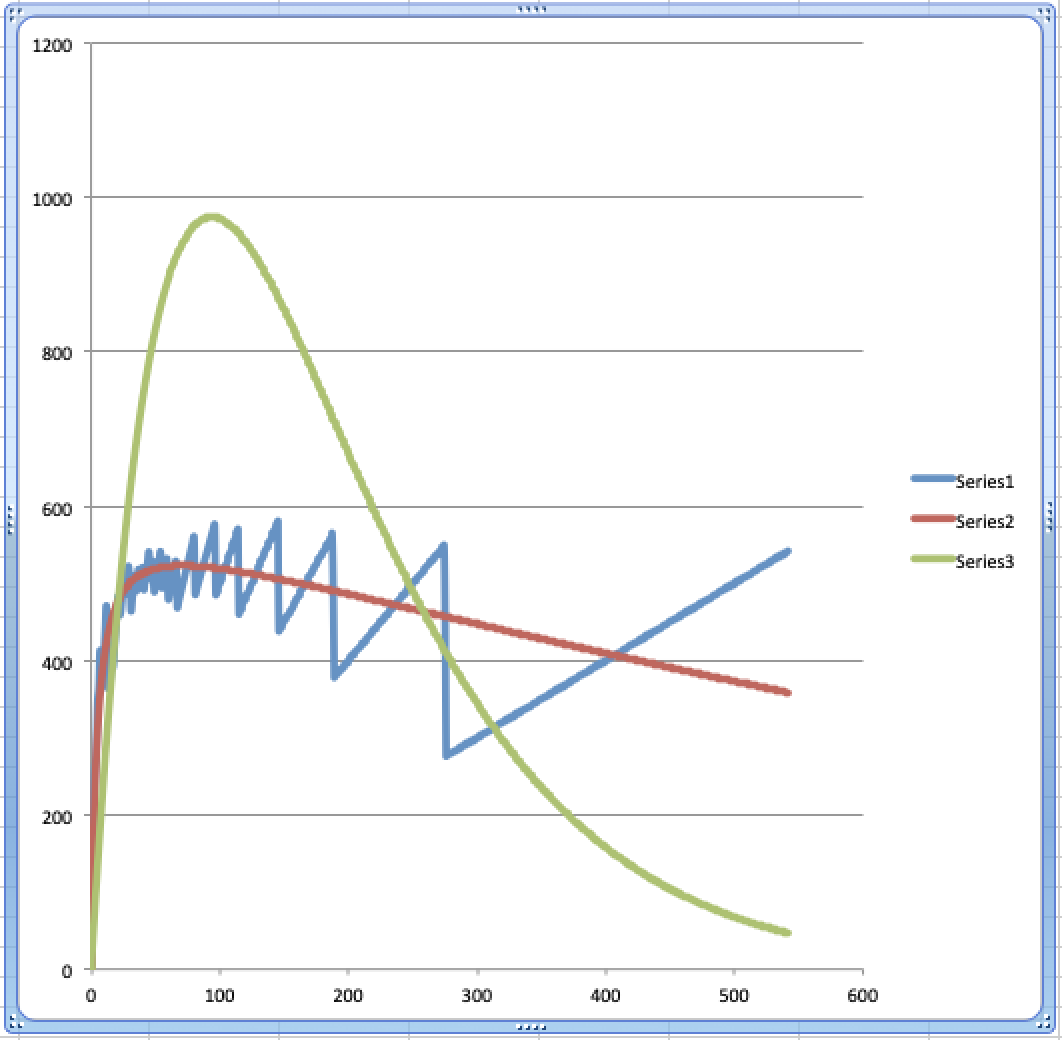}
\caption{A representation of the `energy distribution' of the Winnie the Pooh story `In Which Piglet Meets a Haffalump' \citep{milne1926} as listed in Table \ref{piglethaffalunmp}. The blue graph (Series 1) represents the energy radiated by the story per energy level (column `Energies from data $E(E_i)$' of Table \ref{piglethaffalunmp}), the red graph (Series 2) represents the energy radiated by the Bose-Einstein model of the story per energy level (column `Energies Bose-Einstein' of Table \ref{piglethaffalunmp}), and the green graph (Series 3) represents the energy radiated by the Maxwell-Boltzmann model of the story per energy level (column `Energies Maxwell-Boltzmann' of Table \ref{piglethaffalunmp}).}
\label{piglethaffalunmpenergygraph}
\end{figure}

\begin{quotation}
\noindent
{\it When we determine the two constants A and B, respectively C and D, in the Bose-Einstein distribution function (\ref{boseeinsteindistribution}) and Maxwell-Boltzmann distribution function (\ref{maxwellboltzmanndistribution}), by putting the total number of particles of the model equal to the total number of words of the considered piece of text, (\ref{BEconstraint01}) and (\ref{MBconstraint01}), and by putting the total energy of the model to the total energy of the considered piece of text, (\ref{BEconstraint02}) and (\ref{MBconstraint02}), we find a remarkable good fit of the Bose-Einstein modeling function with the data of the piece of text, and a big deviation of the Maxwell-Boltzmann modeling function with respect to the data of the piece of text.} 
\end{quotation}
The result is expressed in the graphs of Figure \ref{piglethaffalunmpgraphpiglethaffalunmploggraph} (a), where the blue graph represents the data, hence the numbers in column `Energies from data $E(E_i)$' of Table \ref{piglethaffalunmp}, the red graph represents the quantities obtained by the Bose-Einstein model, hence the quantities in column `Bose-Einstein modeling' of Table \ref{piglethaffalunmp}, and the green graph represents the quantities obtained by the Maxwell-Boltzmann model, hence the quantities of column `Energies Maxwell-Boltzmann' of Table \ref{piglethaffalunmp}. We can easily see in Figure \ref{piglethaffalunmpgraphpiglethaffalunmploggraph} (a) how the blue and red graphs almost coincide, while the green graph deviates abundantly from the two other graphs which shows how Bose-Einstein statistics is a very good model for the data we collected from the Winnie the Pooh story, while Maxwell-Boltzmann statistics completely fails to model these data.

To construct the two models, we also considered the energies, and expressed as a second constraint the conditions (\ref{BEconstraint02}), (\ref{MBconstraint02}), that the total energy of the Bose-Einstein model and the total energy of the Maxwell-Boltzmann model are both equal the total energy of the data of the Winnie the Pooh story. The result of both constraints, (\ref{BEconstraint01}), (\ref{MBconstraint01}) and (\ref{BEconstraint02}), (\ref{MBconstraint02}) on the energy functions that express the amount of energy per energy level -- or, to use the language customarily used for light, the frequency spectrum of light -- can be seen in Figure \ref{piglethaffalunmpenergygraph}. We see again that the red graph, which represent the Bose-Einstein radiation spectrum, is a much better model for the blue graph, which represents the experimental radiation spectrum, as compared to the green graph, which represents the Maxwell-Boltzmann radiation spectrum.

Both solutions, the Bose-Einstein shown in the red graph, and the Maxwell-Boltzmann shown in the green graph, have been found by making use of a computer program calculating the values of $A$, $B$, $C$ and $D$ such that (\ref{BEconstraint01}), (\ref{BEconstraint02}), (\ref{MBconstraint01}) and (\ref{MBconstraint02}) are satisfied, which gives the approximate values
\begin{eqnarray} \label{gastemperature}
A  \approx 1.0078 \quad B  \approx 593.51 \quad C  \approx 0.0353 \quad D \approx 93.63
\end{eqnarray}
In the graphs of Figure \ref{piglethaffalunmpenergygraph}, we can see that a maximum is reached for the energy level $E_{71}$, corresponding to the word {\it First}, which appears seven times in the Winnie the Pooh story. If we use the analogy with light, we can say that the radiation spectrum of the story `In Which Piglet Meets a Haffalump' has a maximum at {\it First}, which would hence be, again in analogy with light, the dominant color of the story \footnote{We are happy, although it is of course a coincidence, that it is also the `first' story we analyzed and also use in this article.}. We have indicated this radiation peak in Table \ref{piglethaffalunmp}, where we can see that the amount of energy the story radiates, following the Bose-Einstein model, is 522.79.

Due to their shape, the graphs in Figure \ref{piglethaffalunmpgraphpiglethaffalunmploggraph} (a) are not easily comparable, and although quite obviously the blue and red graphs are almost overlapping, while the blue and green graphs are very different, which shows that the data are well modeled by Bose-Einstein statistics and not well modeled by Maxwell-Boltzmann statistics, it is interesting to consider a transformation where we apply the $\log$ function to both the $x$-values, i.e. the domain values, and the $y$-values, i.e. the image values, of the functions underlying the graphs. This is a well-known technique to render functions giving rise to this type of graphs more easily comparable. 

In Figure \ref{piglethaffalunmpgraphpiglethaffalunmploggraph} (b), the graphs can be seen where we have taken the $\log$ of the $x$-coordinates and also the $\log$ of the $y$-coordinates of the graph representing the data, which is again the blue graph in Figure \ref{piglethaffalunmpgraphpiglethaffalunmploggraph} (b), of the graph representing the Bose-Einstein distribution model of these data, which is the red graph in Figure \ref{piglethaffalunmpgraphpiglethaffalunmploggraph} (b), and of the graph representing the Maxwell-Boltzmann distribution model of the data, which is the green graph in Figure \ref{piglethaffalunmpgraphpiglethaffalunmploggraph} (b). For readers acquainted with Zipf's law as it appears in human language, they will recognize Zipf's graph in the blue graph of Figure \ref{piglethaffalunmpgraphpiglethaffalunmploggraph} (b). It is indeed the $\log/\log$ graph of `ranking' versus 'numbers of appearances' of the text of the Winnie the Pooh story `In Which Piglet Meets a Haffalump', which is the `definition' of Zipf's graph. As to be expected, we see Zipf's law being satisfied, the blue graph is well approximated by a straight line with negative gradient close to -1. We see that the Bose-Einstein graph still models very well this Zipf's graph, and what is more, it also models the (small) deviation from Zipf's graph of the straight line. Zipf's law and the corresponding straight line when a $\log/\log$ graph is drawn is an empirical law. Intrigued by the modeling of the Bose-Einstein statistics by the Zipf graph, we have analyzed this correspondence in detail in Section \ref{boseeinsteinzipf}.

In the next section, however, we want to describe what a Bose gas is in physics, when it is brought nearby its state of Bose-Einstein condensate, with the aim of identifying the physical equivalent to the Winnie the Pooh story `In Which Piglet Meets a Haffalump' and other pieces of texts which we will also consider.

\section{The Bose-Einstein condensate in physics \label{storiesboseeinsteincondensates}}
We will explain in this section different aspects related to the experimental realization of a Bose gas near to it being a Bose-Einstein condensate where most of the bosons are in the lowest energy state. The awareness of the existence of this special state of a Bose gas came about as a consequence of a peculiar exchange between the Indian physicist Satyendra Nath Bose and Albert Einstein \citep{bose1924,einstein1924,einstein1925}. Bose actually devised a new way to derive Planck's radiation law for light -- which has the form of a Bose-Einstein statistics, hence, like we now know, being a consequence of the indistinguishability of the photon as a boson, but that was not known in these pre-quantum theory times -- and sent the draft of his calculation to Einstein. Although what Bose did was far from being fully understood in that time, the new method of calculation must have caught right away the full attention of Einstein, because he translated the article from English to German and supported its publication in one of the most important scientific journals of that time \citep{bose1924}. Einstein himself then, inspired by Bose's method, worked our a new model and calculation for an atomic gas consisting of bosons, and predicted the existence of what we now call a Bose-Einstein condensate, an amazing accomplishment, taken into account that the difference between bosons and fermions and the Pauli exclusion principle were not yet known \citep{einstein1924,einstein1925}. Because of the intense study of Bose-Einstein condensates that took off after their first experimental realizations \citep{andersonetal1995,bradleyetal1995,davisetal1995}, a lot of new knowledge, experimental, but also theoretical, has been obtained, material on which we built upon for some of the details of the present article \citep{ketterleandvandruten1996,parkinsandwalls1998,dalfovoetal1999,ketterledurfeestamper-kum1999,gorlitzetal2001,hennetal2008}.

The principle idea is still the one foreseen by Einstein, namely to take a dilute gas of boson particles and then stepwise lower its temperature and as a consequence its total energy such that at a certain moment there is so little energy in the gas that all boson particles are forced to transition to the lowest energy state. At that moment, all boson particles are in the same state, namely this lowest energy state, and the gas behaves then in a way for which there is no classical equivalent -- we will see that given our conceptuality interpretation of quantum theory and the boson gas model we built here for human language, we will be able to put forward a new way to view the indistinguishability that lies at the heart of a Bose-Einstein condensate (see Section \ref{identifyindistinguishability}).

The Bose-Einstein condensates that have been realized so far all consist mainly of massive boson particles, hence generally atoms with integer spins, which makes them bosons. Indeed, the situation of the bosons of light, i.e. of photons, is more complicated, because photons interact so abundantly with matter that their number is never constant, which makes it difficult to realize a thermal equilibrium in this case, albeit not impossible \citep{klaersverwingerweitz2010a,klaersverwingerweitz2010b,klaersetal2011,klaesweitz2013}.
We do want to keep using our analogy of language with light, although of course the pieces of texts that we will study contain a fixed number of words, but a dynamic use of human language will also give rise to a continuous coming into existence of new words, which means that for such a dynamic situation the example of light is probably even more representative than gases with a fixed number of atoms.
In this stage of our analysis, also because they are the more easy to realize Bose-Einstein condensates, we however focus on massive bosons, hence atoms with integer spins.

The underlying idea is that the gas consists of atoms in a good approximation not interacting with each other, hence only carrying the kinetic energy $K = p^2 / 2m$ generated by random movements due to the temperature $T$. It can be shown that in this situation the average kinetic energy of a free particle equals $K= \pi kT$, where $k$ is Boltzmann's constant, hence we have
\begin{eqnarray} \label{thermaldebroglie01}
{\overline{p^2} \over 2m} = \pi kT
\end{eqnarray}
where $m$ is the mass of the atoms and $p$ the absolute value of their momentum. From (\ref{thermaldebroglie01}) and de Broglie's formula $\lambda = h / p$ we can calculate the `thermal de Broglie wave length' $\lambda_{th}$ of the atoms of the gas
\begin{eqnarray} \label{thermaldebroglie02}
\lambda_{th} = {h \over \sqrt{2 \pi m k T}}
\end{eqnarray}
Let us make things more concrete and calculate this thermal de Broglie wave lengths for the atoms that were used in the Bose-Einstein condensates realized by Eric Cornell and Carl Wieman at the University of Colorado at Boulder in their NIST-JILA lab \citep{andersonetal1995}, and by the group led by Wolfgang Ketterle at MIT, for which they jointly were attributed the Nobel Prize in physics in 1999. At Cornell they used a vapor of rubidium 87 atoms in a number density of $2.5 \times 10^{12}$ atoms per cubic centimeter, cooled down to a temperature of 170 nanokelvin, to see the condensate fraction appear containing an estimated 2000 atoms and be preserved for more than 15 seconds. At MIT, they used a dilute gas of sodium atoms in a number density higher than $10^{14}$ atoms per cubic centimeter to realize the formation of a condensate containing up to $500000$ atoms at a temperature of 2 microkelvin, with a lifetime of 2 seconds.

Let us calculate $\lambda_{th}$ for both these condensate formations. Next to the values of Planck's and Boltzmann's constants, and the value of $\pi$, we only need the value of the mass of a rubidium 87 atom and of a sodium atom to do the calculation. The atomic mass of a rubidium 87 atom and of a sodium atom are, respectively, $86.909180527$ and $22.989769$ unified atomic mass units, and given that one such unified atomic mass unit is $1.66053904 \times 10^{-27} {\rm kg}$ we get
\begin{eqnarray}
&&m_{Rb} \approx 1.44316 \times 10^{-25} {\rm kg} \\
&&m_{Na} \approx 3.81754 \times 10^{-26} {\rm kg} \\
&&h \approx 6.62607004 \times 10^{-34} {\rm kg m}^2 / {\rm s} \\
&&k \approx 1.38065 \times 10^{-23} {\rm kg m}^2 / {\rm s}^2 {\rm K} \\
&&\pi \approx 3.14159
\end{eqnarray} 
Using the above values into (\ref{thermaldebroglie02}), we obtain for the rubidium gas at 170 nanokelvin and the sodium gas at 2 microkelvin
\begin{eqnarray} \label{phasedensity170}
&&\lambda_{thRb} \approx 4.54195 \times 10^{-7} {\rm m} \approx 454\, {\rm nm} \\
&&\lambda_{thNa} \approx 2.57465 \times 10^{-7} {\rm m} \approx 257\, {\rm nm}
\end{eqnarray}
Often one can read that in states of the Bose gas that are `nearing the Bose-Einstein condensate', the `de Broglie waves' of the particles start to `overlap', and that this is the reason why quantum effects become dominant. There is an interesting measure to express in a quantitative way this notion of `overlapping de Broglie waves' and it is called the `phase space density' $\rho_{ps}$ of the boson gas 
\begin{eqnarray} \label{phasespacedensity}
\rho_{ps} = n \times \lambda_{th}^3
\end{eqnarray}
where $n$ is the `atom density' of the gas expressed in `number of atoms per cubic centimeter'. From (\ref{phasespacedensity}) follows that $\rho_{ps}$ corresponds to the number of atoms in a region of space of the `de Broglie wave' cube size. If this number is much smaller than 1, this means that the de Broglie wave length is much smaller than the distance between the atoms, hence there will be no overlapping and the gas will behave classically. The more this number is greater than 1, the more the de Broglie waves of the atoms are overlapping, hence quantum behavior will increase. It has been shown \citep{bagnatopritchardkleppner1987} that independent of the trapping device used for the atoms, a box, or a magnetic trap -- which is the one used in actually realized Bose-Einstein condensates -- the condensate starts to form whenever the value of $\rho_{ps}$ is such that
\begin{eqnarray} \label{quantumregime}
2.612 \le \rho_{ps}
\end{eqnarray}
Considering (\ref{thermaldebroglie02}) and (\ref{phasespacedensity}), the value of $\rho_{ps}$ in the process of formation of a Bose-Einstein condensate is determined by the temperature $T$ and number density $n$ of the atom gas. In the last stage of the formation, the temperature is lowered by a technique called `evaporative cooling under influence of a radio frequency field'. The effect is that also the number density decreases, hence to attain the quantum regime of overlapping de Broglie wave lengths it is necessary to lower the temperature faster than diluting the gas. The group at MIT mentions explicitly the number density that they reached when the Bose-Einstein condensate is formed, namely, between $10^{14}$ and $4 \times 10^{14}$ atoms per cubic centimeter \citep{davisetal1995}. The Boulder group, since they identified the formation of their rubidium Bose-Einstein condensate at a temperature of $170\, {\rm nK}$, taking into account (\ref{quantumregime}), we can calculate that the number density of the rubidium gas must have been around $2.8 \times 10^{13}$ atoms per cubic centimeter.

We give in Table \ref{energylengthtablesodium} an overview of the energies and lengths that are characteristic for the realizations of the sodium condensate in MIT \citep{ketterledurfeestamper-kum1999}. Because the gas is very dilute and the temperature is very low, the size of the atoms is very small compared to the distance between the atoms, while the thermal de Broglie wave lengths are large, such that they are overlapping. With each length scale $l$ there is an associated energy scale which is the kinetic energy $K = \pi kT$ of a particle with a de Broglie wavelength $l$, that is
\begin{eqnarray} \label{debrogliewavelengthenergy}
K \approx {h^2 \over 2ml^2}
\end{eqnarray}
gives a good indication of the relation between sizes and energies.

A good measure for the size of atoms which are diluted like in the considered boson gas is the so-called elastic s-scattering length $a = l/2\pi$. For sodium this has been measured to be 3 nanometers, which using (\ref{debrogliewavelengthenergy}) corresponds to an energy of 1 millikelvin in temperature \citep{marteetal2002}. Around this temperature elastic s-wave scattering between the atoms will be dominant. 
\begin{table}[h]
\begin{center}
\begin{tabular}{p{4.5cm}p{2cm}p{4.5cm}p{3.5cm}p{2cm}}
  Energy scale $E$ & $ \approx h^2 / 2ml^2$ &  Length scale $l$  & $ \approx h / \sqrt{2mE}$ &   \\
  \hline
  limiting temperature for s-wave scattering &  $\approx 1\, {\rm mK}$ & scattering length &$a \approx l / 2\pi$ & $ \approx 3\, {\rm nm}$  \\
  Bose-Einstein condensate transition temperature &  $\approx 2\, \mu{\rm K}$ & separation between atoms & $n^{-{1 \over 3}} \approx l / \sqrt{\pi}(2.612)^{1 \over 3}$ & $\approx 200\, {\rm nm}$ \\
  Temperature $T$ & $\approx 1\, \mu {\rm K}$ & thermal de broglie wave length & $\lambda_{th} = l / \sqrt{\pi}$ & $\approx 300\, {\rm nm}$ \\
  harmonic oscillator level spacing $h\nu$ &  $\approx 0.5\, {\rm nK}$ & oscillator length $\nu = 10\, {\rm Hz}$ & $a_{HO}= l / \sqrt{2}\pi$ & $\approx 6.5\, \mu{\rm m}$ 
\end{tabular}
\end{center}
\caption{Energy and length scales of the sodium Bose-Einstein condensate}
\label{energylengthtablesodium}
\end{table}
The separation between the atoms in the gas can be estimated by considering the cubic root $n^{1 \over 3}$ of the number density, which gives us the number of atoms spread out over 1 centimeter. For sodium, with a number density higher than $10^{14}$ atoms per cubic centimeter, this gives rise to a spacing between the atoms of around 200 nanometers. The length $l$ can be calculated by making use of (\ref{quantumregime}) which gives us the following estimate for $l$
\begin{eqnarray}
2.612 \approx n \times \lambda_{th}^3 \Leftrightarrow (2.612)^{1/3} \approx n^{1/3} \times {l \over \sqrt{\pi}} \Leftrightarrow l \approx {\sqrt{\pi} \times (2.612)^{1/3} \over n^{1/3}}
\end{eqnarray}
and hence, by making use of (\ref{debrogliewavelengthenergy}) we find that $E$ is around $2\, \mu {\rm K}$. 

A temperature of around $1\, \mu {\rm K}$ gives rise to a thermal de Broglie wavelength of around $300\, {\rm nm}$.

The largest length scale is related to the confinement characterized by the size of the box potential or by the oscillator length $a_{HO} = {1 \over 2\pi}\sqrt{h / m \nu}$, which is the typical size of the ground state wave function in a harmonic oscillator potential of frequency $\nu$ (see Appendix \ref{appendixquantumharmonicoscillator}). With $\nu = 10\, {\rm Hz}$, we get a value for $a_{HO}$ of about $6.5\, \mu {\rm m}$. The energy scale related to the confinement is characterized by the harmonic oscillator energy level spacing, given by $h \nu$. Again, for $\nu = 10\, {\rm Hz}$ we get an energy value for the spacing of about $0.5\, {\rm nK}$. 

In Table \ref{energylengthtablerubidium}, we made the calculations of length and energy scales for the rubidium 78 Bose-Einstein condensate, taking into account that a density of around $2.8 \times 10^{13}$ atoms per cubic centimeter was realized within the condensate of 2000 atoms.
\begin{table}[h]
\begin{center}
\begin{tabular}{p{4.5cm}p{2cm}p{4.5cm}p{3.5cm}p{2cm}}
  Energy scale $E$ & $ \approx h^2 / 2ml^2$ &  Length scale $l$  & $\approx h / \sqrt{2mE}$ &   \\
  \hline
  limiting temperature for s-wave scattering &  $\approx 0.1\, {\rm mK}$ & scattering length &$a=l / 2\pi$ & $ \approx 5\, {\rm nm}$  \\
  Bose-Einstein condensate transition temperature &   $\approx 170\, {\rm nK}$ & separation between atoms & $n^{-{1 \over 3}} \approx l / \sqrt{\pi}(2.612)^{1 \over 3}$ & $ \approx 300\, {\rm nm}$ \\
  Temperature $T$ & $ \approx 50\, {\rm nK}$ & thermal de broglie wave length & $\lambda_{th} = l / \sqrt{\pi}$ & $ \approx 800\, {\rm nm}$ \\
  harmonic oscillator level spacing $h\nu$ &  $ \approx 1\, {\rm nK}$ & oscillator length $\nu \approx 10\, {\rm Hz}$ & $a_{HO}= l / \sqrt{2}\pi$ & $ \approx 4\, \mu {\rm m}$ 
\end{tabular}
\end{center}
\caption{Energy and length scales of rubidium Bose-Einstein condensate}
\label{energylengthtablerubidium}
\end{table}
We want to show now that our Bose-Einstein distribution model of the Winnie the Pooh story `In Which Piglet Meets a Haffalump' is well modeled by a Bose gas close to the Bose-Einstein condensate of this gas, and will take the rubidium and sodium gases that we described in as inspiration. What is important to notice is the difference in order of magnitude between the energy level spacings of the harmonic trap oscillator, they are of the order of $1\, {\rm nK}$, and the energies involved with the gas itself, of the order of $1\, \mu {\rm K}$. The Winnie the Pooh story `In Which Piglet Meets a Haffalump' is not in a Bose-Einstein condensate state, because then all the words of the story should be the word {\it And}, populating the zero energy level. So, it is in a state which is close to a Bose-Einstein condensate.

We have not yet explained what the parameters $A$ and $B$ of (\ref{boseeinsteindistribution}) are for the situation of a physical boson gas, for which the Bose-Einstein distribution is often written as
\begin{eqnarray} \label{boseeinsteindistributionphysics}
N(E_i) = {g_i \over {e^{{E_i-\mu \over kT}}-1}}
\end{eqnarray}
where $\mu$ is called the `chemical potential', and $g_i$ the `multiplicity'. The multiplicity $g_i$ of a specific energy level $E_i$ is the number of states that are different but have this same energy $E_i$. That different states can have the same energy is connected to the symmetries of the configuration, often spatial ones. For example, for the most simple model of the harmonic trap, the one of a quantum harmonic oscillator, the multiplicity in $s$ dimensions equals 
\begin{eqnarray}
{(n + s -1)! \over n!(s-1)!}
\end{eqnarray}
which becomes $(n+1)(n+2) / 2$ in 3 dimensions, $(n+1)$ in 2 dimensions, and $1$ in the one-dimensional situation. The different dimensions are relevant for the Bose-Einstein condensates realized in laboratories, because, although the boson gas exists always in 3 dimensions, often the harmonic traps give rise to very elongated cigar-like configurations, such that a quantum description in terms of an effective one-dimensional harmonic oscillator is a better model. Anyhow, for the text of the Winnie the Pooh story we do not have to hesitate about its dimension, pronouncing a text while reading it is certainly one-dimensional. Also a written text, although materialized on a page which is two dimensional, is a one-dimensional structure. This means that in the formula for the Bose-Einstein distribution we have rightly taken $g_i = 1$ for every energy level $E_i$.

What about the `chemical potential' $\mu$? There is another quantity which is introduced with respect to it which is called the `fugacity'
\begin{eqnarray}
f = e^{\mu \over kT} = {1 \over A}
\end{eqnarray}
If we look at (\ref{boseeinsteindistributionphysics}), taking into account that $g_i = 1$ and $E_0 = 0$, we get
\begin{eqnarray} \label{numbercondensateparticles}
&&N(E_0) = N_0 = {1 \over {e^{{-{\mu \over kT}}}-1}} =  {f \over 1 - f} \\ \label{fugacity}
&\Leftrightarrow& f = {N_0 \over 1+ N_0} \\ \label{chemicalpotential}
&\Leftrightarrow& \mu = kT \log{N_0 \over 1+ N_0}
\end{eqnarray}
which means that the chemical potential and the fugacity are determined by the number $N_0$ of particles that are in the lowest energy state, hence the number of particles that are in the condensate state. More specifically, for the Winnie the Pooh story we find
\begin{eqnarray}
f \approx 0.9923 \quad \mu \approx -4.581
\end{eqnarray}
Let us note that from (\ref{fugacity}) follows that the fugacity is a number contained between 1/2 and 1, in case we have at least one particle in the condensate state, and the chemical potential is a negative number, they respectively approach 1 and 0 when the condensate grows in terms of number of particles in the lowest energy level. For what concerns the second constant $B$, we have
\begin{eqnarray}
B = kT 
\end{eqnarray}
which means that the second constant $B$ is given by the temperature of the Bose gas.

The rubidium condensate is a better example for the Winnie the Pooh story, as also the number of atoms, 2000, is of the same order of magnitude as the number of words, 2655, of the Winnie the Pooh story. The energy levels of the trap for the rubidium condensate are of the order of $1\, {\rm nK}$, while the temperature of the gas is $170\, {\rm nK}$ (Table \ref{energylengthtablerubidium}), which is 170 times bigger. We see for the Winnie the Pooh story that if we take 1 unit of energy for the energy level spacings, we have $B = kT = 593$, following (\ref{gastemperature}), and hence ${1 \over 2} kT$, being a good estimate for the average energy per atom of a one-dimensional gas, gives for the latter $271$, which means that we are in this respect also in the same order of magnitude for the Winnie the Pooh story and the rubidium condensate. Hence, we can say that the Winnie the Pooh story can be looked at as behaving similarly to a Bose gas of rubidium 87 atoms in one-dimension at a temperature of $170\, {\rm nK}$. We will see in Section \ref{boseeinsteinzipf}, where we consider the text of the novel `Gulliver's Travels' of Jonathan Swift \citep{swift1726}, that the sodium condensate is a better example for this text. 

Let us introduce a second piece of text in Table \ref{magicshop}, namely a story entitled `The magic shop' written by Herbert George Wells \citep{wells1903}, with which we want to illustrate an aspect of our `Bose gas representation of human language' that we have not yet touched upon. For the Winnie the Pooh story, If we look at Figure \ref{piglethaffalunmpenergygraph} and Table \ref{piglethaffalunmp}, we can see that the `energy spectrum' does not cover the whole range of possible energy values. Indeed, the red graph of Figure \ref{piglethaffalunmpenergygraph} on the right hand side of the graph has still a substantial value, and is not at all close to zero. Hence one can wonder what happens further on for higher energy spectrum with this graph?

On the low energy spectrum, the amount of radiation increases starting from zero radiation for energy level $E_0$, hence for the words that are captured in the zero energy level of the Bose-Einstein condensate, there is no radiation emerging from them following the considered choice of zero in the energy scale -- for the case of the Winnie the Pooh story, the zero level energy state puts the cogniton in state {\it And} -- and then the amount of radiation increases steeply -- we have already a radiation of 111 energy units (and 105.84 in the Bose-Einstein model) for $E_1$ for the Winnie the Pooh story and the cogniton in state {\it He}. The energy radiation keeps increasing steeply -- 182 for $E_2$ (179.36 for the Bose-Einstein model) for the cogniton in state {\it The}, 255 for $E_3$ (233.36 for the Bose-Einstein model) for the cogniton in state {\it It}, 280 for $E_4$ (274.65 for the Bose-Einstein model) for the cogniton in state {\it A}, 345 for $E_5$ (307.23 for the Bose-Einstein model) for the cogniton in state {\it To}, etc. -- to reach a maximum at $E_{71}$ with a radiation level of 522.79 energy units for the cogniton in state {\it First}. Then the radiation starts to decrease slowly. But, remark that at energy level $E_{542}$, with the cogniton in state {\it You've}, which is the highest energy level of Table \ref{piglethaffalunmp}, we still have a radiation of 385.55 energy units, which is more than half of the maximum radiation reached at energy level $E_{71}$ for the cogniton in state {\it First}.

How can we understand this, because we have in Table \ref{piglethaffalunmp} exhausted all the words of the Winnie the Pooh story and hence seemingly represented all possible energy levels. But is this true? To see clear in this, we have to reflect about the difference of the numbers in the third and the fourth column of Table \ref{piglethaffalunmp}, respectively the `numbers of appearances' of the specific words in the Winnie the Pooh story and the `values of the Bose-Einstein distribution that we used to model these numbers of appearances'. The values in the fourth column are of a probabilistic nature and express averages of stories `similar' to the one of Winnie the Pooh with respect to the numbers of appearances of the specific words, while the values in the third column express real counts for one specific story. More concretely, by `similar' we actually mean `containing the same total number of words, and containing the same total amount of energy'. Remember indeed that the Bose-Einstein distribution function only contains two parameters, which hence will be determined by the total number of words and the total amount of energy. Or to put it even more concretely, suppose we would collect a vast number of pieces of `meaningful' text all containing the same total number of words $N$ and the same amount of total energy $E$, the Bose-Einstein distribution function (\ref{boseeinsteindistribution}) is then supposed to model a specific type of average that can be obtained for all these texts, and the more numerous these texts the better this average will correspond with the Bose-Einstein distribution function. The reason is that this function is the consequence of the limit process in statistical mechanics  of a  micro-canonical ensemble of states of particles with the same $N$ and $E$ \citep{bose1924,einstein1924,einstein1925,huang1987}.

The above reasoning indicates that we can consider to introduce a `place for words that do no appear in the considered text but could have appeared'. Remark that these new words do not add to the sum $N$ of all words, since they have `number of appearance zero', which means that this operation of `adding new words' leaves $N$ unchanged. In the ranking of energy levels, they have to be classified by `additional energy levels higher than the highest one we now identified with respect to the last alphabetically classified word that appears one time in the text'. Remark that also $E$ remains unchanged by this adding of words that could have appeared. Indeed, although these new added words carry high energies, since all of them have appearance number zero, they do not add to the total amount of energy because the product of the energy of an even very high energy level with the zero of its number of appearances equals zero. Since $N$ and $E$ are left unchanged by the adding of these new words that could have appeared also the micro-canonical ensemble and its thermodynamical equilibrium remain unchanged. However the adding of the new words does alter substantially the Bose-Einstein distribution function and the Maxwell-Boltzmann distribution function calculated to model the data, because they both do not have appearance values equal to zero for these words, which means that there will be contributions to the total number of words and the total energy of their modeling. Hence, this operation of adding words such that the energy spectrum completes itself over the whole range is a necessary operation in the modeling with Bose-Einstein or Maxwell-Boltzmann.

Again more concretely, let us consider the words that appear one time in the Winnie the Pooh story, and look for synonyms of these words, then the word that appears now one time could not have appeared and instead its synonym could then have appeared. So, the synonyms can be listed in a new set of words to add with zero appearance, as `could have appeared', and indeed, the Bose-Einstein distribution function will not be zero for them, which expresses exactly this `they could have appeared'.

To illustrate the above, we consider the H. G. Wells story `The magic shop' \citep{wells1903} for which we have classified its words in energy levels in Table \ref{magicshop}. As we can see, the energy level $E_{1153}$ corresponding to the state of the cogniton characterized by the word {\it Youngster}, would have been the highest energy level in case we had stopped, like we did for the Winnie the Pooh story, to add energy levels at the `one word appearance number'. For this new story `The magic shop' we have however added the `zero word appearance number' explicitly, starting with {\it Garden}, which is a word that does not appear in the story, synonym of {\it Yard} of energy level $E_{1149}$ and we attributed energy level $E_{1154}$ to the cogniton in a state characterized by {\it Garden}. And indeed, in the third column in the row where {\it Garden} appears in Table \ref{magicshop} there is 0, indicating that {\it Garden} does not appear in the story `The magic shop'. In the fourth column, in the row of {\it Garden} in Table \ref{magicshop}, we however have 0.25, which is the value of the Bose-Einstein distribution function at energy level $E_{1154}$, and in the fifth column, in the row of {\it Garden} in Table \ref{magicshop}, we have 0.07, which is the value of the Maxwell-Boltzmann distribution function at energy level $E_{1154}$. Both numbers indicate that `{\it Garden} could have appeared in a story similar to the H. G. Wells story', because they are not zero. These numbers are linked to the probability of {\it Garden} to appear in a similar story than the story of `The magic shop' in the way we explained above. And indeed there should be not zeros in these places because there is a probability that {\it Garden} would appear in such a similar story. We added the word {\it Okay} at energy level $E_{1155}$ as synonym of {\it Yes} at energy level $E_{1150}$, as a new not appearing state of the cogniton, however potentially appearing in a similar story. We continued in the same way adding {\it Junior} as synonym of {\it Youngster}, but there are no synonyms of {\it You'd} and {\it You're}, which gives us the occasion to mention that the added words that could appear in a similar story do not have to be synonyms.

\small
\begin{longtable}{p{1.5cm}p{1.5cm}p{1.5cm}p{1.5cm}p{1.5cm}p{1.5cm}p{1.5cm}p{1.5cm}} 
\label{magicshop}
Words concepts cognitons & Energy levels $E_i$ & Appearance numbers $N(E_i)$ & Bose-Einstein modeling &  Maxwell-Boltzmann modeling & Energies from data $E(E_i)$ & Energies Bose-Einstein & Energies Maxwell-Boltzmann  \\
 \hline
{\it The} & 0 & 202 & 201.4 & 18.84 & 0 & 0 & 0 \\
{\it And} & 1 & 176 & 157.28 & 18.75 & 176 & 157.28 & 18.75 \\
{\it A} & 2 & 125 & 128.99 & 18.66 & 250 & 257.97 & 37.33  \\
{\it I} & 3 & 113 & 109.3 & 18.57 & 339 & 327.89 & 55.72  \\
{\it Of} & 4 & 95 & 94.81 & 18.48 & 380 & 379.22 & 73.94 \\
{\it Was} & 5 & 72 & 83.69 & 18.4 & 360 & 418.46 & 91.98 \\
{\it To} & 6 & 71 & 74.9 & 18.31 & 426 & 449.41 & 109.85 \\
{\it He} & 7 & 67 & 67.77 & 18.22 & 469 & 474.41 & 127.54 \\
{\it In} & 8 & 67 & 61.87 & 18.13 & 536 & 495.00 & 145.06 \\
{\it It} & 9 & 63 & 56.92 & 18.05 & 567 & 512.24 & 162.41 \\
{\it Said} & 10 & 59 & 52.69 & 17.96 & 590 & 526.86 & 179.59 \\
{\it That} & 11 & 51 & 49.04 & 17.87 & 561 & 539.42 & 196.61 \\
{\it Gip} & 12 & 48 & 45.86 & 17.79 & 576 & 550.29 & 213.45 \\
{\it With} & 13 & 45 & 43.06 & 17.7 & 585 & 559.80 & 230.13 \\
{\it His} & 14 & 43 & 40.58 & 17.62 & 602 & 568.16 & 246.65 \\
{\it My} & 15 & 36 & 38.37 & 17.53 & 540 & 575.58 & 263.00 \\
{\it You} & 16 & 33 & 36.39 & 17.45 & 528 & 582.18 & 279.19 \\
{\it Had} & 17 & 31 & 34.59 & 17.37 & 527 & 588.10 & 295.22 \\
{\it Shopman} & 18 & 27 & 32.97 & 17.28 & 486 & 593.42 & 311.09 \\
{\it There} & 19 & 27 & 31.49 & 17.2 & 513 & 598.22 & 326.80 \\
{\it As} & 20 & 25 & 30.13 & 17.12 & 500 & 602.58 & 342.35 \\
{\it At} & 21 & 25 & 28.88 & 17.04 & 525 & 606.54 & 357.74 \\
{\it Magic} & 22 & 25 & 27.73 & 16.95 & 550 & 610.16 & 372.98 \\
{\it But} & 23 & 24 & 26.67 & 16.87 & 552 & 613.46 & 388.07 \\
{\it Little} & 24 & 23 & 25.69 & 16.79 & 552 & 616.49  & 403.00 \\
{\it One} & 25 & 22 & 24.77 & 16.71 & 550 & 619.27 & 417.78 \\
   \ldots &  \ldots &  \ldots   &  \ldots    & \ldots &  \ldots &  \ldots &  \ldots  \\
                                   &     &      &      &     &      &      &       \\
                                   &     &      &      &     &      &      &       \\
       \ldots &  \ldots &  \ldots   &  \ldots    & \ldots &  \ldots &  \ldots &  \ldots  \\
  {\it What} & 65 & 9 & 10.04 & 13.79 & 585 & 652.41 & 896.44 \\
{\it Which} & 66 & 9 & 9.89 & 13.73 & 594 & 652.47 & 905.87 \\
{\it Behind} & 67 & 8 & 9.74 & 13.66 & 536 & 652.51 & 915.19 \\
{\it Boy} & 68 & 8 & 9.6 & 13.59 & 544 & 652.54 & 924.40 \\
{\it Do} & 69 & 8 & 9.46 & 13.53 & 552 & 652.55201 & 933.50 \\
{\it Door} & {\bf 70} & 8 & 9.32 & 13.46 & 560 & {\bf 652.55204} & 942.50 \\
{\it Genuine} & 71 & 8 & 9.19 & 13.4 & 568 & 652.54 & 951.38 \\
{\it Glass} & 72 & 8 & 9.06 & 13.34 & 576 & 652.51 & 960.16 \\
{\it Hat} & 73 & 8 & 8.94 & 13.27 & 584 & 652.48 & 968.83 \\
{\it Moment} & 74 & 8 & 8.82 & 13.21 & 592 & 652.43 & 977.40 \\
{\it More} & 75 & 8 & 8.7 & 13.14 & 600 & 652.37 & 985.87 \\
   \ldots &  \ldots &  \ldots   &  \ldots    & \ldots &  \ldots &  \ldots &  \ldots  \\
                                   &     &      &      &     &      &      &       \\
                                   &     &      &      &     &      &      &       \\
       \ldots &  \ldots &  \ldots   &  \ldots    & \ldots &  \ldots &  \ldots &  \ldots  \\
{\it Yard} & 1149 & 1 & 0.25 & 0.08 & 1149 & 292.03 & 87.03 \\
{\it Yes} & 1150 & 1 & 0.25 & 0.08 & 1150 & 291.78 & 86.68 \\
{\it You'd} & 1151 & 1 & 0.25 & 0.08 & 1151 & 291.53 & 86.34 \\
{\it You're} & 1152 & 1 & 0.25 & 0.07 & 1152 & 291.28 & 86.01 \\
{\it Youngster} & 1153 & 1 & 0.25 & 0.07 & 1153 & 291.02 & 85.67 \\
{\it  Garden}& 1154 & 0 & 0.25 & 0.07 & 0 & 290.77 & 85.33 \\
{\it  Okay}& 1155 & 0 & 0.25 & 0.07 & 0 & 290.52 & 85.00 \\
{\it  Store}& 1156 & 0 & 0.25 & 0.07 & 0 & 290.27 & 84.66 \\
{\it  Meter}& 1157 & 0 & 0.25 & 0.07 & 0 & 290.02 & 84.33 \\
{\it Junior}& 1158 & 0 & 0.25 & 0.07 & 0 & 289.76 & 84.00 \\
   \ldots &  \ldots &  \ldots   &  \ldots    & \ldots &  \ldots &  \ldots &  \ldots  \\
                                   &     &      &      &     &      &      &       \\
                                   &     &      &      &     &      &      &       \\
       \ldots &  \ldots &  \ldots   &  \ldots    & \ldots &  \ldots &  \ldots &  \ldots  \\
{\it  Continued}& 3494 & 0 & 0.01 & 0 & 0 &  &  \\
{\it  Adding}& 3495 & 0 & 0.01 & 0 & 0 & 27.71 & 0.003 \\
{\it  Mention}& 3496 & 0 & 0.01 & 0 & 0 & 27.68 & 0.003 \\
{\it  Similar}& 3497 & 0 & 0.01 & 0 & 0 & 27.65 & 0.003 \\
{\it  Criterion}& 3498 & 0 & 0.01 & 0 & 0 & 27.61 & 0.003 \\
{\it  Obviously}& 3499 & 0 & 0.01 & 0 & 0 & 27.58 & 0.003 \\
{\it  Appearing}& 3500 & 0 & 0.01 & 0 & 0 & 27.55 & 0.003 \\
   \hline
{\it Totalities} &       &3934   &3934.00    &3934.00 & 817415 & 817415.00 & 817414.18  \\
\caption{An energy scale representation of the words of the story `The magic shop' by H. G. Wells as published in \citet{wells1903}. The words are in the column `Words concepts cognitons' and the energy levels are in the column `Energy levels $E_i$', and are attributed according to the `numbers of appearances' in the column `Appearance numbers $N(E_i)$', such that  lower energy levels correspond to higher order of appearances, and the value of the energy levels is determined according to (\ref{Ei}). The `amounts of energies radiated by the words of energy level $E_i$' are in the column `Energies from data $E(E_i)$'. In the columns `Bose-Einstein modeling', `Maxwell-Boltzmann modeling', `Energies Bose-Einstein' and `Energies Maxwell-Boltzmann' are respectively the predicted values of the Bose-Einstein and the Maxwell-Boltzmann model of the `numbers of appearances', and of the `radiated energies'. Words and their corresponding energy levels were added with zero number of appearances to complete the energy spectrum for the high energy region as shown in Figure \ref{magicshopenergygraph}.}
\end{longtable}
\normalsize
\noindent
The only criterion is that `they appear in a meaningful story with the same total number of words and the same total energy'. Hence, adding synonyms is a simple way to ensure that the whole story remains meaningful, but also a completely new meaningful part to the story can be added with words that are no synonyms'.

So, we added many more energy levels, namely till the cogniton being in energy level $E_{3500}$. We have only shown the seven last ones of these words in Table \ref{magicshop}, namely {\it  Continued}, {\it  Adding}, {\it  Mention}, {\it  Similar}, {\it  Criterion}, {\it  Obviously} and {\it  Appearing},
having zero number of appearances in the H. G. Wells story, but their Bose-Einstein value in the Bose-Einstein model, as well as their Maxwell-Boltzmann value in the Maxwell-Boltzmann model, being not zero.

\begin{figure}%
    \centering
    \subfloat[Numbers of appearances distribution graphs]{{\includegraphics[width=8cm]{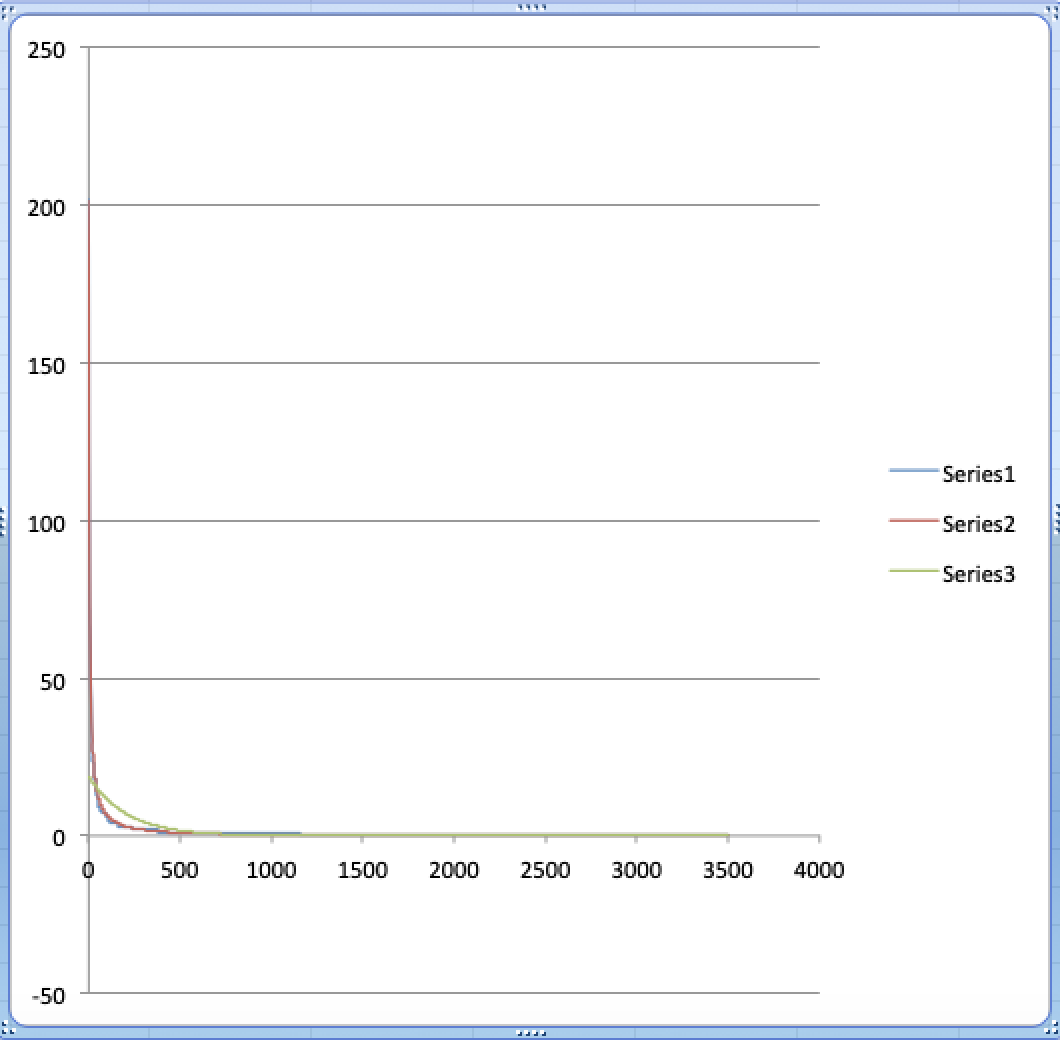} }}%
    \qquad
    \subfloat[$\log/\log$ graphs of numbers of appearances distributions]{{\includegraphics[width=8cm]{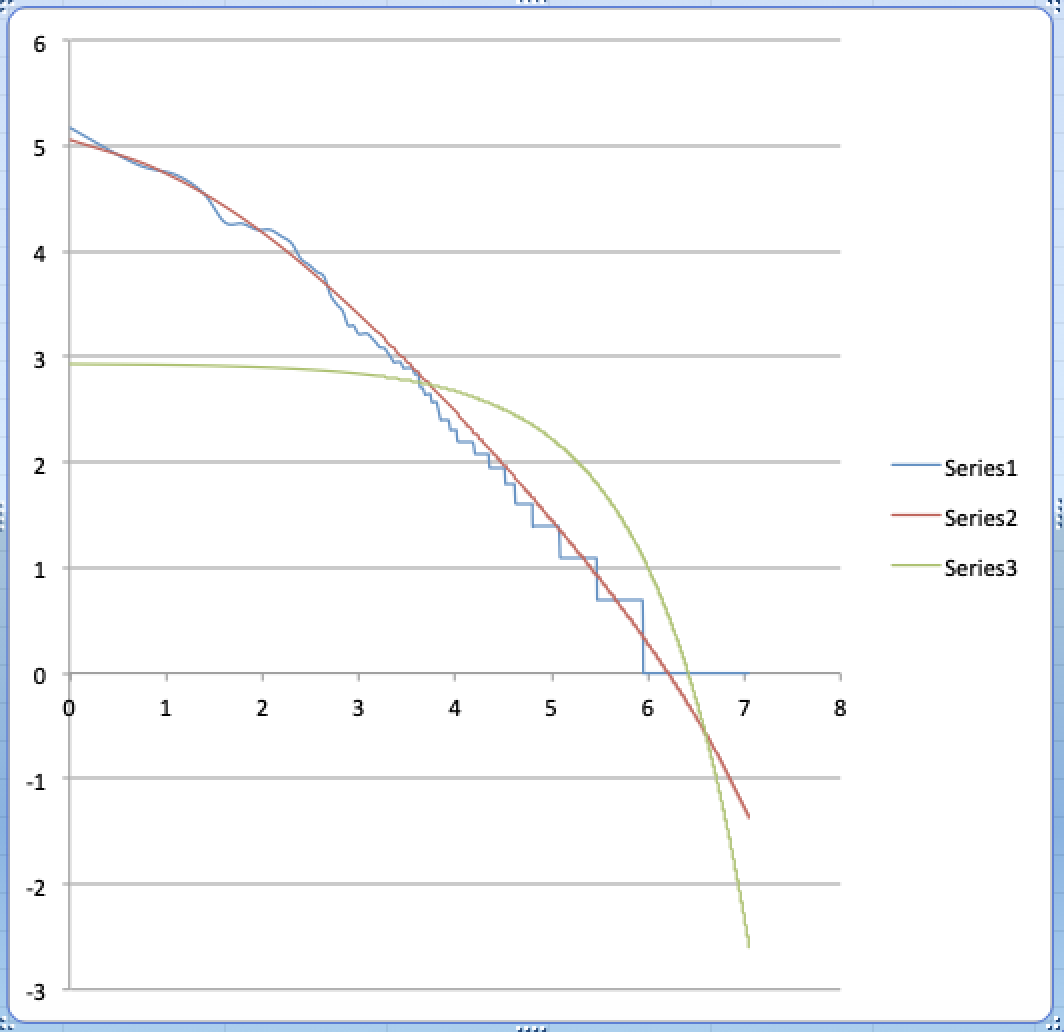} }}%
    \caption{In (a) the numbers of appearances of words in the H. G. Wells story `The magic shop' \citep{wells1903} is represented, ranked from lowest energy level, corresponding to the most often appearing word, to highest energy level, corresponding to the least often appearing word, as listed in Table \ref{magicshop}. The blue graph (Series 1) represents the data, i.e. the collected numbers of appearances from the story (column `Appearance numbers $N(E_i)$' of Table \ref{magicshop}), the red graph (Series 2) is a Bose-Einstein distribution model for these numbers of appearances (column `Bose-Einstein modeling' of Table \ref{magicshop}), and the green graph (Series 3) is a Maxwell-Boltzmann distribution model (column `Maxwell-Boltzmann modeling' of Table \ref{magicshop}). In (b) the $\log / \log$ graphs of the appearance numbers distributions are represented. The red and blue graphs coincide almost completely in both (a) and (b) while the green graph does not coincide at all with the blue graph of the data. This shows that the Bose-Einstein distribution is a good model for the numbers of appearances while the Maxwell-Boltzmann distribution is not.}%
    \label{magicshopgraphmagicshoploggraph}%
\end{figure}

\begin{figure}
\centering
\includegraphics[width=8cm]{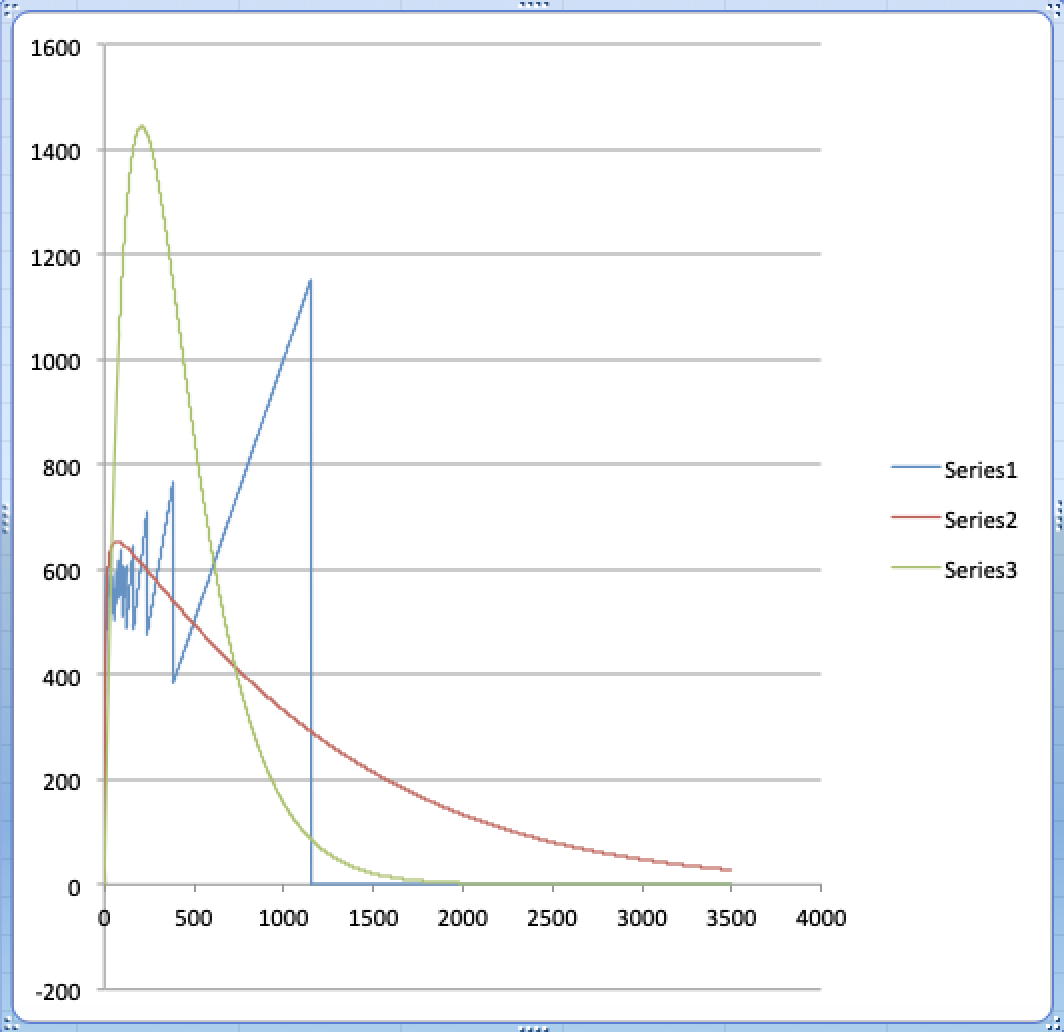}
\caption{A representation of the `energy distribution' of the H. G. Wells story `The magic shop' \citep{wells1903} as listed in Table \ref{magicshop}. The blue graph represents the energy radiated by the story per energy level (column `Energies from data $E(E_i)$' of Table \ref{magicshop}), the red graph represents the energy radiated by the Bose-Einstein model of the story per energy level (column `Energies Bose-Einstein' of Table \ref{magicshop}), and the green graph represents the energy radiated by the Maxwell-Boltzmann model of the story per energy level (column `Energies Maxwell-Boltzmann' of Table \ref{magicshop}).}
\label{magicshopenergygraph}
\end{figure}

In Figure \ref{magicshopgraphmagicshoploggraph} (a) and Figure \ref{magicshopgraphmagicshoploggraph} (b), we have represented, respectively, the numbers of the appearing and not appearing words with respect to the energy levels, a graph very steeply going down, and the $\log/\log$ graphs of these numbers of appearances, where we take the logarithm of both $y$ and $x$. In Figure \ref{magicshopenergygraph}, we have represented the amounts of radiated energy with respect to the energy levels, and we see that this time the red graph representing the Bose-Einstein model of the data, after steeply going up and reaching a maximum, goes slowly down to touch closely the zero level of amount of energy radiated for high energy level cognitons. We see again, like in Figure \ref{piglethaffalunmpgraphpiglethaffalunmploggraph}, that the Bose-Einstein distribution function, the red graph, gives an almost complete fit with the data, the  blue graph, and gives definitely a much better fit than the Maxwell-Boltzmann distribution function, the green graph, does. Let us look more carefully to the amounts of energy graphs in Figure \ref{magicshopenergygraph}. Also here we see that the red graph, which is the Bose-Einstein distribution, is a much better fit for the blue graph of the data, than the green graph, which is the Maxwell-Boltzmann distribution. We see that the maximum amount of radiation is reached at energy level $E_{70}$ in the state of the cogniton characterized by {\it Door} and the amount is $652.55204$ energy units. So the frequency of {\it Door} would be the dominant color with which the story `The magic shop' shines.

Comparing with the Winnie the Pooh story, we have a higher temperature, $kT$ equals 722 instead of 593, a higher fugacity, $f$ equals 0.9951 instead of 0.9923, and a higher chemical potential, $\mu$ is $-3.576$ instead of $-4.581$. This will be generally so when we consider longer texts like again will be illustrated by the text of `Gulliver's Travels' considered in Section \ref{boseeinsteinzipf}. We mentioned already that the sodium condensate realized at MIT, which we described above in detail, is a better model for the `magic shop' story, and indeed, in Table \ref{energylengthtablesodium} we can see that the harmonic oscillator level spacing for the sodium condensate is around $0.5\, {\rm nK}$ while the temperature of the sodium gas is $1\, {\rm mK}$, which is a factor 2000 in difference of size. In Table \ref{magicshop}, we see that we have 3500 energy levels for the story `The magic shop', which is of the same order of magnitude. The number of atoms in the MIT sodium condensate was estimated to be 500000, which is way more still than the number of words in the H. G. Wells story `The magic shop', which is 3934. When we analyze larger texts that come closer to this size, such as the text of Gulliver's Travels in Section \ref{boseeinsteinzipf}, we find an even better correspondence in magnitudes with the data of the sodium condensate. But before showing this, we have to investigate more in depth another aspect of our modeling, namely the aspect related to the `global energy level structure'.

We have not yet revealed the parameters $A$, $B$, $C$ and $D$ for the story `The magic show', they have the following values
\begin{eqnarray}
&&A  \approx 1.0005 \quad B  \approx 722.05 \quad f  \approx 0.9951 \quad \mu  \approx -3.576 \quad C  \approx 0.0531 \quad D  \approx 208.28
\end{eqnarray}
There are two quantum models that also in physics are used as an inspiration for the energy level structure of the trapped atoms, one is the `harmonic oscillator and its variations' (Appendix \ref{appendixquantumharmonicoscillator}) and the other is the `particle in a box and its variations' (Appendix \ref{appendixparticleinabox}). From the harmonic oscillator model follows that the energy levels are equally (linearly) spaced, which is also the way we have modeled them for the two examples that we have considered, the Winnie the Pooh story and the H. G. Wells story. However, the energy levels of the particle in a box are quadratically spaced. We will see in the following of our analysis that in view of our experimental findings in analyzing numerous texts in all generality, the energy levels of the cognitons, depending on the story considered, are spaced following a power law, with a power coefficient which is in principle between 0 and 2, but for all the stories that we investigated was between 0.75 and 1.25. This indicates that different energy situations on both sides of the `harmonic oscillator' are at play, from the `anharmonic oscillator', with converging spacings between energy levels, to the `particle in a box', with  quadratic spacings between energy levels. We will show in next section how this generalization for the energy spacings strengthens the correspondence with Zipf's law in human language.
\section{Zipf's law and the Bose gas of human language \label{boseeinsteinzipf}}
\begin{figure}
\centering
\includegraphics[width=8cm]{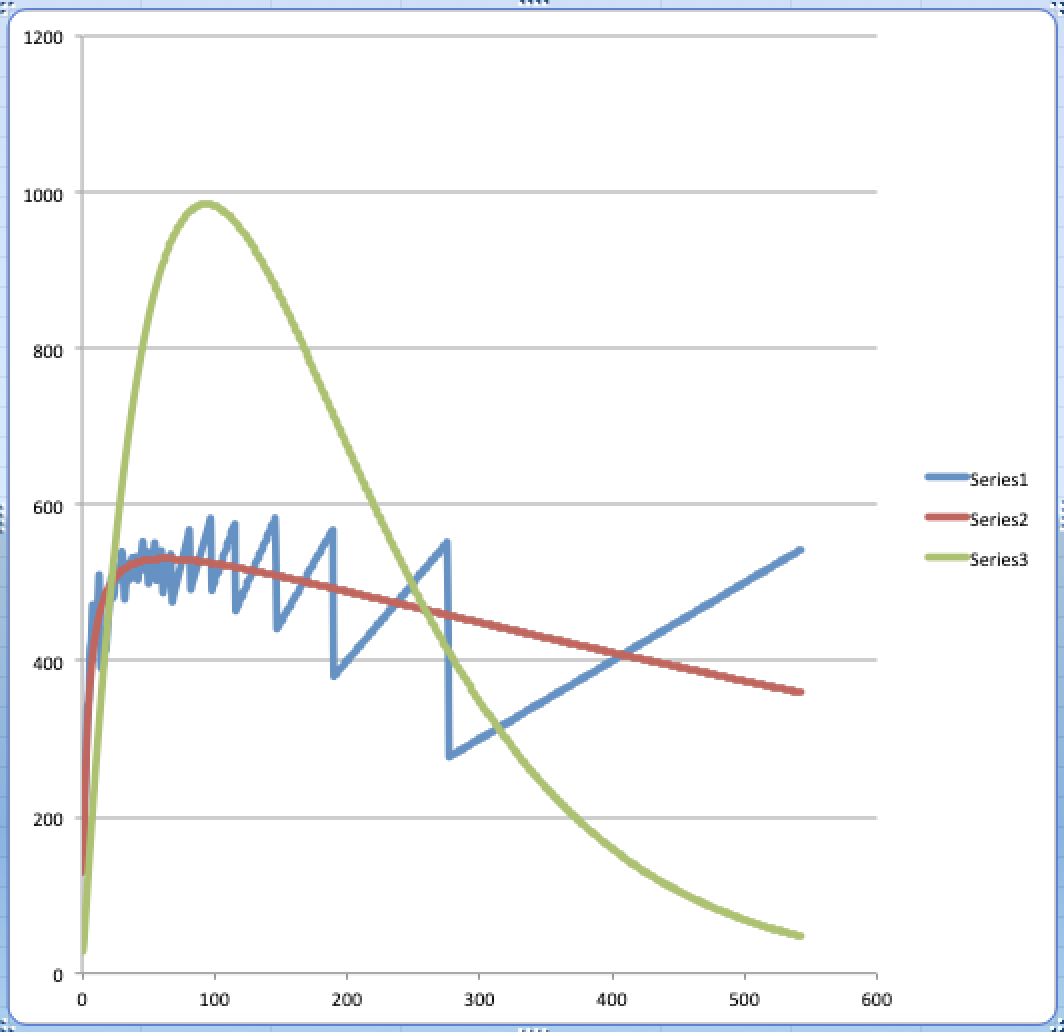}
\caption{The blue graph (Series 1) is a representation of the products $R_i \times N_i$ for the text of the Winnie the Pooh story that we have investigated in Section \ref{languagebosegas}, where $R_i$ is the $i$-th rank in Zipf's ranking and $N_i$ is the number of appearances corresponding to this ranking. The $x$-ccordinate represents the ranks $R_i$, and the $y$-coordinate represents the products $R_i \times N_i$. For the red graph (Series 2) and the green graph (Series 3) the values of respectively the Bose-Einstein distribution and the Maxwell-Boltzmann distribution which we developed in section \ref{languagebosegas} were used as a comparison with the graph in Figure \ref{piglethaffalunmpenergygraph}.}
\label{zipfpiglethaffalunmpgraph}
\end{figure}

Zipf's law is considered to be one of the mysterious structures encountered in language \citep{zipf1935,zipf1949}. It was originally noted in its most simple form in the following way. When ranking words according to their numbers of appearances in a piece of text, the product of the rank with the number of appearances is a constant. Hence Zipf's law was originally stated mathematically as follows
\begin{eqnarray} \label{zipfslaw}
R \times N = c
\end{eqnarray}
where $R$ is the rank, $N$ the number of appearances, and $c$ is a constant.
We have presented in Figure \ref{zipfpiglethaffalunmpgraph} the products $R_i \times N_i$ for the text of the Winnie the Pooh story that we have investigated in Section \ref{languagebosegas}, where $R_i$ is the $i$-th Zipf's ranking and $N_i$ is the number of appearances corresponding to this ranking. The $x$-coordinate of the graphs in Figure \ref{zipfpiglethaffalunmpgraph} represents the ranks $R_i$, and the $y$-coordinate represents the products $R_i \times N_i$ for the blue graph, and the values of respectively the Bose-Einstein distribution, and the Maxwell-Boltzmann distribution for the red and green graphs.

It is not a coincidence that there is a striking resemblance between the graphs shown in Figure \ref{zipfpiglethaffalunmpgraph} and the energy distribution graphs of the Winnie the Pooh story as a boson gas shown in Figure \ref{piglethaffalunmpenergygraph}. Indeed, the energy levels $E_i$ that we introduced are very simply related to the Zipf rankings $R_i$, the only difference being that we started with value zero for the lowest energy level, while Zipf started with value 1 for his first rank. Hence, more concretely, we have
\begin{eqnarray}
R_i = E_i +1
\end{eqnarray}
This means that although none of the values of the Zipf products in Figure \ref{zipfpiglethaffalunmpgraph} is equal to the energies in Figure \ref{zipfpiglethaffalunmpgraph}, the differences are small, because $R_i$ equals $E_i +1$.
Consulting Table \ref{piglethaffalunmp}, we can see that the biggest difference is at the zero point of the graph, where on the $x$-axis $E_0 = 0$ and $R_0 = 1$, hence between the product $R_0 \times N_0$, which equals $(E_0+1) \times N_0$, that is between $1 \times 133 = 133$ and $E_0 \times N_0 = 0 \times 133 = 0$. This can not easily be seen as a difference between the graphs of Figure \ref{zipfpiglethaffalunmpgraph} and the graphs of Figure \ref{piglethaffalunmpenergygraph}, since 133 is still little compared to the values the functions take at $R_1$ and $E_1$. Again consulting Table \ref{piglethaffalunmp}, we indeed see that $R_1 \times N_1 = (E_1+1) \times N_1 = 2 \times 111 = 222$, while $E_1 \times N_1 = 1 \times 111 = 111$. This means that both the `product graph' of Figure \ref{zipfpiglethaffalunmpgraph} and the `energy distribution graph' of Figure \ref{piglethaffalunmpenergygraph} go quickly up between $R_0$ and $R_1$ and between $E_0$ and $E_1$, the first from value 113 to value 222, and the second from value 0 to value 111, which is almost with the same steepness. Both graphs will then remain increasing quite quickly and then slowly flatten till they reach their maxima at Zipf rank $R_{70}$ and energy level $E_{71}$. Then, from this maximum on, both the Zipf product and the energy distribution slowly decrease from their maxima to a lower value. More specifically, the maximum value is $522.79$ in both cases, and for the last considered Zipf rank $R_{542}$ and energy level $E_{542}$ we find values 359.22 and 358.55 respectively. This shows that there is a decreasing for the Zipf products and not constancy like Zipf's law predicts.

In the foregoing reasoning on Zipf's law, we have always considered the two graphs, the blue and the red one, in both Figure \ref{zipfpiglethaffalunmpgraph} and Figure \ref{piglethaffalunmpenergygraph}. Of course, Zipf did not know of the Bose-Einstein distribution that is represented by the red graph in both figures, and which we used to model the data, represented by the blue graph in both figures. Hence Zipf only had the blue graph in Figure \ref{zipfpiglethaffalunmpgraph} available to come up with the hypothesis that the product of rank and number of appearances is a constant. If one considers the blue graph in Figure \ref{zipfpiglethaffalunmpgraph}, one could indeed imagine it to vary around a constant function, certainly in the middle part of the graph. The beginning part can then be considered as a deviation, which is also what Zipf did when noting that in the first ranks the law did not hold up well. It was also known to Zipf that the end part of the graph, as a consequence of how ranks and numbers of appearances behave there, making the product go up and down heavily, did not behave very well with respect to his law either, and the slight downward slope all at the end was identified by Zipf as well. We see it explicitly pictured by the red graph, representing the Bose-Einstein distribution modeling of the data.

There is however another aspect of the situation which was overlooked by Zipf. It is self-evident that `if Zipf's law is a law, it has to be a probabilistic law'. Let us specify what we mean by this. Suppose we had a large number of texts available with exactly the same number of different words in it, such that a Zipf analysis would lead to the same total number of ranks for each of the texts. Zipf's graphs, including the `product graph', i.e. the blue graph in Figure \ref{zipfpiglethaffalunmpgraph}, will then show a statistical pattern for the set of texts where it is tested on. Suppose we make averages for the numbers of appearances pertaining to the same rank over the available texts, then the function representing these averages of the numbers of appearances for the different texts will be a distribution function with a steep upward slope in the first ranks going towards a maximum and then a slow downwards slope in the ranks after this maximum. It will be a function similar to the Bose-Einstein distribution we have used to model texts as Bose gases, i.e. the red graph. This will be even more so when we add the two constraints that in our case follow naturally from our modeling, namely that the different texts need to count the same total number of words, and the sum of the products, which in our interpretation of the Bose gas model is the total energy, needs to be the same for each one of the texts. What is however more important still is that `if Zipf's law is a probabilistic law, we should also introduce rankings that represent words with a zero number of appearances', exactly like what we have done for the H. G. Wells story `The magic shop', for which we have represented the data and the Bose-Einstein model in Table \ref{magicshop}, and the graphs representing these data in Figure \ref{magicshopgraphmagicshoploggraph} (a), in Figure \ref{magicshopgraphmagicshoploggraph} (b) and in Figure \ref{magicshopenergygraph}. 

If we look carefully at the energy distribution graph in Figure \ref{magicshopenergygraph}, we can understand again somewhat better why Zipf came to believe that the products of the ranks and the numbers of appearances are a constant. Indeed, having added the zero number of appearance till the energy distribution becomes close to zero in the high energy levels, like shown in Figure \ref{magicshopenergygraph}, we can see how the blue graph goes first far up where the one word appearance cases are, to compensate the long row of zero appearance cases that take a great part of the $x$-axis. So, if one leaves out the zero appearance part, one easily can get the impression that the blue graph represents a constant on average, at least when neglecting the low energy levels at the start, where it goes steeply up.
\begin{figure}
\centering
\includegraphics[width=8cm]{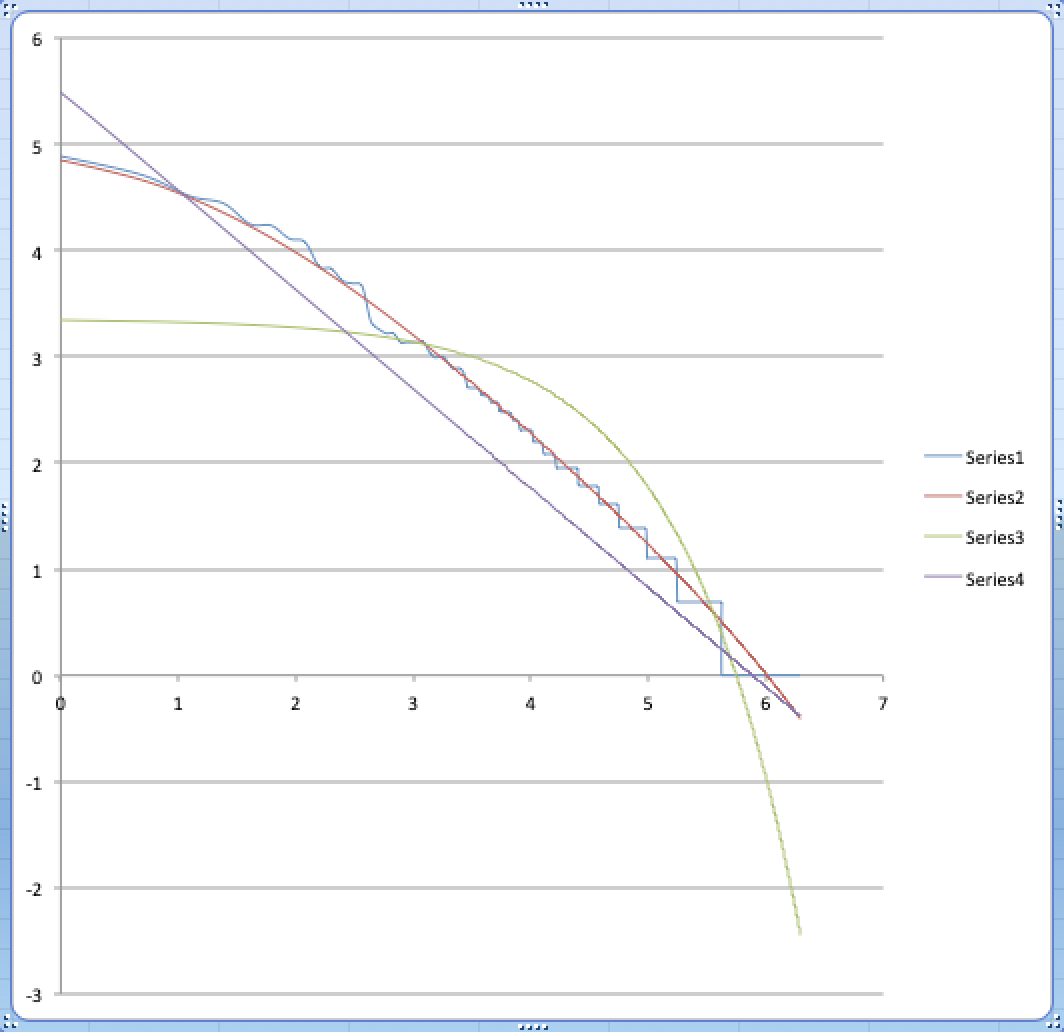}
\caption{Representation of the $\log/\log$ graphs of the Zipf data. The blue graph represents that data (Series 1), the red graph represents the Bose-Einstein model (Series 2), the green graph represents the Maxwell-Boltzmann model (Series 3) and the purple graph represents a straight line (Series 4) that is an `as good as possible approximation' of the other graphs to illustrate that the gradient of the `straight line approximation' is not equal to $-1$.}
\label{zipfpiglethaffalunmploggraphstraightline}
\end{figure}
Most of the investigations of Zipf's findings afterwards concentrated on the $\log/\log$ graph representation, where the $\log$ is taken for the rank as well as for the numbers of appearances, hence the Zipf equivalents for the $\log/\log$ graphs we considered for our Bose gas modeling represented in Figure \ref{piglethaffalunmpgraphpiglethaffalunmploggraph} (b) and in Figure \ref{magicshopgraphmagicshoploggraph} (b). For what concerns Zipf's law expressed in (\ref{zipfslaw}), the $\log/\log$ graph of the Zipf product gives rise to a straight line with gradient equal to $-1$. Indeed, when we take the $\log$ of both sides of (\ref{zipfslaw}) we get
\begin{eqnarray} \label{zipfsloglaw}
\log R + \log N = \log c
\end{eqnarray}
which graph, with $\log R$ on the $x$-axis and $\log N$ on the $y$-axis, is a straight line with gradient equal to $-1$. It is indeed much more easy to see by the naked eye that such a $\log/\log$ graph like those in Figure \ref{piglethaffalunmpgraphpiglethaffalunmploggraph} (b) and in Figure \ref{magicshopgraphmagicshoploggraph} (b) can be approximated well by a straight line as compared to seeing the constancy of the Zipf's products in a graph like the one in Figure \ref{zipfpiglethaffalunmpgraph}, where the constancy needs to be approximated to the up and down moving blue graph. However, the focus of all Zipf's investigations on the $\log/\log$ graphs also has its down side, in the sense that the upper and lower parts of the graph will be more easily considered as slight deviations of the straight line, while, as we see with our Bose-Einstein distribution modeling in its energy graph version, they really represent essential and significant deviations from Zipf's original product law (\ref{zipfslaw}). That in both Figure \ref{piglethaffalunmpgraphpiglethaffalunmploggraph} (b) and in Figure \ref{magicshopgraphmagicshoploggraph} (b) the graphs are slightly bent towards a concave form is the expression of Zipf's law essentially not being satisfied for low ranks and high ranks. 

The foregoing analysis is meant to provide evidence to the Bose-Einstein distribution being a better model for the Zipf data than a constant, or also still than later more complex versions of Zipf's law along the lines of still believing that the product graph is in good approximation a constant, and the $\log/\log$ version in good approximation a straight line. There is however another aspect of Zipf's finding that we want to put forward here, since it will be important for our model of a Bose gas for human language. 

In Figure \ref{zipfpiglethaffalunmploggraphstraightline}, we represented the $\log/\log$ graphs of the Zipf data (blue graph) and the Bose-Einstein (red graph) and Maxwell-Boltzmann (green graph) distributions which we used to model them, and we added a straight line (purple graph) that approximates the other graphs as good as possible. We can see that the gradient of the straight line is not equal to $-1$, but to $-0.94$. Although Zipf himself kept focusing on the straight line with gradient $-1$, it was noted by many who studied Zipf's law that a generalization was needed to take into account the gradient of the straight line usually being smaller than $-1$, hence the $\log/\log$ version of law was generalized to
\begin{eqnarray}
p\log R + \log N = \log c
\end{eqnarray}  
which made the original product of rank and frequency be generalized to
\begin{eqnarray}
R^p \times N = c
\end{eqnarray}
where $p$ is called the `power coefficient' of Zipf's law.

We will apply this `power coefficient' in Zipf's law also in our modeling. Let us explain why and how we will do so. First of all, there is no a priori reason why the energy levels would be as simple as we presented it in the two examples that we considered, namely such that 
\begin{eqnarray} \label{linearsystemenergylevels}
E_i = i (E_1 - E_0)+E_0
\end{eqnarray}
where $E_1 - E_0$ is the unit of energy that we introduced. Of course, we have systematically taken $E_0 = 0$, see (\ref{Ei}), which makes the energy levels we have introduced in both stories even more simple, but it is not necessarily so that $E_0 = 0$ as a rule, which is why we now formulate the `linear system of energy levels' as in (\ref{linearsystemenergylevels}). This simple linear system is inspired by the energy levels of the quantum harmonic oscillator (Appendix \ref{appendixquantumharmonicoscillator}), where we have
\begin{eqnarray}
E_i = {h \nu \over 2} + i h \nu
\end{eqnarray}
with $\nu$ being the frequency of the oscillator. But that energy spacings between consecutive energy levels are the same, like in the case of the harmonic oscillator, is a very exceptional situation of quantization. For general quantized systems the spacings between consecutive energy levels will not be the same, and both cases exist, for not confined quantized situations the spacings will decrease, while for confined situations the spacings will increase. For example, for the quantized energy levels of the `particle in a box' (Appendix \ref{appendixparticleinabox}), we have
\begin{eqnarray}
E_i = {h^2 \over8mL^2} + {h^2 \over8mL^2} i^2
\end{eqnarray}
which means that the energy levels change quadratically in function of the unit of energy
\begin{eqnarray}
E_i = i^2 (E_1 - E_0) + E_0
\end{eqnarray}
Remark that in Appendices \ref{appendixparticleinabox} and Appendix \ref{appendixquantumharmonicoscillator} we have used $n$ to indicate the `quantum numbers', because that is the traditional letter used for quantum numbers within standard quantum theory. In the approach we followed we have used $i$ to indicate the `energy levels', because we do not want to make a direct and exclusive reference to standard quantum theory alone, since our aim is to also make a connection with Zipf's law in language. More generally, we want to elaborate a `quantum cognition theory' for `human language and cognition' from basic principles on a more foundational level than the one where standard quantum theory is situated, building on earlier work in quantum cognition and quantum computer science \citep{aertsaerts1995,khrennikov1999,atmanspacher2002,gaboraaerts2002,vanrijsbergen2004,aertsczachor2004,widdows2004,bruzacole2005,busemeyeretal2006,pothosbusemeyer2009,lambertmoglianskizamirzwirn2009,bruzaetal2009,busemeyerbruza2012,dallachiaraetal2012,dallachiaraetal2015,havenkhrennikov2013,melucci2015,pothosetal2015,blutnerbeimgraben2016,moreirawichert2016,broekaertetal2017,gaborakitto2017,busemeyerwang2018}. 

In this we will also be inspired by the global foundational work we have done in our Brussels group \citep{aerts1986,aerts1990,aerts1999,aerts2009b,aertsetal2010,aertsetal2012,aertsetal2013,aertsetal2018a,aertsetal2019a,aertsbroekaertgabora2011,aertsgabora2005a,aertsgabora2005b,aertsgaborasozzo2013,aertssassolidebianchi2014,aertssassolidebianchi2017,aertssassolidebianchisozzo2016,aertssozzo2011,aertssozzo2014,aertssozzoveloz2015a,aertssozzoveloz2016,sassolidebianchi2011,sassolidebianchi2013,sassolidebianchi2014,sassolidebianchi2019,sozzo2014,sozzo2015,sozzo2017,sozzo2019,velozzhaoaerts2014,velozdujardin2015}, and by the more specific work on the `conceptuality interpretation' \citep{aerts2009a,aerts2010a,aerts2010b,aerts2013,aerts2014,aertsetal2018d,aertsetal2019c}. To mention a concrete aspect in need of a more foundational approach, there is yet no well identified spatial domain for human language, which means that we will have to build a `quantum cognition' without reference to space \citep{aerts1999,sassolidebianchi2019}.

The `harmonic oscillator' and the `particle in a box' are both special cases where the one-dimensional Schr\"odinger equation can be solved analytically, but for boson gases power law potentials have been studied as more general models \citep{bagnatopritchardkleppner1987}, and hence we will also introduce in our approach a more general variation of the energy levels than the linear one, namely one of a `power law change'
\begin{eqnarray} \label{energyspacingwithpower}
E_i = i^p (E_1 - E_0) + E_0
\end{eqnarray}
Let us show right away how the introduction of a power law for the energy level spacings gives extra strength to the Bose-Einstein modeling of the texts of stories expressed in human language. This time we choose a much larger text than the two ones we investigated before, namely the text of the satirical work Gulliver's Travels by Jonathan Swift \citep{swift1726}, which contains in total 103184 words, hence of the order of 40 times more than the Winnie the Pooh story and 25 times more than the H. G. Wells story. When analyzed as the Winnie the Pooh and the H. G. Wells story, with the hypothesis of equally spaces energy levels, or, which is equivalent, with a power coefficient spacing of the energy levels with power coefficient equal to 1, we find a total of 8294 energy levels without adding the zero number of appearances levels, and the ten highest numbers of appearances and their corresponding words are {\it The}, 5838, {\it Of}, 3791, {\it And}, 3633, {\it To}, 3400, {\it I}, 2852, {\it A}, 2442, {\it In}, 1976, {\it My}, 1593, {\it That}, 1280 and {\it Was}, 1263.
\begin{table}
\small
\centering
\begin{tabular}{p{2.2cm}p{2.2cm}p{2.2cm}p{2.2cm}}
\multicolumn{4}{c}{(a) Gulliver's Travels without power coefficient} \\
\hline
Cogniton state & Energy level & Appearance number & Bose-Einstein value  \\
\hline
{\it The} & $E_0 = 0$  & 5838 & 16454.07 \\
{\it Of}    & $E_1 = 1$  & 3791 & 6297.00 \\
{\it And}  & $E_2 = 2$  & 3633 & 3893.39 \\
{\it To}     & $E_3 = 3$  & 3400 & 2817.73 \\
{\it I}        & $E_4 = 4$  & 2852 & 2207.73 \\
{\it A}       & $E_5 = 5$  & 2442 & 1814.80 \\
{\it In}      & $E_6 = 6$   & 1976 & 1540.59 \\
{\it My}    & $ E_7 = 7$  & 1593 & 1338.35 \\
{\it That}  & $E_8 = 8$  &  1280 & 1183.03 \\
{\it Was}  & $E_9 = 9$   & 1263 & 1060.00 \\
{\it Me}    & $E_{10}=10$ & 991   & 960.14  \\
\multicolumn{4}{c}{(b) Gulliver's Travels with power coefficient} \\
\hline
Cogniton state & Energy level & Appearance number & Bose-Einstein value  \\
\hline
{\it The} & $E_0 = 0$  & 5838 & 5305.75 \\
{\it Of}    & $E_1 = 1$  & 3791 & 4164.08 \\
{\it And}  & $E_2 = 2.11$  & 3633 & 3358.88 \\
{\it To}     & $E_3 = 3.28$  & 3400 & 2795.26 \\
{\it I}        & $E_4 = 4.47$  & 2852 & 2384.16 \\
{\it A}       & $E_5 = 5.69$  & 2442 & 2073.04 \\
{\it In}      & $E_6 = 6.92$   & 1976 & 1830.30 \\
{\it My}    & $ E_7 = 8.18$  & 1593 & 1636.12 \\
{\it That}  & $E_8 = 9.45$  &  1280 & 1477.55 \\
{\it Was}  & $E_9 = 10.73$   & 1263 & 1345.80 \\
{\it Me}    & $E_{10}=12.02$ & 991   & 1234.70  \\
\end{tabular}
\caption{The eleven lowest energy levels of the novel Gulliver's Travels by Jonathan Swift \citep{swift1726}. The values of the Bose-Einstein model are compared with the data, i.e. the numbers of appearances of the words in the text in (a) without the introduction of a power coefficient and in (b) with the introduction of a power coefficient. The comparison for all energy levels can be seen for (a) in Figure \ref{gulliverstravelsloggraphgulliverstravelspowerloggraph} (a) and for (b) in Figure \ref{gulliverstravelsloggraphgulliverstravelspowerloggraph} (b).}
\label{gulliverstravelswithpower}
\label{gulliverstravels}
\end{table}
\normalsize

\begin{figure}%
    \centering
    \subfloat[With power coefficient $p = 1$]{{\includegraphics[width=8cm]{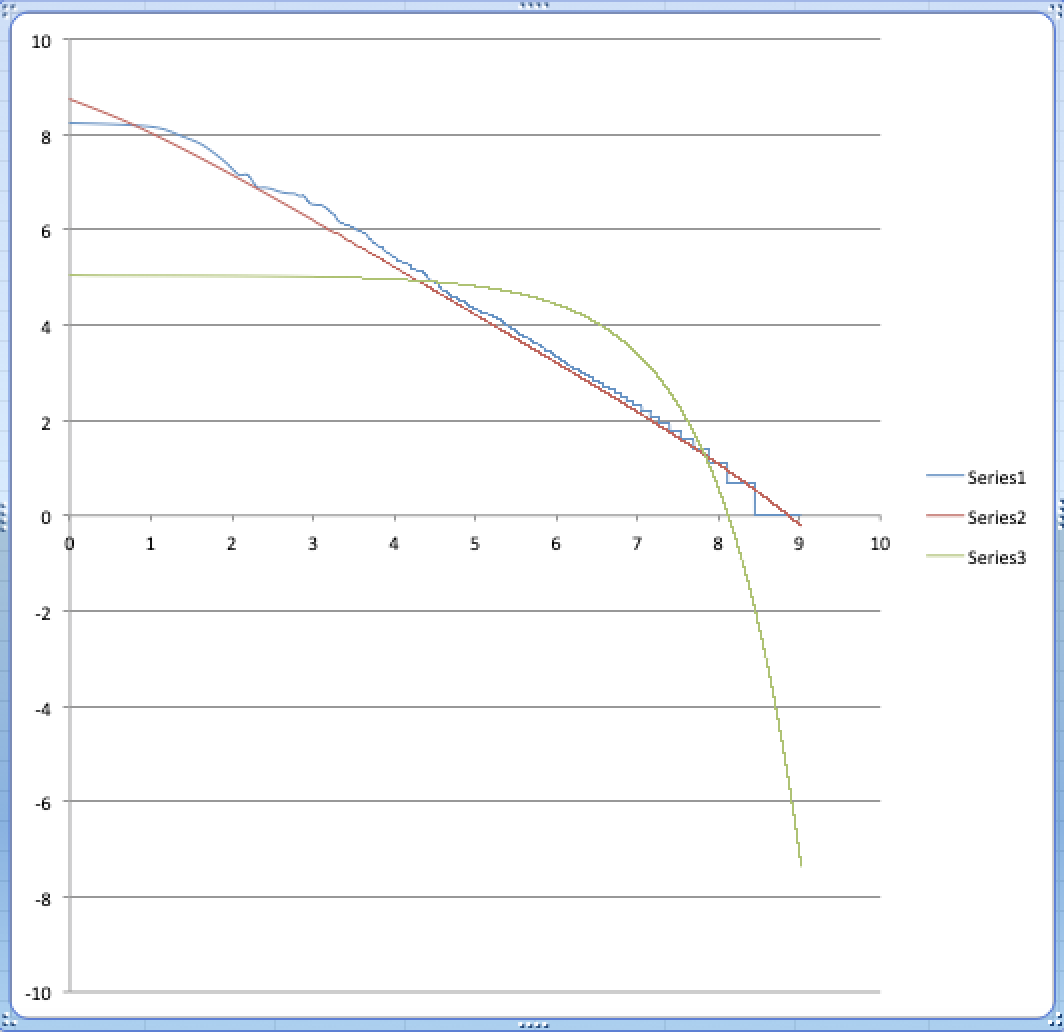} }}%
    \qquad
    \subfloat[With power coefficient $p = 1.08$]{{\includegraphics[width=8cm]{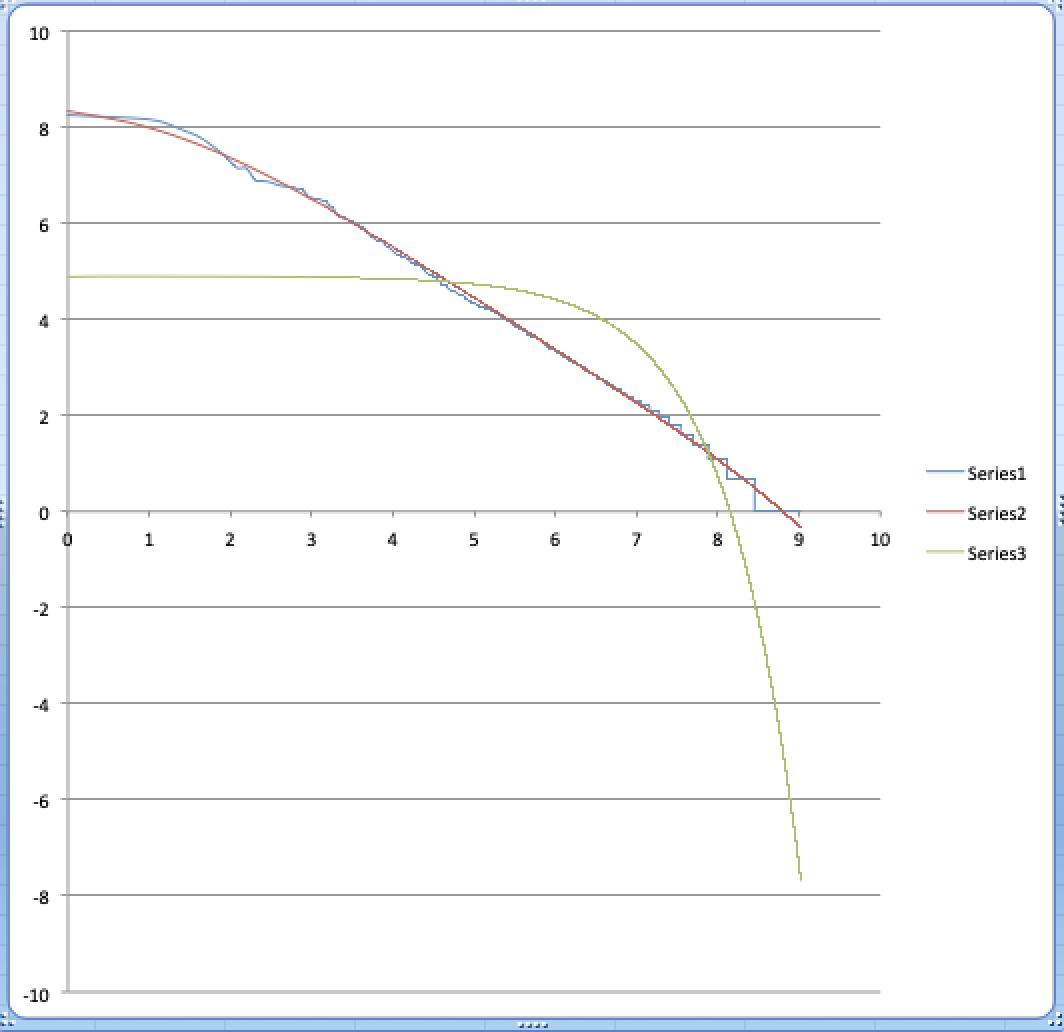} }}%
    \caption{The $\log / \log$ graph of the frequency distributions of  the novel `Gulliver's Travels' \citep{swift1726}. In (a) it is shown how the Bose-Einstein distribution represented by the red graph (Series 2), although still a much better model than the Maxwell-Boltzmann distribution represented by the green graph (Series 3), fails to be as good a model when compared with the Winnie the Pooh story and the H. G. Wells story (Figure \ref{piglethaffalunmpgraphpiglethaffalunmploggraph} (b) and Figure \ref{magicshopgraphmagicshoploggraph} (b)). Indeed, its values (Table \ref{gulliverstravels}) are too high in the lowest energy levels and too low in the middle energy levels, when compared to the data represented by the blue graph (Series 1). However, with addition of the power coefficient 1.08, applied to the spacings between energy levels, in (b) it is shown how the Bose-Einstein distribution model is again a very good model for the data. See Table \ref{gulliverstravelswithpower} for the explicit values of the eleven lowest energy levels.}%
    \label{gulliverstravelsloggraphgulliverstravelspowerloggraph}%
\end{figure}

In Figure \ref{gulliverstravelsloggraphgulliverstravelspowerloggraph} (a), we represented the $\log/\log$ version of the `numbers of appearances' graphs for the Gulliver's Travels story, the blue graph representing the data, the red graph the Bose-Einstein model, and the green graph the Maxwell-Boltzmann model. We can see right away that again the Bose-Einstein model is a much better representation of the data than the Maxwell-Boltzmann model, but we can also see that it is a less good representation of the data than it was the case for the Winnie the Pooh story and the H. G. Wells story. Indeed, the red graph indicates noticeably too high values in the low energy levels and for a large region in the middle energy levels it has values that are too low. In Table \ref{gulliverstravels} (a) we give the eleven lowest energy levels values of the Bose-Einstein distribution model corresponding to the states of the cognitons, i.e. the corresponding words, and compare with the data, and see that the first ones are too high, while the following ones are too low.
For the lowest energy level, with cognitons in state {\it The}, we find the Bose-Einstein distribution to have a value of 16454.07 while {\it The} appears only 5838 times in the Gulliver's Travels text. This is indeed a big difference, the Bose-Einstein is more than three times the experimental value of the number of appearances. We find a similar too high value for the Bose-Einstein distribution for the two next states of the cognitons, the state {\it Of} has a Bose-Einstein distribution value of 6297.00, while {\it Of} appears only 3791 in the text, the state {\it And} has a Bose-Einstein distribution value of 3893.39, while {\it And} appears only 3633 times in the text. For the next states of the cognitons the Bose-Einstein model, however, gives values too low with respect to the experimental data.  For {\it To} the Bose-Einstein distribution value is 2817.73 while {\it To} appears 3400 times in the text, for {\it I} the Bose-Einstein distribution value is 2207.73 while {\it I} appears 2852 times, for {\it A} the Bose-Einstein distribution value is 1814.80 while it appears 2442 times, for {\it In} the Bose-Einstein distribution value is 1540.59 while it appears 1976 times, for {\it My} the Bose-Einstein distribution value is 1338.35 while it appears 1593 times, for {\it That} the Bose-Einstein distribution value is 1183.03 and it appears 1280 times, for {\it Was} the Bose-Einstein distribution value is 1060.00 and it appears 1263 times, and for {\it Me} the Bose-Einstein distribution value is 960.14 while {\it Me} appears 991 times in the text of the Gulliver's Travels story.

We will now apply a `power law' to the spacings between the energy levels, as per (\ref{energyspacingwithpower}), and will see that we can come to a much better match of the Bose-Einstein distribution with the data. Indeed, after applying the power $p=1.08$ to the energy spacings between the energy intervals, we found an almost perfect match and represented the $\log/\log$ version of the graphs in Figure \ref{gulliverstravelsloggraphgulliverstravelspowerloggraph} (b). The values for the eleven lowest energy levels data compared with the Bose-Einstein model with power coefficient 1.08 are given in Table \ref{gulliverstravels} (b).

\begin{figure}
\centering
\includegraphics[width=8cm]{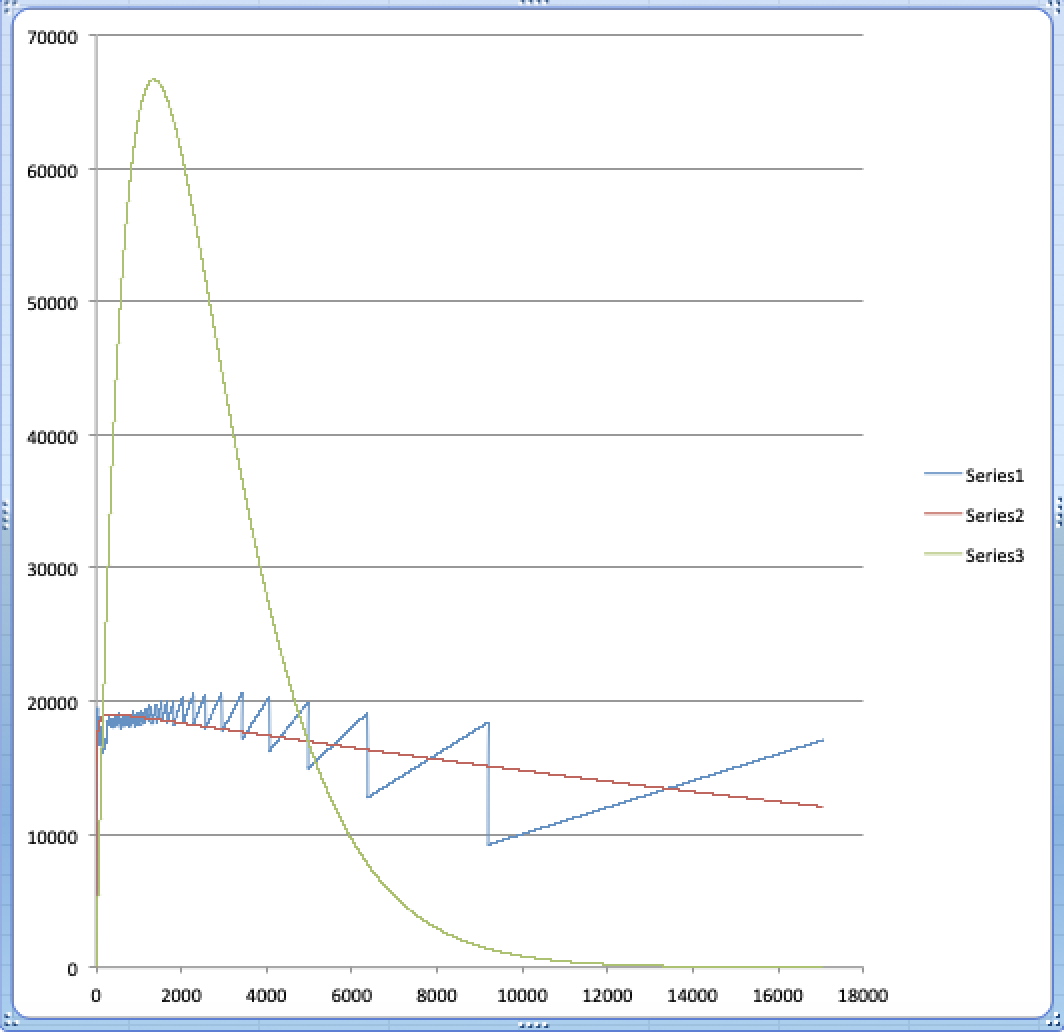}
\caption{A representation of the `energy distribution' of the story of Gulliver's Travels \citep{swift1726}. The blue graph (Series 1) represents the energy radiated by the story per energy level, the red graph (Series 2) represents the energy radiated by the Bose-Einstein model of the story per energy level, and the green graph (Series 3) represents the energy radiated by the Maxwell-Boltzmann model of the story per energy level. We have not added the highest energy levels radiation, but the very slowly descending slope after the maximum 18377.11 has been reached at energy level 43.65, shows that many levels will have to be added with zero number of appearance words for the Bose-Einstein function to approximate zero.}
\label{gulliverstravelsenergygraph}
\end{figure}

We have tested the Bose-Einstein model on a large number of stories, short stories and long stories of the size of  novels, and when we allow the energy spacings between different energy levels to vary according to a power law, we have been able to construct a perfectly matching Bose-Einstein model for the data for all of the considered stories. The power that was each time needed was situated between 0.75 and 1.25. 

We want to emphasize that it is remarkable how the application of the power 1.08 to the linear version of the text of the novel of Gulliver's Travels makes the Bose-Einstein model fit so well the data, and we observed the same effect of the introduction of a power on an original linear version of the model for many of the other example texts that we investigated. We mentioned already how those who studied Zipf's law came to add a power to take into account that the gradient of the best fitting straight line in the $\log/\log$ version of the graphs was not equal to $-1$. However, also the concave slightly curbed nature of the lowest energy level ranks was noticed and tried to be remedied by making the law more general still, however in purely ad hoc ways with the only aim to fit the data \citep{mandelbroth1953,mandelbroth1954,edmundson1972}. That this slight concave curb appears in the Bose-Einstein distribution as a consequence of adding a power to the spacings between energy levels in exactly a way to make it fit with the data is in this sense remarkable, and since we saw it happening in many of the other examples for different values of the power, it is a strong indication of the Bose-Einstein model touching onto a fundamental property of human language.

In Figure \ref{gulliverstravelsenergygraph}, we have represented the low energy part of the `energy distribution' of the story of Gulliver's Travels \citep{swift1726}. The blue graph represents the energy radiated by the story per energy level, the red graph represents the energy radiated by the Bose-Einstein model of the story per energy level, and the green graph represents the energy radiated by the Maxwell-Boltzmann model of the story per energy level. We have not added the highest energy levels radiation because we wanted to show the detail of the low energy distribution, the one where the Bose-Einstein condensate dynamics of the text plays out. The maximum with a value of 18377.11 is reached at energy level 43.65 at quantum number 33, hence very close to the low level energies. The parameters $A$, $B$, $C$ and $D$ of the Bose-Einstein and Maxwell-Boltzmann models are
\begin{eqnarray}
A  \approx 1.00019 \quad B  \approx 19356.22 \quad f  \approx 0.9998 \quad \mu  \approx -3.648 \quad C  \approx 0.0075 \quad D  \approx 1355.31
\end{eqnarray}
Comparing with the Winnie the Pooh story and with the H. G. Wells story we have a higher temperature, $kT$ equals 19356 instead of 722 or 593, a higher fugacity, $f$ equals 0.9998 instead of 0.9951 or 0.9923, and a higher chemical potential, $\mu$ equals $-3.648$ instead of $-3.576$ or $-4.581$. As we remarked already, when we compared the parameters for the Winnie the Pooh story and the H. G. Wells story, this is generally what we expect to happen for longer texts. 

\section{Identity and Indistinguishability \label{identifyindistinguishability}}
We want to reflect now on what can the obtained results teach us about the notions of `identity and indistinguishability' with respect to how they are used in human language and in quantum theory. We also want to reflect on the way in which these results support the `conceptuality interpretation of quantum theory' \citep{aerts2009a,aerts2010a,aerts2013,aerts2014,aertsetal2018d,aertsetal2019c}. Before we start our analysis, we repeat that all the words appearing in the stories that we considered are `states' of the `cogniton', which is the entity that for human language is what a `photon' is for light, or what a `rubidium 87 atom' is for the rubidium gas used to fabricate the Bose-Einstein condensate in \citet{andersonetal1995}. 

Let us first analyze how the issue of `identity and indistinguishability' appears in quantum theory. It is structurally speaking a consequence of the generally adopted mathematical rule that wave functions should be symmetrized or anti-symmetrized, depending of whether the quantum particles in question are bosons or fermions. This entails that a multi-particle wave functions is always a superposition of products of the single particle building blocks of the multi-particle wave function, such that the different product pieces are chosen in a way that the total wave function is symmetric or anti-symmetric, depending on whether the composed quantum entity is a boson or a fermion. Let us make concrete what this means when we apply a quantum model to the text of the Winnie the Pooh story. The set of energy levels $\{E_0, \ldots, E_{542}\}$ shown in Table \ref{piglethaffalunmp} are in principle the energy levels for a one particle situation in quantum theory, and the many particle situation of a text is then described in a Hilbert space which is the tensor product of, in the case of the Winnie the Pooh story, 2655 Hilbert spaces of which each one describes a one particle situation. The symmetrization is obtained by a superposition of all possible permutations of the original products and a renormalization to make the wave function a unit vector.

Let us consider the very simple version of this symmetrization procedure for two boson quantum particles which we call $A$ and $B$, to see how challenging it is to try to understand its meaning. Both particles, when not part of a composite system, are described by their wave functions $\psi_A(x_A)$ and $\psi_B(x_B)$, where $x_A$ and $x_B$ are variables we considered for respectively particle $A$ and particle $B$. When the two particles are joined in a single composite system, the latter is described by the symmetrized wave function
\begin{eqnarray} \label{symmetrystate}
\psi(x_A,x_B)=c(\psi_A(x_A)\psi_B(x_B)+\psi_B(x_B)\psi_A(x_A))
\end{eqnarray}
where $c$ is the renormalization constant. To see to what type of problems this symmetrization procedure leads, suppose for a moment that $x_A$ and $x_B$ are position variables pertaining to separated regions of space $R_A$ and $R_B$, such that for both particles $A$ and $B$ we can understand $\psi_A(x_A)$ and $\psi_B(x_B)$ as being the wave function representing one particle $A$ mainly present in this region of space $R_A$, and another particle $B$ mainly present in this region of space $R_B$ -- $\psi_A(x_A)$ and $\psi_B(x_B)$ are for example wave packets which have negligible values outside respectively regions $R_A$ and $R_B$ of space. The symmetrized wave function $\psi(x_A,x_B)$ describes then a composite quantum entity which however does not consist of one particle pertaining to the region $R_A$ and another particle pertaining to the region $R_B$, because it also predicts the presence of entanglement correlations between measurements performed in both regions $R_A$ and $R_B$. This entanglement was put into evidence originally by Einstein and two of his students, Boris Podolsky and Nathan Rosen, and the correlations it produces are now called EPR correlations \citep{einsteinpodolskyrosen1935}. The theoretical and experimental study of the EPR type of correlations has been one of the major subjects of quantum theory investigation for the last decades and resulted in showing that these correlations are non-local, so there is no longer any doubt in the physics community that the EPR type of correlations predicted by the entanglement carried in symmetrized states such as (\ref{symmetrystate}) constitute an intrinsic reality in the quantum world even if there is still an ongoing debate about how to understand them \citep{bohm1951,bell1964,bell1987,aertsetal2019a}.

Such a symmetrization for bosons and anti-symmetrization for fermions, following quantum theory, exists for all bosons and all fermions, which literally means that all identical quantum particles are entangled in this strong way, giving rise to non-local correlations of the EPR type. This state of affairs is still nowadays a serious unsolved and not understood conundrum for theoretical physics and philosophy of physics \citep{black1952,vanfraassen1984,frenchredhead1988,saunders2003,saunders2006,mullerseevinck2009,krause2010,diekslubberdink2011,diekslubberdink2019}, and this stands in great contrast with how experimentalists go along with it, for example, photons pertaining to different energy levels, hence carrying different frequencies, are treated by them as distinguishable \citep{hongoumandel1987,knilllaflammemilburn2001,zhaoetal2014}. The way in which experimentalists look at the `indistinguishability' of photons was expressed clearly in more recent times, because of the actual importance of the creation of entangled photons for different reasons, e.g. for the fabrication of optically based quantum computers, and hence the focus in quantum optics on how to achieve this. Spontaneous parametric down conversion, which is a nonlinear optical process that converts one photon of higher energy into a pair of photons of lower energy has been historically the process for the generation of entangled photon pairs for the well-known Bell's inequality tests \citep{aspectdalibardroger1982,weihsetal1998}. Parametric down conversion is however an inefficient process because it has a low probability and hence physicists looked for other ways to produce entangled photons. Hence, when a scheme for using linear optics in function of the needs of the production of qubits was presented \citep{knilllaflammemilburn2001}, this made arise an abundance of new research. Most of the applications of this new research rely on the two-photon interference effect with two `iindistinguishable photons' entering from different sides of a beam splitter and leaving in the same direction after undergoing the so called Hong-Ou-Mandel interference effect \citep{hongoumandel1987}. The crucial aspect of Hong-Ou-Mandel interference
is the `indistinguishability of the two photons in the spectral, temporal and polarization degrees of freedom'. 

This stimulated the direct study of the `indistinguishability of photons from different sources', with the finding that `for photons to behave as indistinguishable bosons neither their frequencies nor their arrival times at the beam splitter can be too different, otherwise they behave as distinguishable quantum particles' \citep{lettowetal2010}. What is however most significant for what concerns our take on this, and its value as support of our conceptuality interpretation of quantum theory \citep{aerts2009a,aerts2010a,aerts2010b,aerts2013,aerts2014,aertsetal2018d,aertsetal2019c}, is the result of an amazing experiment that was performed in the series of attempts of quantum opticians to create entanglement within linear optics by making use of the interference due to two photon indistinguishability. In this experiment, photons of different frequencies are used to enter the beam splitter, hence given earlier experiments \citep{lettowetal2010}, these photons should not behave as indistinguishable bosons, but on the outgoing part of the beam splitter a setup is realized that `erases' the information about the different frequencies of the incoming photons. The result of the experiment is that this erasing makes the photons of different frequencies behave as indistinguishable bosons \citep{zhaoetal2014}. This experiment shows that it is sufficient for the photons to be contextually indistinguishable when they are measured, for them to behave as indistinguishable bosons. We should actually not be amazed by this result, because this is what the so called `quantum eraser experiments' are all about \citep{scullydruhl1982,kimetal2000,walbornetal2002}, and if we carefully read the famous analysis of the double-slit experiment by Richard Feynman \citep{feynmanleightonsands1963,feynman1965}, the dependence of interference on the possibility of the measurement apparatus to `know or not know about the available alternatives', was already at the center of his analysis. Hence, given the above analysis and our conceptuality interpretation of quantum theory,  we can now put forward our view on the issue of `identity and indistinguishability' as follows.

\begin{quotation}
\noindent
{\it The way in which we understand in a straightforward way `what identity and indistinguishability are with respect to human language and human mind' teaches us `what identity and indistinguishability are in quantum theory'.}
 \end{quotation}
Let us formulate the reason why it makes sense to state our view as just expressed above given the conceptuality interpretation of quantum theory. The main hypothesis of the latter is that `the role played by the human mind in relation with language is the same as the role played by a measuring apparatus (but also a heat bath and also a context that is perhaps not willingly used by a human being to make a measurement) in relation with a collection of quantum entities'. The statement above in italics follows directly from this hypothesis.

Let us become more concrete and consider the text of the Winnie the Pooh story of which the words can be found in Table \ref{piglethaffalunmp}. We see that -- and the reasoning we develop now can be made for any other of the considered words -- the word {\it Piglet} corresponds to the cogniton being with energy $E_8$, and it appears 47 times in the text of the story. In the quantum wave function that represents the story, which is a multipartite wave function formed by 2655 parts (the total number of words), {\it Piglet} is the state associated with 47 of its parts, or components. It is straightforward that each of the {\it Piglet} in each of the components can be interchanged with each other of the {\it Piglet} in each other of the components without the story being changed even in the slightest way. This means, in physics jargon, that the wave function is symmetric (or anti-symmetric) with respect to the interchange of all these {\it Piglet} components. And, the symmetry (or anti-symmetry) is a consequence of their `absolute indistinguishability'. 
It is also easy to understand that this `absolute indistinguishability' is due to {\it Piglet} being a concept, and not an object. Indeed, let us imagine for a moment, just to make the above more clear still, that the scenery of the story would be pictured in some physical theatrical form with real piglets on the places where now the concept {\it Piglet} appears in the text. If we interchanged these real piglets, of course this would influence the physical scenery of the story. It is indeed not possible to `interchange a real physical piglet with another real physical piglet without changing the whole of the physical scenery'. That is why real piglets when put in baskets will follow a Maxwell-Boltzmann statistics and not a Bose-Einstein statistics as conceptual piglets do. The `interchanging of concepts in a piece of text', hence in the components of the wave function representing this piece of text, is an intrinsically different operation than the `interchange of objects in space', and the basic hypothesis of the conceptuality interpretation of quantum theory consists in believing that quantum particles are like concepts, and that the reason why we find their behavior not understandable is because we think of them as objects. One of the crucial difficulties when thinking of quantum particles as objects comes to the surface exactly in their behavior as indistinguishable entities, as for objects this is something impossible to understand, while for concepts it is something straightforward and natural. 

Let us show now how we can also easily understand the difference we indicated above between theoretical physicists who are struggling with the issue that, following quantum theory, all photons should be identical, in contrast with experimental physicists who pragmatically consider photons of different frequency as distinguishable and hence not identical. Consider again the Winnie the Pooh story, although we all understand right away that all concepts in the {\it Piglet} state are `absolutely indistinguishable', we also are convinced that two different energy states of the cogniton are distinguishable. For example, energy state $E_{43}$, which is the concept {\it Robin}, appearing 12 times in the text, is distinguishable from, {\it Piglet}. It is even very important for the meaning carried by the story that these two states are distinguishable. In a very similar way, for any measuring apparatus that is sensitive to the frequency of light, it is very important that a red photon is distinguishable from a blue photon, e.g. for our eyes, but also, we suppose, for plants practicing photosynthesis. It is even the `essence of the measuring apparatus' to `distinguish these two states'. However, when a special purpose apparatus is fabricated that, when we would read the Winnie the Pooh story, the points where {\it Piglet} appears are made not distinguishable any longer with the points where {\it Robin} appears -- and there is a multitude of ways we can imagine this to be done -- the two cognitons that are still read by us, will be indistinguishable. Again, such an operation consisting of completely erasing the {\it Piglet} nature and {\it Robin} nature of both concepts, can only work `because both are concepts and not objects'. Underneath all of the words of the Winnie the Pooh text is indeed the more abstract notion of {\it Concept}, and hence we can bring all words into this abstract state of just being an unspecified concept in the text, which would make all of them indistinguishable. There are different ways of `erasing', some ways more close to the ontology of the concepts, other more close to the measuring itself, and that is also why the quantum eraser effect can be understood very well within the conceptuality interpretation (see \citet{aerts2009a} Section 4.4).

Does the above mean that `words in different states are distinguishable' and `words in the same state are indistinguishable' and this clarifies all of the issue? Not yet, let us proceed in refining our analysis. It certainly does not mean that `words in different states are objects', they are concepts, and hence behave like concepts, and not like objects. And since they are concepts, when being in different states, their `distinguishability' is not what `distinguishability' means for objects. We have to return to the main subject of our investigation to find this more subtle form of behavior of words in different states distinguishable as concepts and being at the origin of the disagreements between theoreticians and experimentalists when it comes to consider photons of the same frequency and photons of different frequencies. To start with, let is not forget that the radiation law for photons, including photons of different frequencies, is derived in statistical mechanics by considering these photons to obey Bose-Einstein statistics, and since in the foregoing sections we showed that Bose-Einstein statistics is valid for pieces of texts of stories containing a mixture of distinguishable and indistinguishable words, it should be possible to identify what happens differently with distinguishable concepts as compared to distinguishable objects which can lead to distinguishable concepts obeying Bose-Einstein statistics while distinguishable objects obey Maxwell-Boltzmann statistics. Let us start our analysis considering a very typical and simple situation used commonly to illustrate the difference between Bose-Einstein statistics and Maxwell-Boltzmann statistics.
\begin{figure}
\centering
\includegraphics[width=11cm]{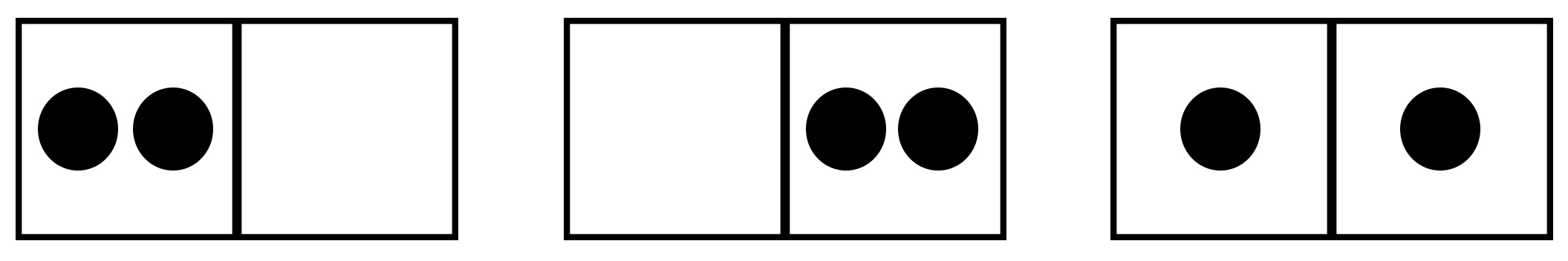}
\caption{Three typical configurations of two particles in two states}
\label{quantumindistinguishability}
\end{figure}
In Figure \ref{quantumindistinguishability} we have represented two particles, the balls, in two states, the boxes, and three different configurations of this situation. The first configuration consists of the two particles in the first state, the second configuration of the two particles in the second state, and the third configuration consists of one particle in one state and the other particle in the other state. If the two particles are indistinguishable in the way that customarily is looked upon quantum indistinguishability, which is also the reason that this example is often displayed, the probabilities that are attached within a Bose-Einstein statistics model are 1/3, 1/3 and 1/3 for each of the configurations. However, if the the two particles are indistinguishable classically, the probabilities that are attached within a Maxwell-Boltzmann statistics are 1/4, 1/4 and 1/2. The reason is that the last configuration of one particle in one state and the other particle in the other state is realized in two ways classically, one way, and its permuted way are different realities. Within the `quantum indistinguishability' these two are not different realities, and given our conceptuality interpretation this would be explained by them indeed not being different realities if they are concepts. What however in case we consider the three configurations of Figure \ref{quantumindistinguishability} for distinguishable states of the cogniton, hence for distinguishable concepts? To make things more concrete, suppose we consider the concepts {\it Cat} and {\it Dog} and the configurations {\it Two Cats}, {\it Two Dogs} and {\it A Cat And A Dog}. Let us remark that this is exactly the situation we have studied already in great detail showing Bose-Einstein statistics to be a better representation as compared to Maxwell-Boltzmann statistics \citep{aerts2009a,aertssozzoveloz2015b,beltran2019}. How can we understand that even for distinguishable concepts Bose-Einstein is a better statistics than Maxwell-Boltzmann? The reason is the presence of `entanglement' and `superposition' also for distinguishable concepts like {\it Cat} and {\it Dog}. Indeed, the probabilities 1/3, 1/3, 1/3 with Bose-Einstein, versus 1/4, 1/4, 1/2 with Maxwell-Boltzmann, actually mean that for Maxwell-Boltzmann there are much more microstates in the third configuration than there are in the first two configurations, actually the double amount. When there is no entanglement and no superposition, and hence {\it Cat} and {\it Dog} are `separated', we can understand this. This `is' what happens when {\it Cat} and {\it Dog} are objects, hence a real cat and a real dog. Let us make this concrete, suppose we visit a farm with a lot of cats and dogs living at the farm, equal in number, and we receive as a present two of them randomly chosen for us by the farmer, then we will have the double chance that the gift will be a cat and a dog as compared to the gift being two cats or two dogs. What however if we ask a child to which it is promised that he or she can have two pets and he or she can choose for each pet whether it is a cat or a dog. The microstates that come into play in this case exist in the conceptual realm of the child's conceptual world, and there is no reason that within this conceptual world there will be a double amount of microstates for the choice of a cat and a dog as compared to the choices for two cats or two dogs. If there are two children that each apart choose one pet and do this independently of each other Maxwell-Boltzmann statistics will be the better one again, because the amount of microstates of the combination of the two choices will be the double of the amount of microstates playing a role for each child apart. This situation was investigated by us in many different and more complex configurations of this type with the result of Bose-Einstein being a better statistics than Maxwell-Boltzmann to model the situation \citep{aerts2009a,aertssozzoveloz2015b,beltran2019}. Actually, we noticed already in our study of quantum entanglement with concept combinations that the violation of Bell's inequalities comes about due to the combined exemplars (microstates) being exemplars of the combined concept directly (giving rise to the Bose-Einstein situation) and not being exemplars of the concepts apart that then afterwards are combined (giving rise to the Maxwell-Boltzmann situation) \citep{aertssozzo2011,aertssozzo2014,aertsetal2019b,aertsetal2019a}. In our investigation of the quantum superposition with concept combinations the situation is even more Bose-Einstein, because the exemplars of the combined concepts that play a role (microstates) are no longer combinations of exemplars of the single concepts, which means that their amount in average will be equal to the amount of exemplars of the single concepts, the situation hence fulfilling the basic requirement to be modeled by Bose-Einstein statistics \citep{aertsgabora2005b,aertsetal2010,aertsetal2012,sozzo2014,aertssozzoveloz2015a,sozzo2015,aertsetal2017}. The insight that also combined distinguishable concepts tend to give rise to Bose-Einstein rather than Maxwell-Boltzmann statistics explains why it is so important for the thermal de Broglie wave-lengths to be large with respect to the distance between the quantum particles, the equivalent for human language always being fulfilled, for the Bose-Einstein statistics to be applicable and why the original Rayleigh Jeans radiation law for light, which is the Maxwell-Boltzmann version of the Planck radiation law, is satisfied for low frequencies.

We have not yet reflected about `identity' in itself. With respect to `the identity' of a quantum particle, it can be proven that when the wave function of two identical quantum particles is considered, there does not exist a self-adjoint operator in the Hilbert space of their states that can represent a measurement that would identify one of the quantum particles \citep{frenchredhead1988,butterfield1993}. Can a concept be said to have an identity? Not in the way we understand identity for an object. What can be attributed to a concept is a `number' indicating `the number of times it is', and that, one could say, is what can be seen as substituting what identity is for an object. The fact that also a `number of times it is' can be attributed to a quantum particle is again a support for the hypothesis of our conceptuality interpretation.

Taking into account our above analysis, what we can understand about the nature of reality goes further than what we have formulated till now, in case we interpret quantum theory following the conceptuality interpretation. Like we mentioned already, we showed in earlier work that `combinations of concepts' give rise to quantum superposition \citep{aertssozzoveloz2015a}. Every sentence in a text is a combination of concepts. Also every paragraph in a text is a combination of concepts, since sentences, as combinations of concepts, combine amongst each others to form paragraphs. Depending on the nature of the text, this process, of increasingly larger pieces of the text being essentially `combinations of concepts', keeps going on, certainly up to the level of stories, where the overall meaning content of a story glues all its concepts together in specific combinations. This implies that superpositions will also form for large subsets of combined concepts, and we believe that this is exactly the mechanism which we call `understanding' when the human mind is engaging in these pieces of text. More concretely, suppose the human mind reads a piece of text. When reading, there is no direct focus on single words as a collection, on the contrary, when the words are read, a `new state is being formed', which integrates `the meaning carried by the combination of all the concerned concepts'. This new state carrying the meaning of the piece of text formed by the combination of these words is exactly the superposition state which we identified already in earlier work \citep{aertssozzoveloz2015a}, and it are these superposition states that form again and again by combining concepts of sentences or paragraphs that again superpose in the course of the reading of the whole text, and lead to the understanding of the whole piece of text. A similar process takes place when talking, thinking or writing, albeit in general in a more discontinuous and complex way than when reading. We believe that what happens with a physical Bose gas close to its Bose-Einstein condensate state can be understood similarly. The role played by the human mind with respect to the text is now played by the heat bath and the measuring apparatuses applied with respect to the Bose gas. When the temperature is low enough and the diluteness of the gas is such that the phase space density (\ref{phasespacedensity}) satisfies (\ref{quantumregime}), hence the thermal de Broglie wave length (\ref{thermaldebroglie02}) is larger than the distance between the atoms, this process of superposition formation starts to happen. Indeed, the de Broglie waves of the different atoms will overlap heavily and give rise to these superpositions, which means that the process which we call `understanding' when the human mind and text are involved takes place in the Bose gas with the heat bath. These superpositions are new emergent states that do not pertain to one of the atoms any longer, but represent several atoms joining in a new entity, just like the several combined concepts represent an emergent meaning. The more the temperature is lowered and the density of the gas is kept such that the de Broglie waves overlap on larger and larger regions of the gas, the more new states are formed containing a synthetic material reality different from single atoms. The Bose-Einstein condensate is an ultimate state where all the atoms have been gathered in the lowest energy state so that for the whole gas a single new state has emerged. The stories that we have studied are in states close to this Bose-Einstein condensate state, where synthetic parts of combined concepts emerge in superposition states and the sizes of these parts are determined by the state of understanding of the human mind of the stories.

\appendix
\appendixpage
\section{The Particle in a Box \label{appendixparticleinabox}}
\begin{figure}
\centering
\includegraphics[width=10cm]{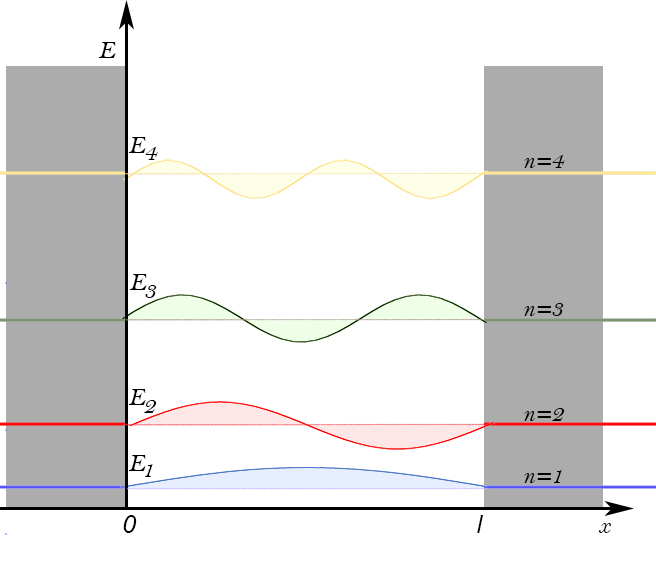}
\caption{A graphical representation of the `particle in a box' as solution of the time independent Schr\"odinger equation with infinite potential well between $0$ and $l$. The wave functions are quantized standing waves inside the box with wave lengths inversely proportional to the width $l$ of the box, and also the energies are quantized in this inversely proportional way, i.e. smaller boxes give rise to larger wave lengths and higher energies. The energy spacings between consecutive quantizations are quadratic in the quantum numbers. We present here the four lowest energy levels.}
\label{particleinabox}
\end{figure}

Schr\"odinger's equation is the fundamental equation of quantum theory and we are specifically interested in its time independent form, because that is the form which gives rise to the quantum eigenstates of the energy, hence states with a predictable fixed energy for a specific energetic situation. How this energetic situation is, we can take inspiration of what we know from classical physics, hence constituting the situation with the energy equal to a part of kinetic energy $K$ plus a part of potential energy $U$, and hence the total energy $E$ is the sum of both
\begin{eqnarray}
E = K + U
\end{eqnarray}
For the specific energetic situation of a `particle in a box', we treat the particle as a free particle as long as it is inside the box, which means that its kinetic energy $K$ equals $p^2/2m$ and the potential energy is a potential which is zero inside the box, and infinite in the region outside of the box. The Schr\"odinger equation `inside the box', where the potential equals zero, becomes the equation for a free particle with mass $m$, hence
\begin{eqnarray} \label{freeparticleschrodingerequation}
-{h^2 \over 8\pi^2m} {d^2 \psi(x) \over dx^2} = E \psi(x)
\end{eqnarray}
which is equivalent to the equation
\begin{eqnarray}
{d^2 \psi(x) \over dx^2} + {8\pi^2mE \over h^2} \psi(x) = 0
\end{eqnarray}
When we put 
\begin{eqnarray} \label{energywavevector}
k^2= {8\pi^2mE \over h^2}
\end{eqnarray}
the Schr\"odinger equation becomes
\begin{eqnarray}
{d^2 \psi(x) \over dx^2} +  k^2 \psi(x) = 0
\end{eqnarray}
which is a second order differential equation of which the general solution is well known
\begin{eqnarray} \label{freeparticlesolution}
\psi(x) = a \sin(kx) + b \cos(kx)
\end{eqnarray}
where $a$ and $b$ are constants, which can be complex numbers, that can be chosen depending on extra conditions to be satisfied. Remark that (\ref{freeparticlesolution}) is the wave function representing a free quantum particle in one dimension because we have not yet expressed in any way the presence of the infinite potential representing the box. Suppose we place the box between $x=0$ and $x=l$, where $l$ is the width of the box as we have shown in Figure \ref{particleinabox}. Hence, this means that at $x = 0$ and $x = l$ we need to have $\psi(0)=\psi(l)=0$, expressing that the walls of the potential representing the box are infinite. Making use of (\ref{freeparticlesolution}) this gives
\begin{eqnarray}
&&0 = \psi(0) = b \Leftrightarrow \psi(x) = a \sin kx \\
&&0 = \psi(l) = a \sin kl \Leftrightarrow \sin kl = 0  \Leftrightarrow k = {n \pi  \over l} \quad n = 1, 2, \ldots
\end{eqnarray}
This means that $k$ is quantized, and the wave functions which are solutions of the Schr\"odinger equation for different quantum numbers $n = 1, 2, \ldots$ are given by
\begin{eqnarray}
\psi_n(x) = a\sin {n \pi \over l} x
\end{eqnarray}

We still have to calculate the value of $a$ by expressing that the probabilities to find the particle at a specific point $x$, given by $|\psi_n(x)|^2$ sums up to 1, hence
\begin{eqnarray} \label{normalization}
1 = a^2 \int_0^l \sin^2 {n \pi \over l} x dx \Leftrightarrow 1 = a^2 {l \over n\pi} \int_0^{n\pi} \sin^2 y dy \Leftrightarrow 1 = a^2 {l \over n\pi}[{y \over 2} - {1 \over 4}\sin 2y]_0^{n\pi} \Leftrightarrow 1 = a^2 {l \over n\pi}{n\pi \over 2} \Leftrightarrow a = \sqrt{{2 \over l}}
\end{eqnarray}
where we make use of $\int sin^2xdx = x/2 - 1/4 \sin 2x$. This gives us the final solution of the Schr\"odinger equation for the particle in the box
\begin{eqnarray}
\psi_n(x) = \sqrt{{2 \over l}} \sin {n \pi \over l} x \quad n = 1, 2, \ldots
\end{eqnarray}
If we use (\ref{energywavevector}) we can calculate the energy of the particle, and see that it is also quantized
\begin{eqnarray}
E_n = {h^2k^2 \over 8\pi^2 m} =n^2  {h^2 \over 8 m l^2} \quad n = 1, 2, \ldots
\end{eqnarray}
We remark that for $n = 1$, hence the lowest energy level, corresponding to the ground state wave function, we have
\begin{eqnarray}
\psi_1(x) =  \sqrt{{2 \over l}} \sin {\pi \over l} x \quad E_1 = {h^2 \over 8 m l^2}
\end{eqnarray}
which means that the energy of the particle is different from zero even in the ground state. This energy is called the `zero point energy', it means that quantum mechanically the particle is unable to `not move', complete lack of motion would indeed violate the Heisenberg uncertainty relations. In Figure \ref{particleinabox} we have represented the energetic situation of the box described by an infinite potential well and drawn the wave functions corresponding to the first four quantum numbers $n = 1, 2, 3$ and $4$. We can see that the wave functions are `standing waves'  that can be imagined to be the wave modes in a string which outer ends are fixed to the walls of the potential well. Remark that the wave lengths and energies are inversely proportional to the width $l$ of the box, i.e. smaller boxes give rise to larger wave lengths and higher energies. This explains some of the differences between the macro-world, where $l$ is large, and hence energies and wave lengths are small, such that no overlapping exists, and typical quantum superposition effects are absent, and the micro-world where energies and wave lengths are large with substantial overlapping such that quantum superposition effects can be abundant \citep{aerts2014}.

\section{The Quantum Harmonic Oscillator \label{appendixquantumharmonicoscillator}}
The potential energy of a harmonic oscillator is traditionally written as follows $U(x) = {1 \over 2}kx^2$
where $k$ is the force constant, which is  is a measure of the stiffness of the spring, in case we realize the harmonic oscillator by means of a particle with a mass attached to a spring. We also can write the potential energy in function of the frequency of the oscillator and the mass of the particle by using that $k = 4\pi^2 \nu^2 m$, and hence the potential energy becomes then $U(x) = 2 \pi^2 \nu^2 x^2$. This gives rise to the following Schr\"odinger equation
\begin{eqnarray} \label{schrodingerharmonicoscillator}
-{h^2 \over 8\pi^2m} {d^2 \psi(x) \over dx^2} + 2\pi^2 \nu^2 x^2 \psi(x) = E \psi(x)
\end{eqnarray}
The `particle in a box' Schr\"odinger equation' which we considered in Appendix \ref{appendixparticleinabox} was easy to solve, and hence we constructed explicitly its solution. The `quantum harmonic oscillator Schr\"odinger equation' is less straight forward to solve and hence we will give its solutions directly. They are again quantized and to write them in a more simple form we introduce
\begin{eqnarray}
\alpha = 4\pi^2 {m \nu \over h}  \quad y = \alpha x
\end{eqnarray}
The general normalized solutions of the Schr\"odinger equation are then
\begin{eqnarray}
\psi_n(y) = ({\alpha \over \pi})^{1 \over 4} {1 \over \sqrt{2^n n!}} H_n(y) e^{-{y^2 \over 2}}
\end{eqnarray}

\begin{figure}
\centering
\includegraphics[width=14cm]{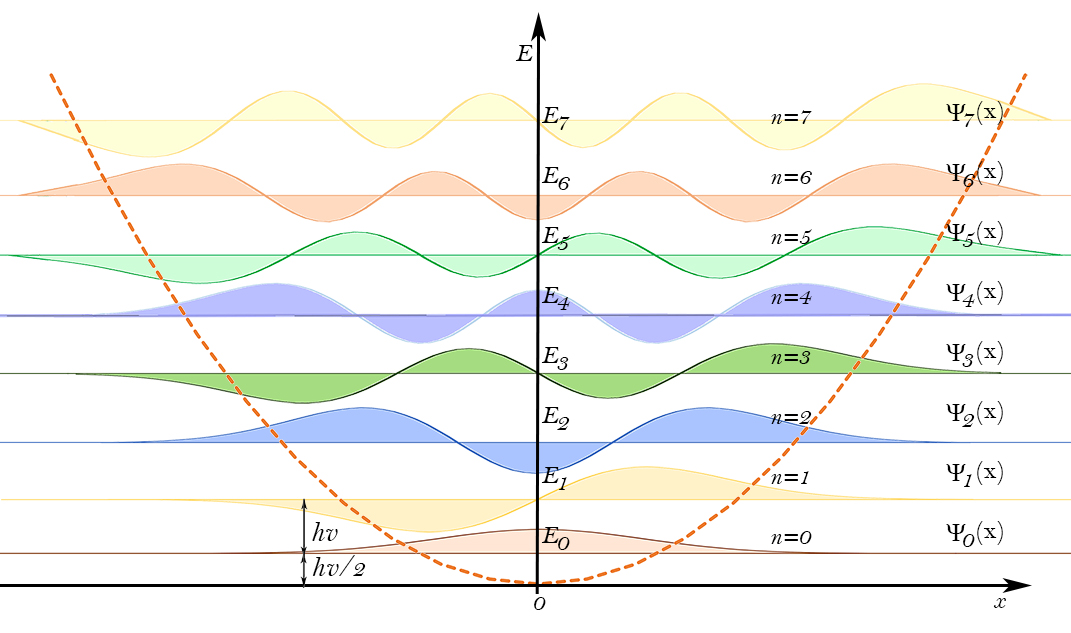}
\caption{A graphical representation of a `quantum harmonic oscillator' as solution of the time independent Schr\"odinger equation with the harmonic oscillator potential. The wave functions are quantized and also the energies are quantized. The energy spacings between consecutive quantizations are linear in the quantum numbers. We present here the seven lowest energy levels.}
\label{harmonicoscillator}
\end{figure} 

where $H_n(y)$ is the Hermite polynomials of grade $n$
\begin{eqnarray}
H_n(y) = (-1)^n e^{y^2} {d^n \over dy^n} (e^{-y^2})
\end{eqnarray}
and hence for the seven lowest energy levels, the ones illustrated in Figure \ref{harmonicoscillator}, these polynomials are the following
\begin{eqnarray}
&&H_0(y) = 1  \quad H_1(y) = 2y \quad H_2(y) = 4y^2-2 \quad H_3(y) = 8y^3 - 12y \\
&&H_4(y) = 16y^4 -48y^2 + 12 \quad H_5(y) = 32y^5 -160y^3 + 120y \\
&&H_6(y) = 64y^6 - 480y^4 + 720y^2 - 120 \quad H_7(y) = 128y^7 - 1344y^5 + 3360y^3 - 1680y
\end{eqnarray}
These solutions of the Schr\"odinger equation lead to a sequence of evenly spaced energy levels characterized by the quantum number n
\begin{eqnarray}
E_n = {h \nu \over 2} + nh\nu
\end{eqnarray}
and, like for the particle in a box, we have a zero point energy different from zero, namely $E_0 = h \nu / 2$. The energy spectrum is reminiscent of the energy spectrum of electromagnet radiation, and indeed, this is a consequence of the traditional way of considering electromagnetic radiation as a collection of harmonic oscillators. The wave functions are essentially Gaussian's multiplied by the Hermite polynomials. Hence, like shown in Figure \ref{harmonicoscillator}, the wave function corresponding to the lowest energy level is a pure Gaussian, since $H_0(y) = 1$, and the higher levels have a positive and negative fluctuating pattern reaching outside of the parabola representing the harmonic oscillator potential due to the presence of the Hermite polynomials.

The harmonic oscillator is one of the foundation situations of quantum theory. Together with the particle in a box, which we presented in Appendix \ref{appendixparticleinabox}, it can be used in many situations as a first approximation, which however gives usually rise to very trustworthy indications for a more sophisticated solution. When a quantum mechanical particle is confined as a consequence of the presence of a macroscopic system, the particle in a box, treating the macroscopic confinement as a box, will serve very well as a first approximation. For complex molecules that interact quantum mechanically the quantum harmonic oscillator will serve very well as a model in the lowest energy levels where the potential is a good approximation for a description of the vibrations that take place as part of the interaction between the molecules. 

\bigskip
\noindent
{\bf Acknowledgement}

\medskip
\noindent
This work was supported by QUARTZ (Quantum Information Access and Retrieval Theory), the Marie
Sklodowska-Curie Innovative Training Network 721321 of the European Union’s Horizon 2020 research
and innovation program. We thanks Massimiliano Sassoli de Bianchi, Sandro Sozzo and Tomas Veloz for their comments to a first version of the present article which helped improve and fine tune the present version.


\begin{thebibliography}{}
\setlength{\itemsep}{-0.27 mm}

\bibitem[Aerts(1986)]{aerts1986} Aerts, D. (1986). A possible explanation for the probabilities of quantum mechanics. {\it Journal of Mathematical Physics 27}, pp. 202-210.

\bibitem[Aerts(1990)]{aerts1990} Aerts, D. (1990). An attempt to imagine parts of the reality of the micro-world. In J. Mizerski, A. Posiewnik, J. Pykacz and M. Zukowski (Eds.), {\it Problems in Quantum Physics (pp. 3-25)}. Singapore: World Scientific.

\bibitem[Aerts(1999)]{aerts1999} Aerts, D. (1999). The stuff the world is made of: physics and reality. In D. Aerts, J. Broekaert and E. Mathijs (Eds.), {\it Einstein meets Magritte: An Interdisciplinary Reflection (pp. 129-183)}. Dordrecht: Springer.

\bibitem[Aerts(2009a)]{aerts2009a} Aerts, D. (2009a). Quantum particles as conceptual entities: A possible explanatory framework for quantum theory. {\it Foundations of Science 14}, pp. 361-411. doi: \url{10.1007/s10699-009-9166-y}.

\bibitem[Aerts(2009b)]{aerts2009b} Aerts, D. (2009b). Quantum structure in cognition. {\it Journal of Mathematical Psychology 53}, pp. 314-348. doi: \url{10.1016/j.jmp.2009.04.005}.

\bibitem[Aerts(2010a)]{aerts2010a} Aerts, D. (2010a). Interpreting quantum particles as conceptual entities. {\it International Journal of Theoretical Physics 49}, pp. 2950-2970. doi: \url{10.1007/s10773-010-0440-0}.

\bibitem[Aerts(2010b)]{aerts2010b} Aerts, D. (2010b). A potentiality and conceptuality interpretation of quantum physics. {\it Philosophica 83}, pp. 15-52. \url{http://www.philosophica.ugent.be/fulltexts/83-2.pdf}.

\bibitem[Aerts(2011)]{aerts2011} Aerts, D. (2011). Quantum interference and superposition in cognition: Development of a theory for the disjunction of concepts. In D. Aerts, J. Broekaert, B. D'Hooghe and N. Note (Eds.), {\it Worldviews, Science and Us: Bridging Knowledge and Its Implications for Our Perspectives of the World}. Singapore: World Scientific.

\bibitem[Aerts(2013)]{aerts2013} Aerts, D. (2013). The quantum mechanics and conceptuality: matter, histories, semantics, and space-time. {\it Scientiae Studia 11}, pp. 75-100. doi: \url{10.1590/S1678-31662013000100004}. English version \url{https://arxiv.org/pdf/1110.4766.pdf}.

\bibitem[Aerts(2014)]{aerts2014} Aerts, D. (2014). Quantum theory and human perception of the macro-world. {\it Frontiers of Psychology 5}, 554. doi: \url{10.3389/fpsyg.2014.00554}.

\bibitem[Aerts et al.(2018a)]{aertsetal2018a} Aerts, D., Aerts Argu\"elles, J., Beltran, L., Beltran, L., Distrito, I., Sassoli de Bianchi, M., Sozzo, S. and Veloz, T. (2018a). Towards a Quantum World Wide Web. {\it Theoretical Computer Science 752}, pp. 116-131. doi: 

\bibitem[Aerts et al.(2017)]{aertsetal2017} Aerts, D., Aerts Argu\"elles, J., Beltran, L., Beltran, L., Sassoli de Bianchi, M., Sozzo, S. and Veloz, T. (2017). Testing quantum models of conjunction fallacy on the World Wide Web. {\it International Journal of Theoretical Physics 56}, pp. 3744-3756. doi: \url{10.1007/s10773-017-3288-8}.

\bibitem[Aerts et al.(2018b)]{aertsetal2018b} Aerts, D., Aerts Argu\"elles, J., Beltran, L., Geriente, S., Sassoli de Bianchi, M., Sozzo, S. and Veloz, T. (2018b). Spin and wind directions I: Identifying entanglement in nature and cognition. {\it Foundations of Science 23}, pp.  323-335. doi: \url{10.1007/s10699-017-9528-9}.

\bibitem[Aerts et al.(2018c)]{aertsetal2018c} Aerts, D., Aerts Argu\"elles, J., Beltran, L., Geriente, S., Sassoli de Bianchi, M., Sozzo, S. and Veloz, T. (2018c). Spin and wind directions II: A Bell State quantum model. {\it Foundations of Science 23}, pp.  337-365. doi: \url{10.1007/s10699-017-9530-2}.

\bibitem[Aerts et al.(2019a)]{aertsetal2019a} Aerts, D., Aerts Argu\"elles, J., Beltran, L., Geriente, S., Sassoli de Bianchi, M., Sozzo, S. and Veloz, T. (2019a). Quantum entanglement in physical and cognitive systems: a conceptual analysis and a general representation. To appear in {\it European Physical Journal Plus}. \url{arXiv:1903.09103}[q-bio.NC].

\bibitem[Aerts(1995)]{aertsaerts1995} Aerts, D. and Aerts, S. (1995). Applications of quantum statistics in psychological studies of decision processes. {\it Foundations of Science 1}, 85--97. doi: \url{10.1007/BF00208726}.

\bibitem[Aerts et al.(2019b)]{aertsetal2019b} Aerts, D., Beltran, L., Geriente, S. and Sozzo, S. (2019b). Quantum theoretic modeling in computer science. A complex Hilbert space model for entangled concepts in corpuses of documents. \url{arXiv:1901.04299}[cs.CL]

\bibitem[Aerts, Broekaert \& Gabora(2011)]{aertsbroekaertgabora2011} Aerts, D., Broekaert, J. and Gabora, L. (2011). A case for applying an abstracted quantum formalism to cognition. {\it New Ideas in Psychology 29}, pp. 136-146. doi: \url{10.1016/j.newideapsych.2010.06.002}.

\bibitem[Aerts et al.(2013)]{aertsetal2013} Aerts, D., Broekaert, J., Gabora, L. and Sozzo, S. (2013). Quantum structure and human thought. {\it Behavioral and Brain Sciences 36}, pp. 274-276. doi: \url{10.1017/S0140525X12002841}.

\bibitem[Aerts et al.(2012)]{aertsetal2012} Aerts, D., Broekaert, J., Gabora, L., Veloz, T. (2012). The guppy effect as interference. In J. R. Busemeyer, F. Dubois, A. Lambert-Mogiliansky and M. Melucci (Eds), {\it Quantum Interaction. QI 2012. Lecture Notes in Computer Science 7620}, pp. 36-47. Springer, Berlin, Heidelberg. doi: \url{10.1007/978-3-642-35659-9_4}.

\bibitem[Aerts \& Czachor(2004)]{aertsczachor2004} Aerts, D. and Czachor, M. (2004). Quantum aspects of semantic analysis and symbolic artificial intelligence. {\it Journal of Physics A: Mathematical and Theoretical 37},  L123-L132.

\bibitem[Aerts et al.(2010)]{aertsetal2010} Aerts, D., Czachor, M., D'Hooghe, B. and Sozzo, S. (2010). The Pet-Fish problem on the World-Wide Web (2010). In the {\it Proceedings of the AAAI Fall Symposium (FS-10-08), Quantum Informatics for Cognitive, Social, and Semantic Processes (pp. 17-21)}. AAAI Publications: 2010 AAAI Fall Symposium Series.

\bibitem[Aerts \& Gabora(2005a)]{aertsgabora2005a} Aerts, D. and Gabora, L. (2005). A theory of concepts and their combinations I: The structure of the sets of contexts and properties. {\it Kybernetes 34}, pp. 167-191. doi: \url{10.1108/03684920510575799}.

\bibitem[Aerts \& Gabora(2005b)]{aertsgabora2005b} Aerts, D and Gabora, L. (2005). A theory of concepts and their combinations II: A Hilbert space representation. {\it Kybernetes 34}, pp.192-221. doi: \url{10.1108/03684920510575807}.

\bibitem[Aerts, Gabora \& Sozzo(2013)]{aertsgaborasozzo2013} Aerts, D., Gabora, L. and Sozzo, S. (2013). Concepts and their dynamics: a quantum theoretic modeling of human thought. {\it Topics in Cognitive Science 5}, pp. 737-772. doi: \url{10.1111/tops.12042}.

\bibitem[Aerts \& Sassoli de Bianchi(2014)]{aertssassolidebianchi2014} Aerts, D. and Sassoli de Bianchi (2014). The extended Bloch representation of quantum mechanics and the hidden-measurement solution to the measurement problem. {\it Annals of Physics, Volume 351}, pp. 975-1025. doi: \url{10.1016/j.aop.2014.09.020}.

\bibitem[Aerts \& Sassoli de Bianchi(2017)]{aertssassolidebianchi2017} Aerts, D. and Sassoli de Bianchi (2017). Do Spins Have Directions? {\it Soft Computing 21}, pp. 1483-1504. doi: \url{10.1007/s00500-015-1913-0}.

\bibitem[Aerts, Sassoli de Bianchi \& Sozzo(2016)]{aertssassolidebianchisozzo2016} Aerts, D., Sassoli de Bianchi, M. and Sozzo, S. (2016). On the foundations of the Brussels operational-realistic approach to cognition. {\it Frontiers in Physics 4}, 17. doi: \url{10.3389/fphy.2016.00017}.

\bibitem[Aerts et al.(2018d)]{aertsetal2018d} Aerts, D., Sassoli de Bianchi, M., Sozzo, S. and Veloz, T. (2018d). On the conceptuality interpretation of quantum and relativity theories. {\it Foundations of Science}. Online First. doi: \url{10.1007/s10699-018-9557-z}.

\bibitem[Aerts et al.(2019c)]{aertsetal2019c} Aerts, D., Sassoli de Bianchi, M., Sozzo, S. and Veloz, T.  (2019c). From quantum axiomatics to quantum conceptuality. {\it Activitas Nervosa Superior 61}, pp. 76-82. doi: \url{10.1007/s41470-019-00030-7}. 

\bibitem[Aerts \& Sozzo(2011)]{aertssozzo2011} Aerts, D. and Sozzo, S. (2011). Quantum structure in cognition: Why and how concepts are entangled. {\it Quantum Interaction. Lecture Notes in Computer Science 7052}, pp. 116-127. doi: \url{10.1007/978-3-642-24971-6\_12}.

\bibitem[Aerts \& Sozzo(2014)]{aertssozzo2014} Aerts, D. and Sozzo, S. (2014). Quantum entanglement in concept combinations. {\it International Journal of Theoretical Physics 53}, pp. 3587-3603. doi: \url{10.1007/s10773-013-1946-z}.

\bibitem[Aerts, Sozzo \& Veloz(2015a)]{aertssozzoveloz2015a} Aerts, D., Sozzo, S. and Veloz, T. (2015a). Quantum structure of negation and conjunction in human thought. {\it Frontiers in Psychology 6}, 1447. doi: \url{10.3389/fpsyg.2015.01447}.

\bibitem[Aerts, Sozzo \& Veloz(2015b)]{aertssozzoveloz2015b} Aerts, D., Sozzo, S. and Veloz, T. (2015b). The quantum nature of identity in human thought: Bose-Einstein statistics for conceptual indistinguishability. {\it International Journal of Theoretical Physics 54}, pp. 4430-4443. doi: \url{10.1007/s10773-015-2620-4}.

\bibitem[Aerts, Sozzo \& Veloz(2016)]{aertssozzoveloz2016} Aerts, D., Sozzo, S., and Veloz, T. (2016). A new fundamental evidence of non-classical structure in the combination of natural concepts. Philosophical Transactions of the Royal Society A. 374, 20150095. doi: \url{10.1098/rsta.2015.0095}.

\bibitem[Aerts Argu\"elles(2018)]{aertsarguelles2018} Aerts Argu\"elles, J. (2018). The heart of an image: Quantum superposition and entanglement in visual perception. {\it Foundations of Science 23}, pp 757-778. doi: \url{10.1007/s10699-018-9547-1}.

\bibitem[Atmanspacher(2002)]{atmanspacher2002} Atmanspacher, H., R\"omer, H. and Walach, H.  (2002). Weak quantum theory: Complementarity and entanglement in physics and beyond. {\it Foundations of Physics 32}, pp. 379-406. doi: \url{10.1023/A:1014809312397}.

\bibitem[Anderson et al.(1995)]{andersonetal1995} Anderson, M. H., Ensher, J. R., Matthews, M. R., Wieman, C. E. and Cornell, E. A. (1995). Observation of Bose-Einstein condensation in a dilute atomic vapor. {\it Science, New Series 269}, pp. 198-201.

\bibitem[Aspect, Dalibard \& Roger(1982)]{aspectdalibardroger1982} Aspect, A., Dalibard, J. and Roger, G. (1982). Experimental test of Bell's inequalities using time-varying analyzers. {\it Physical Review Letters 49}, 1804. doi: \url{10.1103/PhysRevLett.49.1804}.

\bibitem[Bagnato, Pritchard \& Kleppner(1987)]{bagnatopritchardkleppner1987} Bagnato, V., Pritchard, D. E. and Kleppner, D. (1987). Bose-Einstein condensation in an external potential. {\it Physical Review A 35}, 4354. doi: \url{10.1103/PhysRevA.35.4354}.

\bibitem[Bell(1964)]{bell1964} Bell, J. (1964). On the Einstein Podolsky Rosen paradox. {\it Physics 1}, pp. 195-200. doi: \url{10.1103/PhysicsPhysiqueFizika.1.195}.

\bibitem[Bell(1987)]{bell1987} Bell, J. (1987). {\it Speakable and Unspeakable in Quantum Mechanics}. Cambridge, UK: Cambridge University Press.

\bibitem[Beltran(2019)]{beltran2019} Beltran, L. (2019). Quantum nature of statistical behavior of concepts in human language. {\it Special issue Worlds of Entanglement, Foundations of Science}.

\bibitem[Beltran \& Geriente(2019)]{beltrangeriente2019} Beltran, L. and Geriente, S. (2019). Quantum entanglement in corpuses of documents. {\it Foundations of Science 24}, pp. 227-246. doi: \url{10.1007/s10699-018-9570-2}.

\bibitem[Black(1952)]{black1952} Black, M. (1952). The Identity of Indiscernibles. {\it Mind 61}, pp. 153-164.

\bibitem[Blutner \& beim Graben(2016)]{blutnerbeimgraben2016} Blutner, R. and beim Graben, P. (2016). Quantum cognition and bounded rationality. {\it Synthese 193}, pp. 3239-3291. doi: \url{10.1007/s11229-015-0928-5}.

\bibitem[Bohm(1951)]{bohm1951} Bohm, D. (1951). {\it Quantum Theory}. Englewood Cliffs, New Jersey: Prentice-Hall, Inc.

\bibitem[Bose(1924)]{bose1924} Bose, S. N. (1924). Plancks Gesetz und Lichtquantenhypothese. {\it Zeitschrift f\"ur Physik 26}, pp. 178-181. doi: \url{10.1007/BF01327326}. 

\bibitem[Bradley et al.(1995)]{bradleyetal1995} Bradley, C. C., Sackett, C. A., Tollett, J. J. and Hulet R. G. (1995). Evidence of Bose-Einstein condensation in an atomic gas with attractive interactions. {\it Physical Review Letters 75}, 1687. doi: \url{10.1103/PhysRevLett.75.1687}.

\bibitem[Broekaert et al.(2017)]{broekaertetal2017} Broekaert, J., Basieva, I., Blasiak, P. and Pothos, E. M. (2017). Quantum-like dynamics applied to cognition: a consideration of available options. {\it Philosophical Transactions of the Royal Society A 20160387}. doi: \url{10.1098/rsta.2016.0387}.

\bibitem[Bruza \& Cole(2005)]{bruzacole2005} Bruza, P., and Cole, R. (2005). Quantum logic of semantic space: An exploratory investigation of context effects in practical reasoning. In S. Artemov, H. Barringer, A. S. d?Avila Garcez, L. C. Lamb and J. Woods (Eds.), {\it We Will Show Them: Essays in Honour of Dov Gabbay (pp. 339-361)}, volume 1. London: College Publications.

\bibitem[Bruza et al.(2009)]{bruzaetal2009} Bruza, P., Kitto, K., Nelson, D. and McEvoy, C. (2009). Is there something quantum-like about the human mental lexicon? {\it Journal of Mathematical Psychology 53}, pp. 362-377. doi: \url{10.1016/j.jmp.2009.04.004}.

\bibitem[Busemeyer \& Bruza(2012)]{busemeyerbruza2012} Busemeyer, J. and Bruza, P. (2012). {\it Quantum Models of Cognition and Decision}. Cambridge: Cambridge University Press.

\bibitem[Busemeyer \& Wang(2018)]{busemeyerwang2018} Busemeyer, J. R. and Wang, Z. (2018). Hilbert space multidimensional theory. {\it Psychological Review 125}, pp. 572-591. doi: \url{10.1037/rev0000106}.

\bibitem[Busemeyer et al.(2006)]{busemeyeretal2006} Busemeyer, J. R., Wang, Z. and Townsend, J. T. (2006). Quantum dynamics of human decision making. {\it Journal of Mathematical Psychology 50}, pp. 220-241. doi: \url{10.1016/j.jmp.2006.01.003}.

\bibitem[Butterfield(1993)]{butterfield1993} Butterfield, J. (1993). Interpretation and identity in quantum theory, {\it Studies in History and Philosophy of Science 24}, pp. 443-476.

\bibitem[Dalla Chiara et al.(2012)]{dallachiaraetal2012} Dalla Chiara, M. L., Giuntini, R., Luciani, A. R. and Negri, E. (2012). {\it From Quantum Information to Musical Semantics}. Illustrated by C. Seravalli. London: College Publications.

\bibitem[Dalla Chiara et al.(2015)]{dallachiaraetal2015} Dalla Chiara, M. L., Giuntini, R., Leporini, R., Negri, E. and Sergioli, G. (2015). Quantum information, cognition, and music. {\it Frontiers in Psychology  6}, 1583. doi: \url{10.3389/fpsyg.2015.01583}.

\bibitem[Dalfovo et al.(1999)]{dalfovoetal1999} Dalfovo, F., Giorgini, S., Pitaevskii, L. P. and Stringari, S. (1999). Theory of Bose-Einstein condensation in trapped gases. {\it Review of Modern Physics 71}, pp. 463-512. doi: \url{10.1103/RevModPhys.71.463}.

\bibitem[Davis et al.(1995)]{davisetal1995} Davis, K. B., Mewes, M. -O., Andrews, M. R., van Druten, N. J., Durfee, D. S., Kurn D. M., and Ketterle, W. (1995). Bose-Einstein condensation in a gas of sodium atoms. {\it Physical Review Letters 75}, 3969. doi: \url{10.1103/PhysRevLett.75.3969}.

\bibitem[Dieks \& Lubberdink(2011)]{diekslubberdink2011} Dieks, D. and Lubberdink, A. (2011). How classical particles emerge from the quantum world. {\it Foundations of Physics 41}, pp. 1051-1064. doi: \url{10.1007/s10701-010-9515-2}.

\bibitem[Dieks \& Lubberdink(2019)]{diekslubberdink2019} Dieks, D. and Lubberdink, A. (2019). Identical quantum particles as distinguishable objects. \url{arXiv:1902.09280}[quant-ph].

\bibitem[Edmundson(1972)]{edmundson1972} Edmundson, H. P. (1972). The rank hypothesis: A statistical relation between rank and frequency. {\it Technical report TR-186}. College Park: Computer Science Center, University of Maryland.

\bibitem[Einstein(1924)]{einstein1924} Einstein, A. (1924). Quantentheorie des einatomigen idealen Gases. {\it Sitzungsber. phys.-math. Kl. 1924, Gesamtsitzung vom 10. Juli 1924}, pp. 261-267. 

\bibitem[Einstein(1925)]{einstein1925} Einstein, A. (1925). Quantentheorie des einatomigen idealen Gases. {\it Sitzungsberichte der Preussischen Akademie der Wissenschaften, Berlin, Physikalisch-mathematische Klasse, 1925}, pp. 3-14.

\bibitem[Einstein, Podolsky \& Rosen(1935)]{einsteinpodolskyrosen1935} Einstein, A, Podolsky, B. and Rosen, N. (1935). Can quantum-mechanical description of physical reality be considered complete? Physical Review 47, pp. 777-780. doi: \url{10.1103/PhysRev.47.777}.

\bibitem[French \& Redhead(1988)]{frenchredhead1988} French, S. and Redhead, M. (1988). Quantum physics and the identity of indiscernibles. {\it The British Journal for the Philosophy of Science 39}, pp. 233-246. doi: \url{10.1093/bjps/39.2.233}.

\bibitem[Feynman, Leighton \& Sands(1963)]{feynmanleightonsands1963} Feynman, R. P., Leighton, R. B.  and Sands, M. (1963). {\it The Feynman Lectures on Physics, Vol III, Chapter 1}. Reading, Massachusetts: Addison-Wesley.

\bibitem[Feynman(1965)]{feynman1965} Feynman, R. (1965). {\it The Character of Physical Law}. Cambridge, Massachusetts: MIT Press.

\bibitem[Gabora \& Aerts(2002)]{gaboraaerts2002} Gabora, L. and Aerts, D. (2002). Contextualizing concepts using a mathematical generalization of the quantum formalism. {\it Journal of Experimental and Theoretical Artificial Intelligence 14}, pp. 327-358. doi: \url{10.1080/09528130210162253}. 

\bibitem[Gabora \& Kitto(2017)]{gaborakitto2017} Gabora, L. and Kitto, K. (2017). Toward a quantum theory of humor. {\it Frontiers of Physics 4}, 53. doi: \url{10.3389/fphy.2016.00053}.

\bibitem[G\"orlitz et al.(2001)]{gorlitzetal2001} G\"orlitz, A., Vogels,  J. M., Leanhardt, A. E., Raman, C., Gustavson, T. L., Abo-Shaeer, J. R., Chikkatur, A. P., Gupta, S., Inouye, S., Rosenband,  T. and Ketterle, W. (2001). Realization of Bose-Einstein condensates in lower dimensions. {\it Physical Review Letters 87}, 130402. doi: \url{10.1103/PhysRevLett.87.130402}.

\bibitem[Haven \& Khrennikov(2013)]{havenkhrennikov2013} Haven, E. and Khrennikov, A. (2013). {\it Quantum Social Science}. Cambridge:  Cambridge University Press.

\bibitem[Henn at al.(2008)]{hennetal2008} Henn, E. A. L., Seman, J. A., Seco, G. B., Olimpio, E. P., Castilho, P., Roati, G., Magalhaes, D. V., Magalhaes, K. M. F. and Bagnato, V. S. (2008). Bose-Einstein condensation in 87Rb: Characterization of the Brazilian experiment. {\it Brazilian Journal of Physics 38}, pp. 279-286. doi: \url{10.1590/S0103-97332008000200012}.

\bibitem[Hong, Ou \& Mandel(1987)]{hongoumandel1987} Hong, C. K., Ou, Z.Y. and Mandel, L. (1987). Measurement of subpicosecond time intervals between two photons by interference. {\it Physical Review Letters 59}, 2044. doi: \url{10.1103/PhysRevLett.59.2044}.

\bibitem[Huang(1987)]{huang1987} Huang, K. (1987). {\it Statistical Mechanics}. New York: Wiley.

\bibitem[Ketterle, Durfee \& Stamper-Kum(1999)]{ketterledurfeestamper-kum1999} Ketterle W., Durfee D.S. and Stamper-Kurn D.M. (1999). Making, probing and understanding Bose-Einstein condensates. In M. Inguscio, S. Stringari and C. E. Wieman (Eds.), {\it Bose-Einstein Condensation in Atomic Gases, Proceedings of the International School of Physics `Enrico Fermi', Course CXL, 12  (pp. 67- 176)}. Amsterdam: IOS Press. doi: \url{10.3254/978-1-61499-225-7-67}.

\bibitem[Ketterle \& van Druten(1996)]{ketterleandvandruten1996} Ketterle, W. and van Druten, N. J. (1996). Bose-Einstein condensation of a finite number of particles trapped in one or three dimensions. {\it Physical Review A 54}, 656. doi: \url{10.1103/PhysRevA.54.656}.

\bibitem[Khrennikov(1999)]{khrennikov1999} Khrennikov, A.(1999). Classical and quantum mechanics on information spaces with applications to cognitive, psychological, social and anomalous phenomena. {\it Foundations of Physics 29}, pp. 1065-1098. doi: \url{10.1023/A:1018885632116}.

\bibitem[Kim et al.(2000)]{kimetal2000} Kim, Yoon-Ho, Yu, R., Kulik, S.P., Shih, Y.H. and Scully, M. O. (2000). Delayed `choice' quantum eraser. {\it Physical Review Letters 84}, pp. 1-5. doi: \url{10.1103/PhysRevLett.84.1}.

\bibitem[Klaers et al.(2011)]{klaersetal2011} Klaers, J., Schmitt, J., Damm, T., Vewinger, F. and Weitz, M. (2011). Bose-Einstein condensation of paraxial light. {\it Applied Physics B 105}, pp. 17-33. doi: \url{10.1007/s00340-011-4734-6}.

\bibitem[Klaers, Verwinger \& Weitz(2010a)]{klaersverwingerweitz2010a} Klaers, J., Vewinger, F. and Weitz, M. (2010a). Thermalization of a two-dimensional photonic gas in a `white wall? photon box. {\it Nature Physics 6}, pp. 512-515. doi: 

\bibitem[Klaers, Verwinger \& Weitz(2010b)]{klaersverwingerweitz2010b} Klaers, J., Vewinger, F. and Weitz, M. (2010b). Bose-Einstein condensation of photons in an optical microcavity. {\it Nature 468}, pp. 545-548. doi: \url{10.1038/nature09567}.

\bibitem[Klaers \& Weitz(2013)]{klaesweitz2013} Klaers, J. and Weitz, M. Bose-Einstein condensation of photons. In Karl-Heinz Bennemann and John B. Ketterson (Eds.), {\it Novel Superfluids}. Oxford, UK: Oxford University Press.

\bibitem[Knill, Laflamme \& Milburn(2001)]{knilllaflammemilburn2001} Knill, E., Laflamme, R. and Milburn, G. J. (2001). A scheme for efficient quantum computation with linear optics. {\it Nature 409}, pp. 46-52.

\bibitem[Krause(2010)]{krause2010} Krause, D. (2010). Logical aspects of quantum (non-)individuality. {\it Foundations of Science 15}, pp. 79-94.

\bibitem[Lambert Mogilianski Zamir \& Zwirn(2009)]{lambertmoglianskizamirzwirn2009} Lambert Mogiliansky, A., Zamir, S. and Zwirn, H. (2009). Type indeterminacy: A model of the KT (Kahneman-Tversky)-man. {\it Journal of Mathematical Psychology 53}, 349--361.

\bibitem[Lettow et al.(2010)]{lettowetal2010} Lettow, R., Rezus, Y. L. A., Renn, A., Zumofen, G.,  Ikonen, E.,  G\"otzinger, S. and Sandoghdar, V. (2010). Quantum interference of tunably indistinguishable photons from remote organic molecules. {\it Physical Review Letters 104}, 123605. doi: \url{10.1103/PhysRevLett.104.123605}.

\bibitem[Mandelbrot(1953)]{mandelbroth1953} Mandelbroth, B. (1953). An Informational Theory of the Statistical Structure of Language. In J. Willis (Eds.), {\it Communication Theory: Papers Read at a Symposium on `Applications of Communication Theory' (pp. 486-502)}. London: Butterworths.

\bibitem[Mandelbrot(1954)]{mandelbroth1954} Mandelbrot, B. (1954). Structure formelle des textes et communication. {\it Word 10}, pp. 1-27, pp. 424-425. doi: \url{10.1080/00437956.1954.11659509}.

\bibitem[Marte et al.(2002)]{marteetal2002} Marte, A., Volz, T., Schuster, J., D\"urr, S., Rempe, G., van Kempen, E.G.M. and Verhaar, B. J. (2002). Feshbach resonances in rubidium 87: Precision measurement and analysis. {\it Physical Review Letters 89}, 283202. doi: \url{10.1103/PhysRevLett.89.283202}.

\bibitem[Melucci(2015)]{melucci2015} Melucci, M. (2015). {\it Introduction to Information Retrieval and Quantum Mechanics}. Berlin Heidelberg: Springer. 

\bibitem[Milne(1926)]{milne1926} Milne, A. A. (1926). {\it Winnie-the-Pooh}. London: Methuen \& Co. Ltd.

\bibitem[Moreira \& Wichert(2016)]{moreirawichert2016} Moreira, C. and Wichert, A. (2016). Quantum probabilistic models revisited: the case of disjunction effects in cognition. {\it Frontiers in Physics 4}. 26. doi: \url{10.3389/fphy.2016.00026}.

\bibitem[Muller \& Seevinck(2009)]{mullerseevinck2009} Muller, F. A. and Seevinck, M. (2009). Discerning elementary particles. {\it Philosophy of Science 76}, pp. 179-200.

\bibitem[Parkins \& Walls(1998)]{parkinsandwalls1998} Parkins, A.S. and Walls, D.F. (1998) The physics of trapped dilute-gas Bose-Einstein condensates. {\it Physics Reports 303}, pp. 1-80. doi: \url{10.1016/S0370-1573(98)00014-3}.

\bibitem[Pothos \& Busemeyer(2009)]{pothosbusemeyer2009} Pothos, E. and Busemeyer, J. (2009). A quantum probability explanation for violations of `rational' decision theory. {\it Proceedings of the Royal Society of London B: Biological Sciences 276}, pp. 2171-2178. doi: \url{10.1098/rspb.2009.0121}.

\bibitem[Pothos et al.(2015)]{pothosetal2015} Pothos, E. M., Barque-Duran, A., Yearsley, J. M., Trueblood, J. S., Busemeyer, J. R. and Hampton, J. A. (2015). Progress and current challenges with the quantum similarity model. {\it Frontiers in Psychology 6}, 205. doi: \url{10.3389/fpsyg.2015.00205}.

\bibitem[Sassoli de Bianchi(2011)]{sassolidebianchi2011} Sassoli de Bianchi (2011). Ephemeral properties and the illusion of microscopic particles. {\it Foundations of Science 16}, pp. 393-409. doi: \url{10.1007/s10699-011-9227-x}.

\bibitem[Sassoli de Bianchi(2013)]{sassolidebianchi2013} Sassoli de Bianchi, M. (2013). Quantum dice. {\it Annals of Physics 336}, pp. 56-75. doi: \url{10.1016/j.aop.2013.05.018}.

\bibitem[Sassoli de Bianchi(2014)]{sassolidebianchi2014} Sassoli de Bianchi, M. (2014). A remark on the role of indeterminism and non-locality in the violation of Bell's inequality. {\it Annals of Physics 342}, pp. 133-142. doi: \url{10.1016/j.aop.2013.12.011}.

\bibitem[Sassoli de Bianchi(2019)]{sassolidebianchi2019}
Sassoli de Bianchi, M. (2019). A non-spatial reality. To be published in {\it Foundations of Science}.

\bibitem[Saunders(2003)]{saunders2003} Saunders, S. (2003). Physics and Leibniz?s principles. In K. Brading and E. Castellani (Eds.), {\it Symmetries in Physics: Philosophical Reflections}. Cambridge, UK: Cambridge University Press.

\bibitem[Saunders(2006)]{saunders2006} Saunders, S. (2006). Are quantum particles objects? {\it Analysis 66}, pp. 52-63.

\bibitem[Scully \& Druhl(1982)]{scullydruhl1982} Scully, M. O., Druhl, K. (1982). Quantum eraser: a proposed photon correlation experiment concerning observation and `delayed choice' in quantum mechanics. {\it Physical Review A 25}, 2208. doi: \url{10.1103/PhysRevA.25.2208}.

\bibitem[Sozzo(2014)]{sozzo2014} Sozzo, S. (2014). A quantum probability explanation in Fock space for borderline contradictions. Journal of Mathematical Psychology 58, 2014, pp. 1-12. doi: \url{10.1016/j.jmp.2013.11.001}.

\bibitem[Sozzo(2015)]{sozzo2015} Sozzo, S. (2015). Conjunction and negation of natural concepts: A quantum-theoretic modeling. {\it Journal of Mathematical Psychology 66}, pp. 83-102. doi: \url{10.1016/j.jmp.2015.01.005}.

\bibitem[Sozzo(2017)]{sozzo2017} Sozzo, S. (2017). Effectiveness of the quantum-mechanical formalism in cognitive modeling. {\it Soft Computing 21}, pp. 1455-1465. doi: \url{10.1007/s00500-015-1834-y}.

\bibitem[Sozzo(2019)]{sozzo2019} Sozzo, S. (2019). Explaining versus describing human decisions: Hilbert space structures in decision theory. {\it Soft Computing}. Online First. doi: \url{10.1007/s00500-019-04140-x}.

\bibitem[Swift(1726)]{swift1726} Swift, J. (1726). {\it Travels into several Remote Nations of the World. In four parts. By Lemuel Gulliver, first a surgeon, and then a captain of several ships}. London : Benj. Motte.

\bibitem[Van Fraassen(1984)]{vanfraassen1984} Van Fraassen, B. (1984). The problem of indistinguishable particles. In J. T. Cushing, C. F. Delaney, and G. M. Gutting (Eds.), {\it Science and Reality: Recent Work in the Philosophy of Science: Essays in Honor of Erman McMullin (pp. 153-172)}. Notre Dame: University Notre Dame Press.

\bibitem[Van Rijsbergen(2004)]{vanrijsbergen2004} van Rijsbergen, C. J. (2004). {\it The Geometry of Information Retrieval}. Cambridge: Cambridge University Press.

\bibitem[Veloz \& Dujardin(2015)]{velozdujardin2015} Veloz, T and Desjardins, S. (2015). Unitary transformations in the quantum model for conceptual conjunctions and its application to data representation. {\it Frontiers In Psychology 6}, 1734. doi: \url{10.3389/fpsyg.2015.01734}.

\bibitem[Veloz, Zhao \& Aerts(2013)]{velozzhaoaerts2014}  Veloz, T., Zhao, X. and Aerts, D. (2014). Measuring conceptual entanglement in collections of documents. In H. Atmanspacher, E. Haven, K. Kitto and D. Raine, (Eds), {\it Quantum Interaction. QI 2013. Lecture Notes in Computer Science 8369}, pp. 134-146. Berlin, Heidelberg: Springer.

\bibitem[Walborn et al.(2002)]{walbornetal2002} Walborn, S. P., Terra Cunha, M. O., P\'adua, S. and Monken, C. H. (2002). Double-slit quantum eraser. {\it Physical Review A 65}, 033818. doi: \url{10.1103/PhysRevA.65.033818}.

\bibitem[Weihs et al.(1998)]{weihsetal1998} Weihs, G., Jennewein, T., Simon, C., Weinfurter, H. and Zeilinger, A. (1998). Violation of Bell's inequality under strict Einstein locality conditions. {\it Physical Review Letters 81}, 5039. doi: \url{10.1103/PhysRevLett.81.5039}. 

\bibitem[Wells(1903)]{wells1903} Wells, H. G. (1903). The magic shop. {\it The Strand Magazine}. London: George Newnes.

\bibitem[Widdows(2004)]{widdows2004} Widdows, D. (2004). {\it Geometry and Meaning}. Stanford: CSLI publications.

\bibitem[Zhao et al.(2014)]{zhaoetal2014} Zhao,T-M., Zhang,H., Yang,J., Sang, Z-R., Jiang, X., Bao, X-H. and Pan, J-W. (2014). Entangling different-color photons via time-resolved measurement and active feed forward. {\it Physical Review Letters 112}, 103602. doi: \url{10.1103/PhysRevLett.112.103602}.

\bibitem[Zipf(1935)]{zipf1935} Zipf, G. K. (1935). {\it The Psycho-Biology of Language}. Boston: Houghton Mifflin Co.

\bibitem[Zipf(1949)]{zipf1949} Zipf, G. K. (1949). {\it Human Behavior and the Principle of Least Effort}. Cambridge: Addison Wesley.

\end{thebibliography}
\end{document}